\documentclass[manuscript,screen,nonacm]{acmart}

\usepackage{enumitem}
\usepackage{gensymb}
\usepackage{multirow}
\usepackage{prettyref}
\usepackage[labelfont=bf,format=plain,singlelinecheck=false]{caption}
\usepackage{subcaption}
\usepackage{bm,microtype}
\usepackage{amsmath}
\usepackage{caption}
\usepackage[percent]{overpic}
\usepackage{booktabs}

\definecolor{myblue}{RGB}{31, 119, 180}
\definecolor{lightblue}{RGB}{222,235,247}
\definecolor{myorange}{RGB}{255, 127, 14}
\definecolor{mygreen}{RGB}{44, 160, 44}
\definecolor{myred}{RGB}{214, 39, 40}
\definecolor{mypurple}{RGB}{122,1,119}

\usepackage{soul}
\AtBeginDocument{%
  }

\usepackage[boxed,ruled,linesnumbered]{algorithm2e} 
\SetAlFnt{\small}
\SetAlCapFnt{\small}
\SetAlCapNameFnt{\small}
\SetAlCapHSkip{0pt}

\SetCommentSty{mycommfont}

\acmJournal{TOG}

\usepackage{tikz}
\usetikzlibrary{angles, calc}
\tikzset{
  dot/.style={
    circle, fill=black, inner sep=1pt, outer sep=0pt
  },
  dot label/.style={
    circle, inner sep=0pt, outer sep=1pt
  },
  pics/right angle/.append style={
    /tikz/draw, /tikz/angle radius=5pt
  }
}
\usepackage{xcolor}
\usepackage{tikzducks}
\usepackage{duckuments}

\usepackage[noend]{algpseudocode}

\usepackage{prettyref}
\newrefformat{fig}{Figure~\ref{#1}}
\newrefformat{par}{Section~\ref{#1}}
\newrefformat{appen}{Appendix~\ref{#1}}
\newrefformat{sec}{Section~\ref{#1}}
\newrefformat{sub}{Section~\ref{#1}}
\newrefformat{table}{Table~\ref{#1}}
\newrefformat{ass}{Assumption~\ref{#1}}
\newrefformat{alg}{Algorithm~\ref{#1}}
\newrefformat{def}{Definition~\ref{#1}}
\newrefformat{thm}{Theorem~\ref{#1}}
\newrefformat{cor}{Corollary~\ref{#1}}
\newrefformat{lem}{Lemma~\ref{#1}}
\newrefformat{step}{Step~\ref{#1}}
\newrefformat{ln}{Line~\ref{#1}}
\newrefformat{rem}{Remark~\ref{#1}}
\newrefformat{eq}{Equation~\ref{#1}}
\newrefformat{pb}{Problem~\ref{#1}}
\newrefformat{it}{Item~\ref{#1}}
\newrefformat{te}{Term~\ref{#1}}
\def\Eqref Eq:#1:{\eqref{eq:#1}}
\newrefformat{Eq}{Equation~\Eqref#1:}

\usepackage{rotating}

\newcommand{\GG}{\mathbf{G}}

\newcommand\restr[2]{{\left.\kern-\nulldelimiterspace{}#1\right|_{#2}}}

\usepackage{svg}
\usepackage{xcolor}
\usepackage{wrapfig}
\usepackage[export]{adjustbox}
\usepackage{colortbl}
\usepackage{makecell}

\usepackage{tikz}
\usetikzlibrary{angles, calc}
\tikzset{
  dot/.style={
    circle, fill=black, inner sep=1pt, outer sep=0pt
  },
  dot label/.style={
    circle, inner sep=0pt, outer sep=1pt
  },
  pics/right angle/.append style={
    /tikz/draw, /tikz/angle radius=5pt
  }
}

\definecolor{smGrey}{rgb}{0.8, 0.8, 0.8}  

\definecolor{smOrange}{rgb}{0.0, 0.0, 0.0}  
\definecolor{smBlue}{rgb}{0.0, 0.0, 0.0}  

\begin{document}

\title{Quantized High-Performance High-Order Moment-Encoded LBM}

\author{Yixin Chen}
\email{yixinc.chen@mail.utoronto.ca}
\orcid{0000-0001-7547-9587}
\affiliation{
    \institution{University of Toronto}
    \city{Toronto}
    \country{Canada}
}

\author{Wei Li}
\authornote{Corresponding author.}
\email{1104720604wei@gmail.com}
\orcid{}
\affiliation{
    \institution{Shanghai Jiao Tong University}
    \city{Shanghai}
    \country{China}
}

\author{David I.W. Levin}
\orcid{0000-0001-7079-1934}
\affiliation{%
 \institution{University of Toronto}
 \city{Toronto}
 \state{ON}
 \country{Canada}
}
\affiliation{%
 \institution{Nvidia}
 \city{Toronto}
 \state{ON}
 \country{Canada}
}
\email{diwlevin@cs.toronto.edu}

\author{Kui Wu}
\email{walker.kui.wu@gmail.com}
\orcid{}
\affiliation{
    \institution{LIGHTSPEED}
    \city{Los Angeles}
    \country{USA}
}

\begin{abstract}

Lattice Boltzmann methods (LBM) are widely used for visually rich fluid simulation, but their high memory bandwidth and storage requirements limit GPU performance and scalability, especially when coupled with complex geometry. \textcolor{smOrange}{Recent high-order moment-encoded LBM (HOME-LBM) reduces memory cost by storing compact velocity moments and performing collision directly in moment space, but its tightly coupled boundary treatment introduces divergent control flow.} We present a high-performance, quantized HOME-LBM framework for large-scale turbulent fluid--solid simulations on a single GPU. Our geometry-decoupled computation scheme separates the regular fluid update from a sparse solid-correction stage, isolating geometry-dependent computations to improve GPU efficiency. To further reduce storage and memory traffic, we develop a reduced-precision representation for HOME-LBM by combining von Neumann stability analysis, empirical moment-range estimation, and runtime decomposition of second-order moments to control quantization error. We evaluate our approach on real-time demonstrations and large-scale examples using both D3Q19 and D3Q27 lattices, with additional 2D and 3D numerical experiments assessing accuracy and stability. \textcolor{smOrange}{Across the tested benchmarks, our framework substantially accelerates the original FP32 HOME-LBM implementation, with a maximum speedup of $32.5\times$ on the Hilbert-curve geometry. 
Moment storage is reduced by $50\%$ in fluid-only simulations and $25\%$ in fluid--solid coupling examples, while maintaining numerical stability, fidelity, and high visual quality. 
In the static delta-wing benchmark, our solver achieves a $5.2\times$ speedup over Leapfrog Flow Maps (LFM) under the tested configuration.}

\end{abstract}

\ccsdesc[500]{Computing methodologies~Physical simulation}

%
%

\keywords{fluid simulation; lattice Boltzmann method; GPU computing; quantization; fluid--solid coupling}


\begin{teaserfigure}
\centering
\includegraphics[width=\linewidth,clip,trim= 70 230 40 270]{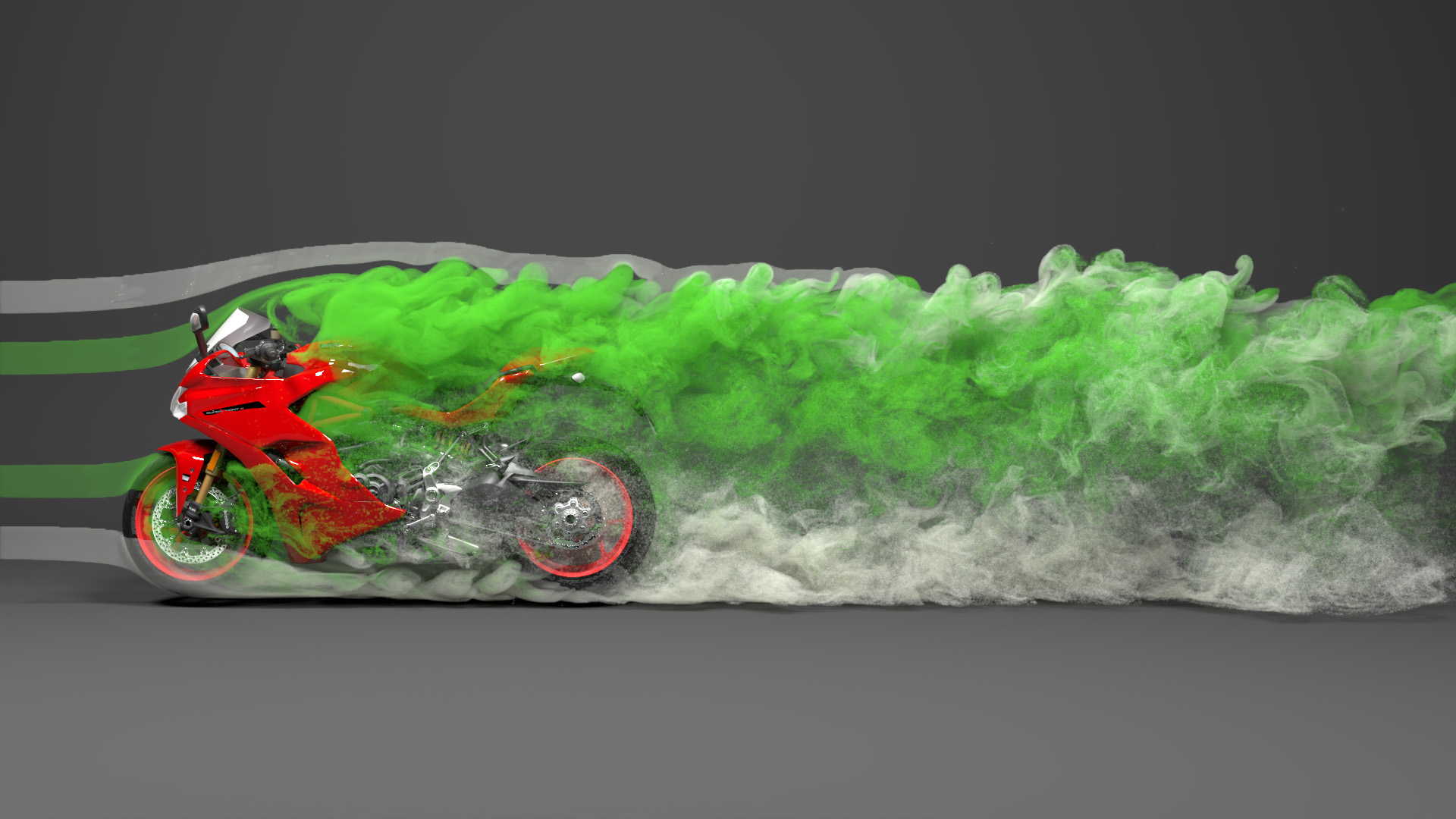}
\caption{
Turbulent smoke interacting with a Ducati motorcycle model with intricate geometry on a $1000\times400\times400$ grid using our GPU-optimized, 16-bit quantized LBM solver. The simulation preserves rich fine-scale vortical structures and remains numerically stable under highly complex boundary conditions. \textcolor{smOrange}{For this scene, our method achieves a $9.0\times$ speedup over the original FP32 HOME-LBM~\cite{li2023high} implementation, while reducing memory usage by $25\%$}.}
\Description{}
\label{fig:teaser}
\end{teaserfigure}


\maketitle
\section{Introduction}
\label{sec:intro}

\begin{figure*}[t]
\centering
\includegraphics[width=\linewidth]{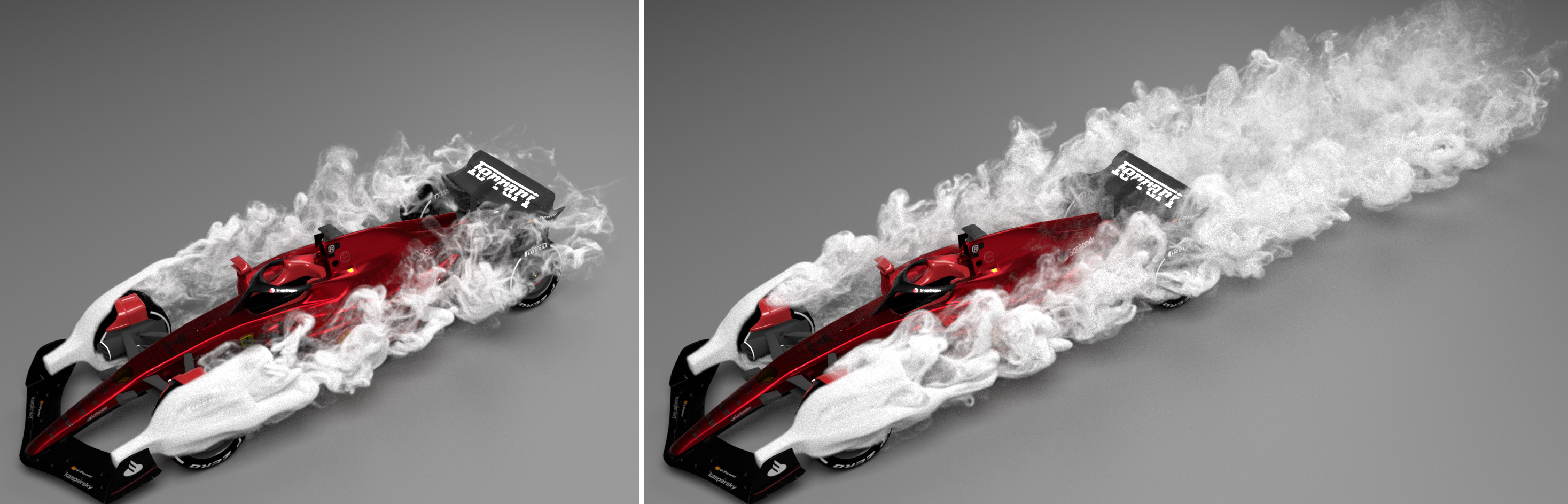}
\caption{Smoke around an F1 car on a $1000 \times 400 \times 400$ grid (\textcolor{smBlue}{left: frame 80; right: frame 160}) highlights a complex, multi-scale wake: strong shear layers from the wheels, rear wing, and diffuser roll up into coherent vortices near the vehicle, then stretch and break down into a turbulent, filamentary plume downstream.}
\label{fig:highres_f1}
\Description{}
\end{figure*}

Fluid simulation plays a central role in computer graphics, enabling the realistic synthesis of natural phenomena such as wind, smoke, and fire. Beyond visual fidelity, high-performance fluid simulation has become increasingly important for downstream applications, including shape design~\cite{Ma2021}, robotic control in virtual environments~\cite{Ma2018,xian2023fluidlab,hu2025learning,Song2025,Liu2022}, and inverse optimization problems~\cite{Du2020,Li2022,chen2024fluid}.

Stable Fluids~\cite{Stam1999}, which solves the Navier--Stokes (NS) equations on Eulerian grids, remains widely used in real-time applications because of its simplicity, efficiency, and stability. While subsequent work has improved computational efficiency and better preserved fine-scale details, achieving stable, high-resolution simulations under strict real-time constraints remains challenging in practice. Grid-based solvers typically rely on global pressure solves and therefore can incur substantial synchronization and communication overhead across the grid. Particle-based methods, by contrast, often require expensive neighbor searches, irregular memory access patterns, and dynamic data structures, all of which reduce throughput and GPU utilization. Together, these bottlenecks limit scalability on modern GPUs and worsen as simulation resolution and geometric complexity grow.
In contrast to these macroscopic solvers, the lattice Boltzmann method (LBM) has emerged as a compelling alternative for turbulent, weakly compressible flows because of its massive parallelism and strong vortex-preserving properties. By operating locally on discretized lattice nodes at the mesoscopic scale, LBM is well suited to GPU acceleration. However, standard LBM schemes impose substantial memory overhead: a conventional double-buffered D3Q27 implementation stores two sets of $27$ floating-point distribution functions per node, leading to heavy memory traffic and limited scalability.

To address this limitation, \citet{li2023high} proposed the high-order moment-encoded LBM (HOME-LBM), which stores a compact set of velocity moments instead of the full lattice distribution functions and performs a lightweight collision operator directly in moment space. This formulation reduces memory and bandwidth requirements and has been shown to improve robustness in turbulent-flow simulations. Nevertheless, several practical bottlenecks remain. Its fluid–solid boundary treatment still relies on costly link–solid intersection tests tightly coupled with the streaming operation, leading to irregular control flow and reduced GPU utilization. Furthermore, although quantization has proven effective in reducing memory footprint and bandwidth in large-scale simulations~\cite{Hu2021QuanTaichi,Liu2022autoquantization}, designing a robust reduced-precision representation for LBM remains nontrivial due to its stability constraints.

To accelerate HOME-LBM with complex fluid--solid interactions and reduce memory consumption for large-scale simulations on a single GPU, we propose a high-performance quantized HOME-LBM framework. Our approach consists of two main components. First, we analyze the computational bottlenecks of the original HOME-LBM on GPUs when handling interactions with solid objects, and then present a geometry-decoupled computation scheme for efficient GPU parallelization. Second, to further reduce memory cost, \textcolor{smOrange}{we develop a 16-bit fixed-point quantization for moment storage (\autoref{sec:stability_quantization})}. We characterize the linear stability of \textcolor{smOrange}{the unquantized moment-encoded updates on D2Q9 and D3Q27 lattices} through von Neumann analysis and combine these results with empirical moment statistics to guide quantization-range selection. For second-order moments, we store and quantize only the non-equilibrium components and reconstruct the equilibrium contributions analytically, narrowing the storage dynamic range.

We validate our approach through real-time demonstrations and large-scale benchmarks under both D3Q19 and D3Q27 configurations, \textcolor{smOrange}{as well as 2D and 3D numerical experiments assessing accuracy and stability. 
At a fixed fluid-grid resolution, mesh-complexity experiments isolate the performance contributions of geometry decoupling and moment quantization. Compared with the original FP32 HOME-LBM, our method achieves $4.1\times$ to $32.5\times$ speedups across the tested configurations (\autoref{table:timing_memory}), with the maximum on the Hilbert-curve benchmark (\autoref{fig:hilbert_3d}). Moment-buffer storage is reduced by $50\%$ in fluid-only simulations and $25\%$ in fluid--solid coupling, where one buffer remains in FP32 for boundary-correction accumulation. On the static delta-wing benchmark, our solver achieves a $5.2\times$ speedup over Leapfrog Flow Maps (LFM)~\cite{Sun2025Leapfrog} under the tested configuration (\autoref{fig:deltawing_comparison}). Large-scale examples, including $1024^3$ fluid-only simulations and fluid--solid interactions involving multimillion-triangle geometries, demonstrate improved scalability on a single GPU while preserving numerical stability, fidelity, and high visual quality.
}

\section{Related Work}
\label{sec:related_work}

\begin{figure*}[ht!]
\centering
\includegraphics[width=0.5\linewidth,clip,trim=13cm 1.cm 15cm 0cm]{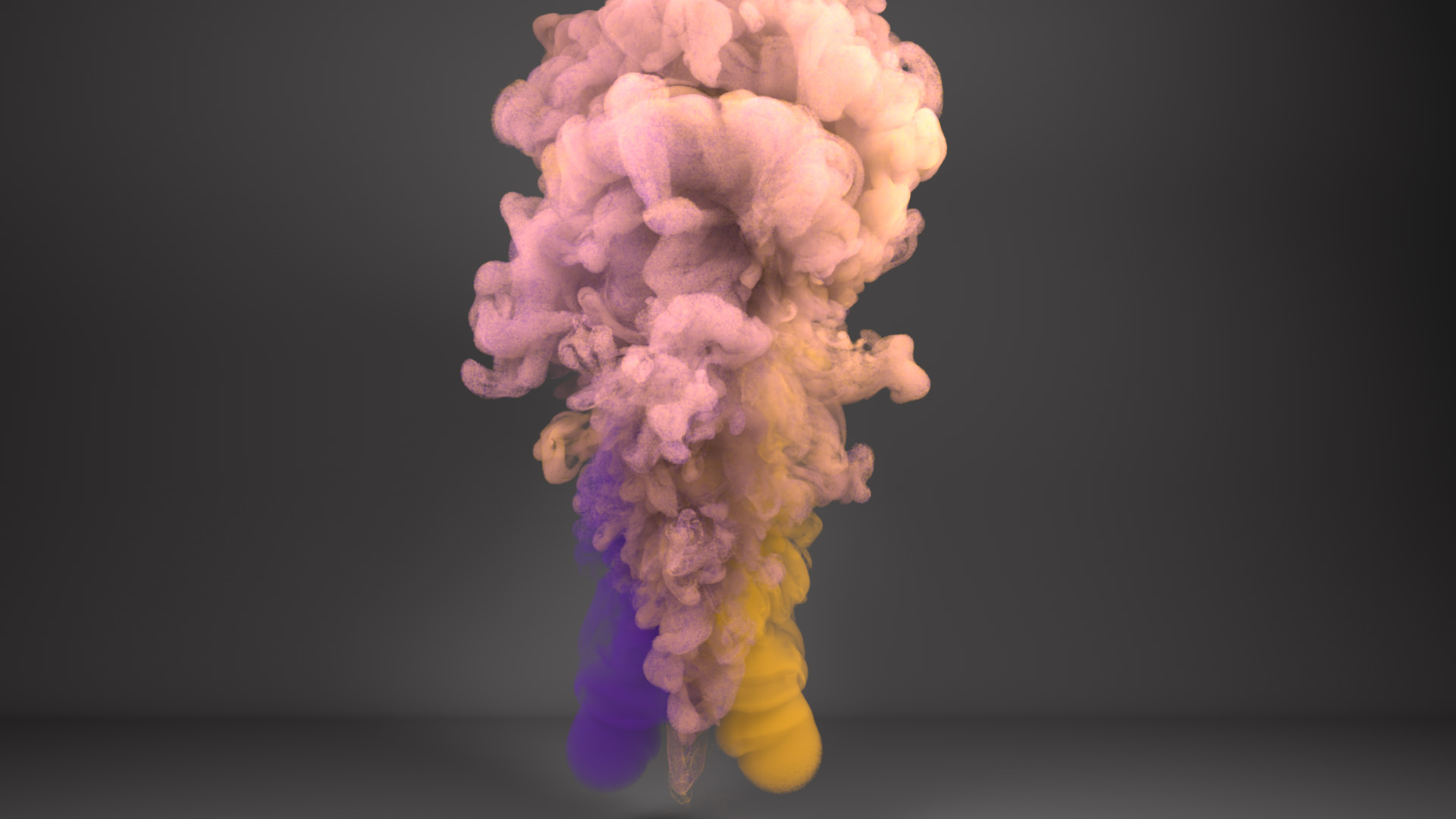}\hfill
\includegraphics[width=0.5\linewidth,clip,trim=13cm 1.cm 15cm 0cm]{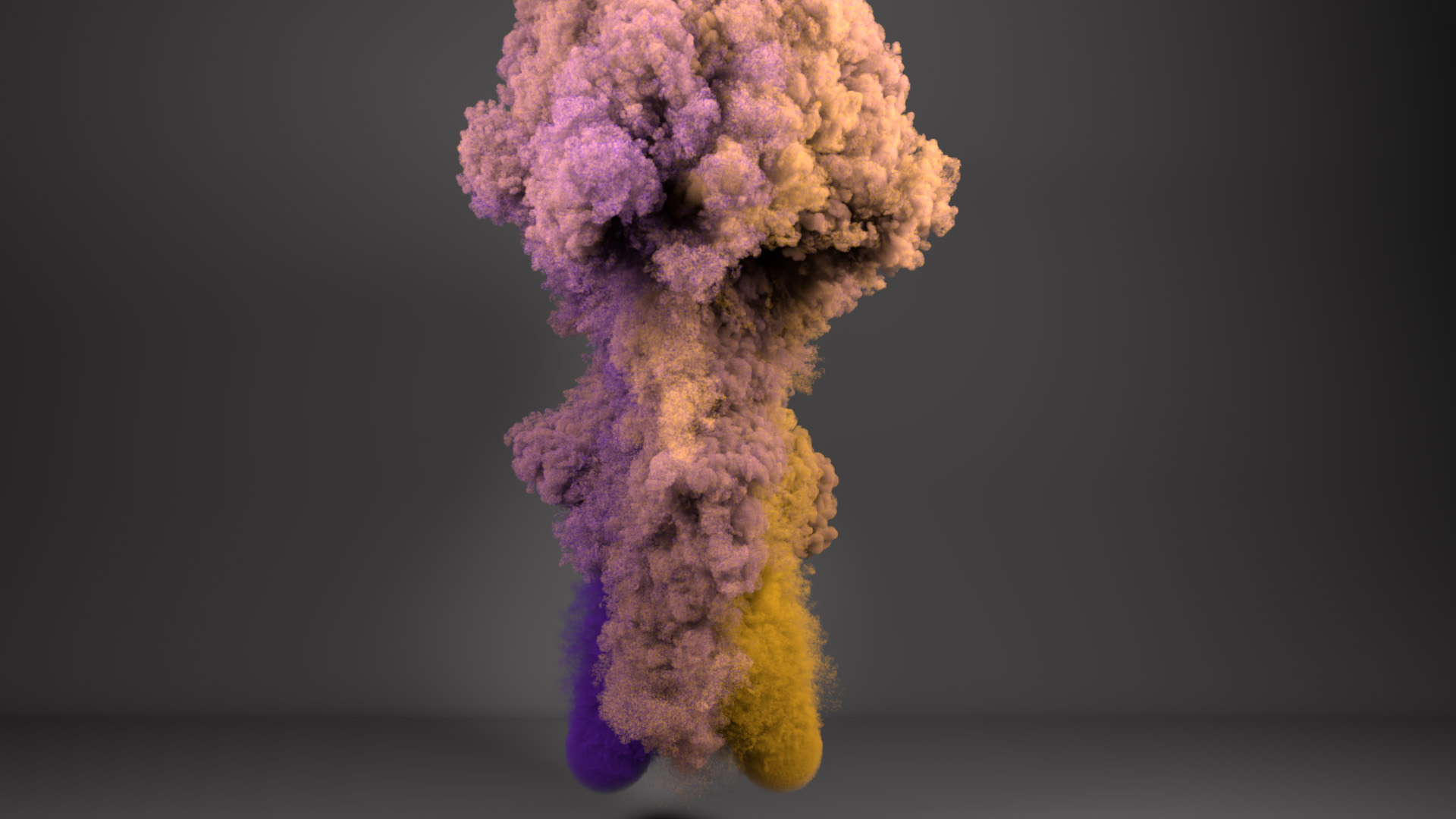}
\caption{Fluid-only plume simulations (\textcolor{smBlue}{frame 180}) using our 16-bit quantized kinetic solver at a grid resolution of $1024^3$. Left: Reynolds number $\mathrm{Re} =$ 40,960. Right: $\mathrm{Re} =$ 2,048,000.}
\label{fig:plume_3d}
\Description{}
\end{figure*}

\paragraph{Navier--Stokes Solvers}
Macroscopic Navier--Stokes solvers have long been the backbone of graphics fluid simulation. As a seminal work, \citet{Stam1999} introduces Stable Fluids to solve the NS equations on Eulerian grids using an unconditionally stable semi-Lagrangian advection scheme. This approach enables fluid simulations with large time steps, significantly improving efficiency for computer graphics applications, but it suffers from excessive numerical diffusion. To mitigate this issue, a variety of improved advection schemes have been proposed, including BFECC~\cite{Kim2005}, the modified MacCormack scheme~\cite{Selle2008}, MacCormack with reflection (MC+R)~\cite{Zehnder2018}, bidirectional flow maps~\cite{Qu2019}, covector advection~\cite{nabizadeh2022covector}, neural flow maps~\cite{Deng2023}, the Eulerian vortex method on flow maps~\cite{wang2024eulerian} and Leapfrog Flow Maps~\cite{Sun2025Leapfrog}, all of which aim to better preserve vortical details over time. Recent work has also explored alternatives to conventional grid-based pressure solvers, including stochastic approaches such as velocity-based Monte Carlo fluids~\cite{sugimoto2024velocity}, neural Monte Carlo fluid simulation~\cite{jain2024neural}, and adaptive mesh-free representations such as Gaussian Fluids~\cite{xing2025gaussian}. These methods offer promising trade-offs in adaptivity, stochastic estimation, and learned acceleration, whereas our work focuses on dense, geometry-rich fluid simulation with highly regular local updates on GPUs.
Beyond Eulerian methods, Lagrangian approaches, \textcolor{smOrange}{such as smoothed particle hydrodynamics (SPH)~\cite{Solenthaler2009, Akinci2012, ihmsen2014sph, Peer2015}, provide particle-based alternatives to Eulerian discretizations}. However, these methods struggle to capture turbulence with few particles. Hybrid Eulerian--Lagrangian approaches combine grid-based computations with particle-based transport to improve accuracy, including FLIP and its extensions~\cite{zhu2005animating, Jiang2015, Zhang2016, Fu2017, Nabizadeh2024}, \textcolor{smBlue}{and recent particle-flow-map formulations designed to improve long-range advection and preserve vortical structures~\cite{zhou2024eulerian, wang2025vortex}}. \textcolor{smBlue}{ST-FLIP~\cite{braun2026} extends particle-to-grid transfer into space--time to support larger time steps in free-surface and two-phase simulations.} Vortex methods, including vortex particles~\cite{Selle2005}, vortex filaments~\cite{Weismann2010}, and vortex sheets~\cite{Pfaff2012}, as well as grid-based vorticity correction~\cite{Zhang2015}, can preserve vortical structures more effectively and provide scalable strategies for maintaining fine-scale flow details.

Alongside advances in flow representation, the computational power of modern GPUs has substantially accelerated fluid solvers through multigrid methods~\cite{Chentanez2011, Sun2025Leapfrog}, fixed-point iterations~\cite{Chen2015}, and compact Poisson filters~\cite{Rabbani2022}, as well as multi-GPU parallelization for large-scale simulations~\cite{Horvath2009, Wang2020}. \textcolor{smBlue}{Fast VEM~\cite{zhang2026fast} addresses complex boundaries using a virtual-element discretization on body-fitted cut-cell grids and a specialized geometric multigrid solver.} Despite these advances, hybrid Eulerian–Lagrangian approaches~\cite{gao2018gpu}, while offering improved fidelity and accuracy, still incur significant memory overhead because they must maintain both grids and particles. Moreover, grid-based Poisson solvers typically rely on iterative methods that require many GPU kernel launches, limiting overall performance.

\paragraph{Boltzmann-based Solvers}
The lattice Boltzmann method, a mesoscopic numerical approach, has recently emerged as a compelling alternative to traditional incompressible NS solvers for both single-phase~\cite{Wei2020, Lyu2021, li2023high} and multiphase~\cite{WeiTVCG2021, Wei2022, Wei2023, Wang2025lbmfoam} flows. Unlike the macroscopic formulations above, LBM evolves distribution functions through purely local collision and streaming operations, avoiding global pressure projection and making it particularly attractive for massively parallel GPU architectures. Early graphics applications~\cite{Li2003, Thurey2009} primarily rely on the low-order Bhatnagar–Gross–Krook (BGK) collision model~\cite{bhathnagor1954model, Chen-1998}, motivating higher-order collision schemes such as central-moment multiple-relaxation-time (CM-MRT) models~\cite{Wei2020}, which reduce both dispersion and dissipation errors. Coupling fluids with solids in the LBM framework has also been extensively investigated, with approaches including the immersed boundary (IB) method~\cite{Peskin-1972, Wei2020}, a variety of bounce-back schemes~\cite{Ladd-1994, Lyu2021, Lyu2023, Liu2023}, and hybrid macroscopic--mesoscopic boundary models~\cite{liu2025hybrid}, enabling applications such as underwater swimming and UAV training~\cite{hu2025learning, Song2025, Liu2022}. \textcolor{smBlue}{Recent work also couples two-phase LBM with the material point method (MPM) to simulate volume-preserving air--water--sand mixtures~\cite{xiao2026volume}.}
Because LBM operates locally at the mesoscopic scale on lattice nodes, it is well suited for GPU acceleration and large-scale simulations~\cite{Rinaldi2012}. \citet{Chen2022} improve GPU efficiency through optimized data layouts, load-balanced immersed-boundary processing, and a multi-kernel execution strategy, while \textcolor{smOrange}{\citet{lehmann2022esoteric} introduces in-place streaming schemes that reduce distribution-buffer storage through direction-dependent read/write scheduling.}
To further reduce memory overhead, HOME-LBM~\cite{li2023high} compresses per-node storage to 10 moment variables for D3Q27 and reconstructs 27 distribution functions on the fly.
However, its original solid-coupling implementation still requires expensive link--solid intersection tests and exhibits divergent branching during solid coupling, reducing parallel efficiency and scalability. \textcolor{smBlue}{For a broader overview of recent advances in LBM and its applications in computer graphics, we refer readers to the SIGGRAPH course by~\citet{li2026lbmcourse}}.

\paragraph{Von Neumann Stability Analysis of LBM}
Von Neumann (VN) stability analysis~\cite{VonNeumann1950} has been widely used to characterize numerical stability, dispersion, and dissipation in LBM. Early studies apply it to the BGK model~\cite{sterling1996stability} and demonstrate the superior stability of the MRT model~\cite{lallemand2000theory}. Subsequent work further analyzed high-order and regularized schemes~\cite{siebert2008lattice, Malaspinas2015, coreixas2017recursive}, as well as cascaded and central-moment MRT schemes~\cite{dubois2015stability, geier2006cascaded, chavez2018improving}. While VN stability analysis has been widely used in LBM, a corresponding VN analysis for moment-encoded formulations remains largely unexplored.

\paragraph{Quantization in Graphics}
Data quantization offers a complementary strategy to reduce storage and memory traffic. In graphics, researchers have explored custom numerical formats and automatic bit-width search~\cite{Hu2021QuanTaichi, Liu2022autoquantization}, but these approaches often rely on heuristic optimization or learning-based search, which are computationally intractable for the chaotic dynamics of turbulent LBM flows. Neural compression methods such as NeuralVDB~\cite{kim2024neuralvdb} and ACORN~\cite{martel2021acorn} achieve impressive compression ratios but incur significant inference overhead, hindering high-frequency real-time updates. \textcolor{smBlue}{Reduced precision has also been studied in LBM~\cite{lehmann2022accuracy},} but directly applying reduced-precision schemes to LBM is particularly challenging because small perturbations may accumulate through repeated collision--streaming updates and destabilize long-horizon simulations. This motivates stability-constrained quantization strategies specifically tailored to the dynamics of moment-encoded LBM.

\section{Background}
\label{sec:background}
Before introducing our method, we briefly review the fundamentals of LBM and high-order moment-encoded formulations, including distribution reconstruction, moment-space collision, and fluid--solid coupling. \textcolor{smBlue}{Bold symbols denote vectors, matrices, or tensors, while non-bold symbols denote scalars or individual components; Latin indices $i,j$ label discrete velocity directions, and Greek indices $\alpha,\beta,\gamma$ label Cartesian components.}

\subsection{Lattice Boltzmann Method (LBM)}
\begin{figure}[ht!]
	\centering
    \includegraphics[width=0.6\columnwidth]{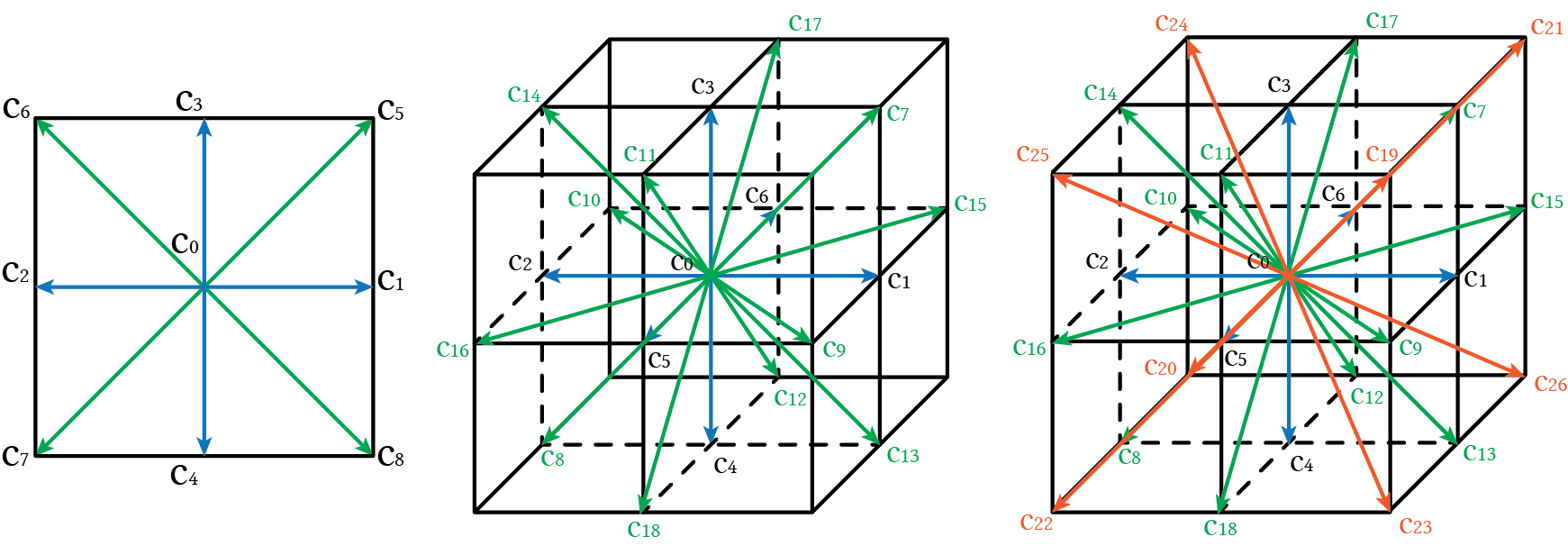}\vspace*{-1mm} 
    \flushleft{\small \hspace*{38mm} D2Q9 \hspace*{20mm} D3Q19 \hspace*{24mm} D3Q27} 
	\caption{\textcolor{smOrange}{Discrete velocity sets used in this
    work: D2Q9 in 2D and D3Q19 and D3Q27 in 3D. } Each discretized distribution function $f_i$ is associated with its corresponding discrete velocity $\bm{c}_i$. 
	\label{fig:lattice_structure}}
    \Description{}
    \vspace{-5mm}
\end{figure}

In the context of single-phase fluid dynamics, the governing kinetic equation for the evolution of the particle distribution function is the Boltzmann equation~\cite{Shan-2006}:
\begin{equation}
\frac{\partial \bm{f}}{\partial t} + \bm{v}\cdot\nabla \bm{f} = \Omega(\bm{f}) + \bm{F}\cdot \nabla_{\bm{v}}\bm{f} \;,
\label{eq:boltzmann_equation}
\end{equation}
where $\bm{f}(\bm{x}, \bm{v}, t)$ denotes the distribution function, $\bm{F}$ represents external forces, and $\Omega$ is the collision operator that relaxes $\bm{f}$ towards the local thermodynamic equilibrium $\bm{f}^\text{eq}$~\cite{coreixas2017recursive}. The collision operator $\Omega$ models local relaxation of the distribution toward equilibrium and determines the viscosity and dissipative behavior of the scheme. In discrete LBM, choices such as BGK~\cite{bhathnagor1954model, Chen-1998} and central-moment MRT~\cite{Wei2020} specify different collision models, while HOME-LBM~\cite{li2023high} evaluates collision
directly in its stored moment representation. 
By discretizing time $t$ into uniform time steps, space $\bm{x}$ into a regular lattice grid, and velocity $\bm{v}$ into a finite set of $q$ discrete directions, the continuous Boltzmann equation (~\autoref{eq:boltzmann_equation}) becomes the lattice Boltzmann equation (LBE) in dimensionless units:
\begin{equation}
f_i(\bm{x}+\bm{c}_i, t+1) - f_i(\bm{x}, t) = \Omega_i + F_i \;,
\label{eq:lbe_normalized}
\end{equation}
where $f_i(\bm{x}, t)$ is the distribution along the $i$-th lattice direction, $\bm{c}_i$ is the corresponding lattice velocity vector, $\Omega_i$ is the discrete collision term, and $F_i$ accounts for the effect of external forces projected onto distribution space. 
In this paper, we consider 9 lattice directions in 2D (D2Q9) and 19 (D3Q19) or 27 (D3Q27) lattice directions in 3D for our LBM fluid simulation, as shown in~\autoref{fig:lattice_structure}.
In lattice units, $\Delta x = \Delta t = 1$.
The evolution in~\autoref{eq:lbe_normalized} can be solved using an operator splitting strategy consisting of two steps:
First, the distribution values are advected via streaming by computing:
\begin{equation}
f^{\text{*}}_i(\bm{x}, t) = f_i(\bm{x}-\bm{c}_i, t)\;,
\label{eq:streaming}
\end{equation}
followed by a collision step expressed as:
\begin{equation}
f_i(\bm{x}, t+1) = f^{\text{*}}_i(\bm{x}, t) + \Omega_i + F_i\;.
\label{eq:collision}
\end{equation}
\textcolor{smBlue}{Here, $f_i^*(\bm{x},t)$ denotes the intermediate post-streaming distribution within the update from $t$ to $t+1$.}

\begin{figure*}[t]
  \centering
  \includegraphics[width=\textwidth]{images/hilbert/hilbert.pdf}
  \caption{Jet flow through a progressively refined 3D space-filling Hilbert curve on a ~$256\times512\times256$ grid. \textcolor{smOrange}{Left: flow visualizations at four levels of detail (LoD 0--3). Right: simulation time per iteration for the original FP32 HOME-LBM, our FP32 variant with geometry decoupling only, and our variant combining geometry decoupling and 16-bit quantization. The inset enlarges the two optimized curves, with labels indicating the speedups over the HOME-LBM baseline, whose per-iteration times are listed in~\autoref{table:timing_memory}.}}
\label{fig:hilbert_3d}
\Description{}
\end{figure*}

The macroscopic quantities, such as density $\rho$, linear momentum $\rho \bm{u}$, and the second-order moment tensor $\rho \bm{S}$, 
can be recovered from the distribution as:
\begin{equation}
\rho=\sum_{i=0}^{q-1} f_i, \; \rho\bm{u} = \textcolor{smOrange}{\frac{1}{2}\bm{F} \, +} \sum_{i=0}^{q-1}\bm{c}_i f_i, \;
\rho {S}_{\alpha \beta}=\sum_{i=0}^{q-1}(c_{i \alpha}c_{i \beta} -c_s^2\delta_{\alpha \beta})\,f_i\,,
\label{eq:rho_u_stress}
\end{equation}
where \textcolor{smBlue}{$q$ is the number of discrete lattice velocities, $\delta_{\alpha\beta}$ is the Kronecker delta, $c_s =\frac{\sqrt{3}}{3}$ is the lattice sound speed (the standard value for the D2Q9, D3Q19, and D3Q27 lattices, shared by all collision models considered
in this paper)}, and the Greek indices $\alpha$ and $\beta$ refer to tensor coordinates, i.e., $\bm{S}\!=\!\{S_{\alpha\beta}\}_{\alpha,\beta}$ for $\alpha,\beta \!\in\!\{x,\!y\}$ in 2D and $\{x,\!y,\!z\}$ in 3D. 

\subsection{High-Order Moment-Encoded LBM (HOME-LBM)}
To reduce the memory footprint and bandwidth cost of standard LBM while preserving physical accuracy, \citet{li2023high} proposed the high-order moment-encoded LBM. 
The key idea of HOME-LBM is to store velocity moments $\rho, \rho\bm{u}$, and $\rho\bm{S}$ in memory and reconstruct distribution functions on the fly using a third-order Hermite basis, balancing accuracy and computational efficiency.

\paragraph{High-order Distribution Reconstruction} 
Given density $\rho$, momentum $\rho \bm{u}$, and the second-order stress tensor $\rho \bm{S}$, distribution functions $\bm{f}$ can be reconstructed via the closed-form function $\bm{h}(\rho,\bm{u},\bm{S})$ as:
\begin{align} \label{eq:three_order_f}
&f_i = {h_i(\rho,\bm{u},\bm{S})} \\ \notag 
&= \rho\, w_i \Biggl[ 1 + \frac{\bm{c}_i\cdot\bm{u}}{c_s^2} + \frac{\bm{H}^{[2]}(\bm{c}_i):\bm{S}}{2c_s^4}
     +\frac{1}{2c_s^6}
\sum_{\alpha\ne\beta}
H^{[3]}_{\alpha\alpha\beta}T_{\alpha\alpha\beta}+\frac{1}{c_s^6}H^{[3]}_{xyz}T_{xyz} 
\Biggr],
\end{align}
where
\begin{align}
T_{\alpha\beta\gamma}
= S_{\alpha\beta}u_\gamma
+ S_{\alpha\gamma}u_\beta
+ S_{\beta\gamma}u_\alpha
- 2 u_\alpha u_\beta u_\gamma.
\end{align}
Here $w_i$ are the lattice weights associated with $\bm{c}_i$, and $H$ denotes the Hermite polynomial basis~\cite{Wang2025lbmfoam}.

\paragraph{Collision and Streaming}
After streaming, three post-streaming temporary moments $\rho^*$, $\rho^*\bm{u}^*$, and $\rho^*\bm{S}^*$ are recovered from the distribution functions based on \autoref{eq:rho_u_stress}. A high-order collision operator is then applied directly in moment space to update the density, velocity, and stress tensor, 
written as:
\begin{align}
&\rho(\bm{x},t+1)
= \rho^*,
\label{eq:ourmomentcollision_density}
\\
&u_\alpha(\bm{x},t+1)
= u_\alpha^*+\frac{F_\alpha}{2\rho^*},
\label{eq:ourmomentcollision_pu}
\\
&S_{\alpha\beta}(\bm{x},t+1)
= \left(1-\tfrac1\tau\right)S_{\alpha\beta}^*
 + \tfrac {u_\alpha^*u_\beta^*} {\tau }
 + \tfrac{2\tau-1}{2\tau\rho^*}
   \left(F_\alpha u_\beta^*+F_\beta u_\alpha^*\right),
\label{eq:ourmomentcollision_mxy}
\\
&S_{\alpha\alpha}(\bm{x},t+1)
= \tfrac{\tau-1}{3\tau}
   \left(2S_{\alpha\alpha}^*
   -S_{\beta\beta}^*-S_{\gamma\gamma}^*\right)
 + \tfrac13\left[
   (u_\alpha^*)^2+(u_\beta^*)^2+(u_\gamma^*)^2
   \right]
\notag\\
&
 + \tfrac1{3\tau}\left[
   2(u_\alpha^*)^2-(u_\beta^*)^2-(u_\gamma^*)^2
   \right]
 + \tfrac{F_\alpha u_\alpha^*}{\rho^*}
 + \tfrac{\tau-1}{3\tau\rho^*}
   \left(2F_\alpha u_\alpha^*
   -F_\beta u_\beta^*-F_\gamma u_\gamma^*\right).
\label{eq:ourmomentcollision_mxx}
\end{align}

\paragraph{Solid Coupling}
HOME-LBM performs fluid--solid coupling by reconstructing boundary distribution functions at solid intersection points and replacing the corresponding streamed distributions on lattice links. Specifically, when a lattice link originating from a grid node $\bm{x}$ intersects a solid boundary at point $\bm{p}$, it reconstructs a boundary distribution at $\bm{p}$ and streams it to the grid node. Under the assumption of a homogeneous weakly compressible fluid, $\rho^{\bm{p}}\!=\!\rho^{\bm{x}}$. 
Velocity $\bm{u}^{\bm{p}}$ can be computed based on the linear velocity and angular velocity of the solid at $\bm{p}$. The second-order tensor $\bm{S}^{\bm{p}}$ is approximated as:
\begin{equation}
\smash{{S}^{\bm{p},\text{eq}}_{\scriptscriptstyle \alpha\beta}= {u}^{\bm{p}}_{\scriptscriptstyle\alpha} {u}^{\bm{p}}_{\scriptscriptstyle \beta}}\;, \quad
{S}^{\bm{p}}_{\alpha\beta} = {u}^{\bm{p}}_\alpha {u}^{\bm{p}}_\beta + \bigl({S}^{\bm{x}}_{\alpha\beta} - {u}^{\bm{x}}_\alpha {u}^{\bm{x}}_\beta\bigr).
\end{equation}
\setlength{\columnsep}{10pt}
\begin{wrapfigure}{r}{0.117\columnwidth}
  \vspace{-10pt}
  \centering
  \includegraphics[width=\linewidth]{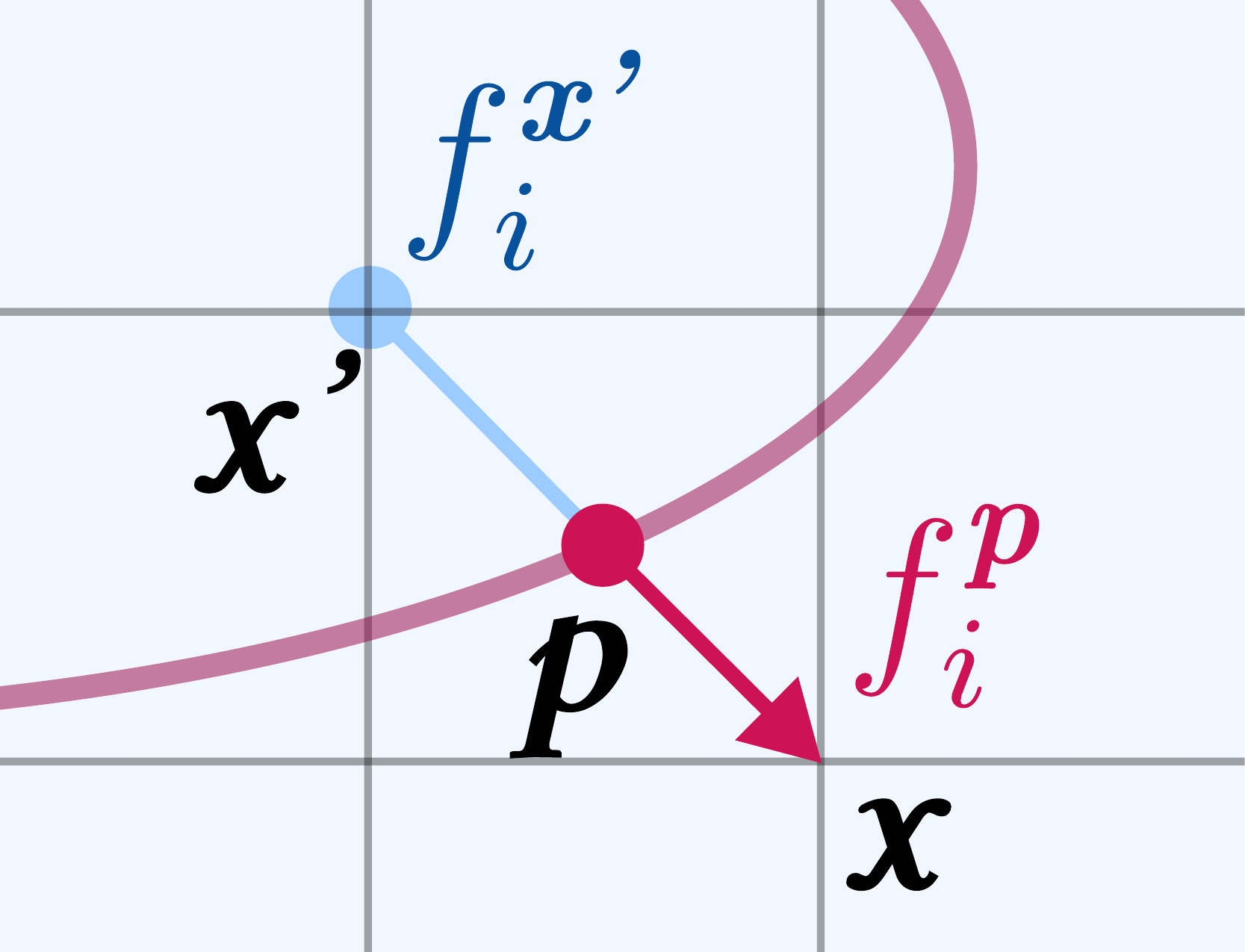}\vspace{-10pt}
\end{wrapfigure}
\textcolor{smOrange}{Given velocity moments at $\bm{p}$, the boundary distribution $f^{\bm{p}}_i$ is reconstructed using \autoref{eq:three_order_f} and then streamed to $\bm{x}$}. Finally, the force and torque exerted by the fluid on the solid are obtained by summing the momentum exchange over all nodes $\bm{x}$ whose links intersect the solid.

\section{Efficient Parallelization with Coupling}
\label{sec:gpu_optimization}

In this section, we first analyze why solid coupling slows down the original HOME-LBM kernel (\autoref{sec:home_bottleneck}), and then present our geometry-decoupled computation scheme for efficient GPU parallelization (\autoref{sec:our_gpu_optimization}).

\subsection{Bottlenecks in HOME-LBM}
\label{sec:home_bottleneck}
HOME-LBM~\cite{li2023high} achieves substantial speedups over earlier GPU-based LBM implementations~\cite{Wei2020} through several key design choices. First, for D3Q27, it reduces per-node storage from $27$ distribution functions to $10$ moment variables, thereby significantly lowering both the memory footprint and memory traffic. Second, by reconstructing distribution functions on the fly from stored moments, the streaming step can be executed more efficiently within each thread block using shared memory. 
In addition, HOME-LBM fuses collision, streaming, and boundary handling into a single GPU kernel, eliminating the overhead of multiple kernel launches and improving overall throughput.

However, with complex solid geometries, this fused formulation introduces a critical GPU bottleneck. In the original update procedure (\autoref{alg:original_lbm}), each lattice thread must determine whether its incoming lattice links intersect the solid surface. For intersecting links, the thread must further reconstruct the boundary distribution function and perform momentum-exchange computations. As a result, the main fluid-update kernel directly incorporates geometric complexity. For high-resolution meshes, the per-step runtime can increase substantially compared with fluid-only simulations (\autoref{fig:hilbert_3d}).

\begin{algorithm}[t!]
\DontPrintSemicolon
\tcp{Each GPU thread processes one lattice node $\bm{x}$}
\tcp{Load moments and reconstruct distribution functions (\autoref{eq:three_order_f})}
$\bm{f}^t \leftarrow \rho^t, \bm{u}^t, \bm{S}^t$\\
\For{each lattice direction $i$}{
    Intersection test between incoming link $l_i$ and obstacle \label{alg:intersect} \\
    \uIf{No intersection}{
        \tcp{Streaming}
        $f^{*}_i(\bm{x}) \leftarrow f^t_i(\bm{x} - \bm{c}_i)$\\
    }\uElse{
        \tcp{Coupling with solid}
        Compute intersection point $\bm{p}$ \\
        Evaluate distribution function $f^t_i(\bm{p})$ at intersection point $\bm{p}$ \\
        $f^{*}_i(\bm{x}) \leftarrow f^t_i(\bm{p})$\\
    }   
}
\tcp{Reconstruct temporary moments (\autoref{eq:rho_u_stress})}
$\rho^{*}, \bm{u}^{*}, \bm{S}^{*} \leftarrow \bm{f}^{*}$\\
\tcp{Moment-based collision (\autoref{eq:ourmomentcollision_density} -- \autoref{eq:ourmomentcollision_mxx}) and write back}
$\rho^{t+1}$, $\bm{u}^{t+1}$, $\bm{S}^{t+1} \leftarrow \rho^{*}, \bm{u}^{*}, \bm{S}^{*}$\\ 
\caption{Original HOME-LBM Kernel}\label{alg:original_lbm}
\end{algorithm}

Although spatial data structures such as bounding volume hierarchies (BVHs) can accelerate these intersection tests, their traversal introduces data-dependent execution paths. Threads near solid boundaries may perform more work than interior threads, causing warp divergence. The additional register requirements and instruction footprint can also reduce occupancy. 

\subsection{Geometry-Decoupled Computation Scheme}
\label{sec:our_gpu_optimization}

To alleviate the solid-coupling bottleneck, we restructure the original HOME-LBM pipeline into two GPU stages: fluid update and solid boundary correction, separated by an intermediate buffer copy.
Rather than embedding boundary processing into every lattice-node update, our method first advances the fluid field uniformly over the entire lattice and then applies only the boundary-induced discrepancies as localized corrections. As a result, the main fluid kernel runtime becomes much less sensitive to solid-geometry complexity.

However, in conventional fluid--solid coupled LBM formulations, boundary processing is mathematically intertwined with streaming. Post-streaming distributions near solid boundaries depend directly on boundary reconstruction or bounce-back updates, so a naive separation of boundary handling is generally inconsistent with the original numerical scheme. A key enabler of this decomposition is a reordering of the update sequence. Instead of the classical {\fontfamily{lmss}\selectfont streaming$\rightarrow$collision} formulation, in which boundary handling is tightly coupled with streaming, we adopt a {\fontfamily{lmss}\selectfont collision$\rightarrow$streaming} scheme. This allows solid effects to be applied afterward as additive moment corrections in a separate sparse stage. 

Consequently, the resulting decomposition is not a straightforward kernel split; instead, a reformulation enables it by approximately preserving the effect of boundary processing on the original update while decoupling geometry-dependent operations from the fluid update. \textcolor{smOrange}{We state where the two differ below and evaluate the quantitative impact of this reformulation on boundary accuracy in~\autoref{sec:boundary_accuracy}}.

\begin{algorithm}[b!]
\DontPrintSemicolon
\tcp{Each GPU thread processes one lattice node $\bm{x}$}
\tcp{Load moments and apply moment-based collision (\autoref{eq:ourmomentcollision_density} -- \autoref{eq:ourmomentcollision_mxx})}
$\rho^{+}, (\rho\bm{u})^{+}, (\rho\bm{S})^{+} \leftarrow \rho^t, (\rho\bm{u})^t, (\rho\bm{S})^t$\\
\tcp{Reconstruct distribution functions (\autoref{eq:three_order_f})}
$\bm{f}^{+} \leftarrow \rho^{+}, (\rho\bm{u})^{+}, (\rho\bm{S})^{+}$\\
\tcp{Streaming}
\For{each lattice direction $i$}{
    $f_i^{\mathrm{I}}(\bm{x}) \leftarrow f_i^{+} (\bm{x} - \bm{c}_i)$\\   
}
\tcp{Reconstruct temporary moments (\autoref{eq:rho_u_stress}) and write back}
$\rho_{\mathrm{out}}^{t+1}$, $(\rho\bm{u})_{\mathrm{out}}^{t+1}$, $(\rho\bm{S})_{\mathrm{out}}^{t+1} \leftarrow \bm{f}^{\mathrm{I}}$\\
\caption{Fluid Update Kernel}\label{alg:homelbm_fluid}
\end{algorithm}

\paragraph{Stage I: Fluid Update.}

In the first kernel (\autoref{alg:homelbm_fluid}), all lattice nodes are processed uniformly to advance the fluid dynamics, independent of the presence of solid boundaries. 
The kernel begins by fetching the stored moment variables $\rho$, $\rho \bm{u}$, and $\rho \bm{S}$, after which the moment-based collision operator \textcolor{smOrange}{(\autoref{eq:ourmomentcollision_density} -- \autoref{eq:ourmomentcollision_mxx})} is applied to obtain the post-collision moments $\rho^+$, $(\rho \bm{u})^+$, and $(\rho \bm{S})^+$. 
These moments are then used to reconstruct the post-collision distribution functions $\bm{f}^+$, which are streamed to neighboring lattice nodes:
\textcolor{smOrange}{
\begin{equation}
f_i^{\mathrm{I}}(\bm{x},t+1)=f_i^+(\bm{x}-\bm{c}_i,t).
\end{equation}
}
\textcolor{smOrange}{The streamed values are subsequently converted back into updated moments at $t+1$ using~\autoref{eq:rho_u_stress} and written to the output buffer in global memory.} By integrating moment-based collision, distribution reconstruction, and streaming into a single regular kernel, this stage removes solid-intersection tests and boundary handling from the main fluid update, avoiding branch-heavy, geometry-related control flow and irregular memory accesses. Consequently, threads within a warp follow nearly identical execution paths and access memory in a regular lattice pattern, maximizing thread efficiency and overall GPU throughput.

\begin{algorithm}[t]
\caption{\textcolor{smBlue}{Moment Buffer Copy Kernel}}
\label{alg:moment_buffer_copy}

\tcp{
Each GPU thread processes one lattice node $\bm{x}$
}
\For{each lattice node $\bm{x}$ in parallel}{
    $\rho_{\mathrm{read}}^{t+1}(\bm{x})
    \gets \rho_{\mathrm{out}}^{t+1}(\bm{x})$\

    $(\rho\bm{u})_{\mathrm{read}}^{t+1}(\bm{x})
    \gets (\rho\bm{u})_{\mathrm{out}}^{t+1}(\bm{x})$\

    $(\rho\bm{S})_{\mathrm{read}}^{t+1}(\bm{x})
    \gets (\rho\bm{S})_{\mathrm{out}}^{t+1}(\bm{x})$\
}
\end{algorithm}

\textcolor{smBlue}{
After the fluid update, a separate kernel (\autoref{alg:moment_buffer_copy}) copies the post-streaming moments at time $t+1$ into a read-only buffer. The solid-correction kernel then reconstructs its inputs from this fixed copy and accumulates moment increments into the output buffer. The copy is required because several triangles may add corrections to the same lattice node through AtomicAdd while other triangles read that node to reconstruct distribution functions; reading from the output buffer would make each reconstruction depend on how many corrections had already been applied,
whereas the fixed copy gives every triangle the same input. The pseudocode describes the FP32 variant, and the reduced-precision storage is introduced in~\autoref{sec:quantization}.
}

\paragraph{Stage II: Solid Boundary Correction.}

\textcolor{smOrange}{After the copy completes, the solid-correction kernel (\autoref{alg:homelbm_solid}) applies corrections induced by solid boundaries.} Rather than launching over all lattice nodes, we assign each triangle face $T$ of the solid geometry to one GPU thread, and examine only the lattice nodes within the bounding box of that triangle.

\begin{algorithm} [t!]
\DontPrintSemicolon
\tcp{Each GPU thread processes one triangle $T$}
\For{each triangle $T$ in parallel}{
    \For{each lattice node ${\bm{x}}$ within $T$'s bounding box}{ \label{alg:goovernodes}
        \For{each lattice direction $i$}{
            Intersection test between link $l_i$ and triangle $T$ \\
            \uIf{No intersection}{
                $\Delta f_i(\bm{x}) \leftarrow 0$\\  
            }\uElse{
                \tcp{Load moments at $\bm{x}$ and recover}
                \tcp{distribution function (\autoref{eq:three_order_f})}
                $f_i^{\mathrm{II}}(\bm{x}) \leftarrow \rho_{\mathrm{read}}^{t+1}, (\rho\bm{u})_{\mathrm{read}}^{t+1}, (\rho\bm{S})_{\mathrm{read}}^{t+1}$\\
                \tcp{Coupling with solid}
                Evaluate $f_i({\bm{p}})$ at intersection point $\bm{p}$ \\
                \tcp{Compute distribution correction}
                $\Delta f_i({\bm{x}}) \leftarrow f_i({\bm{p}}) - f_i^{\mathrm{II}}({\bm{x}})$\\
            }   
        }
        \tcp{Reconstruct moment corrections (\autoref{eq:rho_u_stress})}
        $\Delta \rho$, $\Delta (\rho\bm{u})$, $\Delta (\rho\bm{S}) \leftarrow \Delta \bm{f}$\\
        \tcp{Update moments}
        \texttt{atomicAdd}$(\rho_{\mathrm{out}}^{t+1},\; \Delta \rho)$ \label{alg:add_rho} \\
        \texttt{atomicAdd}$\left((\rho\bm{u})_{\mathrm{out}}^{t+1},\; \Delta (\rho\bm{u})\right)$ \label{alg:add_rhou} \\
        \texttt{atomicAdd}$\left((\rho\bm{S})_{\mathrm{out}}^{t+1},\; \Delta (\rho\bm{S})\right)$ \label{alg:add_rhos}\\
    } 
}
\caption{Solid Boundary Correction Kernel}\label{alg:homelbm_solid}
\end{algorithm}

For each candidate lattice node $\bm{x}$, we test the lattice links associated with $\bm{x}$ for intersection with the triangle $T$. When an intersection is detected, we reconstruct the corresponding boundary distribution functions using the local solid velocity and stress tensor. We then compute the discrepancy between the boundary-consistent distribution and the uniformly streamed value as the correction term
\begin{equation}
\textcolor{smOrange}{
    \Delta f_i(\bm{x})
    =
    f_i(\bm{p}) - f_i^{\mathrm{II}}(\bm{x}),
}
\label{eq:geometry_distribution_correction}
\end{equation}
where $\bm{p}$ denotes the boundary intersection point on the solid surface. \textcolor{smBlue}{For unobstructed links, $\Delta f_i=0$.} These corrections are then projected back into moment space to obtain incremental updates $\Delta \rho$, $\Delta (\rho \bm{u})$, and $\Delta (\rho \bm{S})$, which are added to the global moment fields produced by the fluid update kernel.

Since multiple faces may contribute to the same lattice node, we accumulate these moment corrections in global memory using \texttt{AtomicAdd}. \textcolor{smBlue}{In the quantized implementation, this accumulation
is still performed in FP32, with buffer storage and quantization described in~\autoref{sec:implementation_details}.} In practice, such updates are confined to a thin boundary layer surrounding the solid surface, so the total number of atomic operations remains small relative to the total number of lattice updates.

\paragraph{\textcolor{smOrange}{Relation to the Original Update.}}
\textcolor{smOrange}{
Stage I streams distributions reconstructed from post-collision moments at time $t$, whereas Stage II reconstructs $f_i^{\mathrm{II}}(\bm{x})$ from a fixed copy of the post-streaming moments at time $t+1$. Although both distributions refer to the same target time $t+1$, the boundary treatment uses the updated, post-streaming density $\rho^{t+1}$, so the two reconstructions rely on states from different stages of the time step. 
Consequently, the correction does not generally reproduce the direct boundary replacement in the original HOME-LBM implementation exactly, even without quantization. 
A second, geometric source of discrepancy is that the correction stage treats every triangle independently and assumes a single-layer surface: where a lattice link crosses more than one surface layer within one lattice spacing, as between the upper and lower faces of a thin shell, each intersecting triangle contributes its own correction, whereas the original fused kernel replaces such a link once, at its nearest intersection (see also the
limitations in \autoref{sec:conclusion}).
In \autoref{sec:boundary_accuracy}, we evaluate the resulting discrepancy.
}

\paragraph{Pre-stage.}

As in prior work~\cite{li2023high}, moving solids require an additional preprocessing stage to update the acceleration structure. Without this preprocessing, each GPU thread would need to iterate over all lattice nodes within a triangle's bounding box and perform link--triangle intersection tests, resulting in many redundant evaluations.

To reduce this overhead while keeping kernel launches minimal, we voxelize the solid surface and mark the lattice nodes that contain surface triangles. During traversal of a triangle bounding box, empty lattice nodes can then be efficiently skipped using this occupancy mask. Because this preprocessing is required only when the solid geometry changes, it is negligible for static scenes, where it runs once. For moving solids, the corresponding revoxelization overhead \textcolor{smOrange}{is included in the reported simulation timings (\autoref{table:timing_memory})}.

\section{Stability Analysis and Quantization}
\label{sec:stability_quantization}

To further reduce the memory footprint of HOME-LBM, we introduce a quantization scheme for moment-based representations. Since quantization errors can accumulate over long simulation horizons and potentially destabilize the simulation,
rather than selecting numerical ranges purely through empirical tuning, we perform a von Neumann analysis of the moment-encoded LBM to characterize the stable operating regime (\autoref{sec:home-lbm-analysis}). \textcolor{smBlue}{More details on the von Neumann analysis can be found
in~\autoref{app:standard_lbm_stability_analysis}.} We then combine three complementary sources of constraints to design our quantization scheme (\autoref{sec:quantization}). First, we identify admissible velocity ranges under which perturbations remain bounded. Second, within this regime, we collect conservative runtime statistics from representative simulations to determine practical bounds for density, momentum, and stress-related variables. Third, for second-order moments, we store only their non-equilibrium components and reconstruct the equilibrium contributions analytically, reducing the dynamic range that must be quantized. Together, these ingredients yield a practical $16$-bit fixed-point representation for the stored moments. The resulting format reduces both memory bandwidth and storage cost while preserving long-horizon stability and physical fidelity in challenging turbulent-flow scenarios.

\subsection{Von Neumann Stability Analysis}
\label{sec:home-lbm-analysis}

\paragraph{Mapping Reconstruction.}
Unlike conventional distribution-based LBM, HOME-LBM does not explicitly store the post-collision distribution functions. Instead, it reconstructs them from a reduced set of velocity moments using a third-order Hermite expansion. As a result, the effective update mapping differs from that of standard distribution-based LBM. Specifically, perturbations evolve through a composite mapping that includes collision, moment extraction, and reconstruction, rather than through the standard collision operator alone. We express the HOME-LBM update as
\begin{equation}
\bm{g}^{\mathrm{HOME}}(\bm{f})
= \bm{h}\!\left(\bm{m}(\bm{f}^{+})\right),
\label{eq:home_mapping}
\end{equation}
where $\bm{h}(\cdot)$ denotes Hermite reconstruction of the distribution, $\bm{m}(\bm{f}) \doteq (\rho,\bm{u},\bm{S})$ denotes the moment extraction defined in \autoref{eq:rho_u_stress}, and the intermediate post-collision state $\bm{f}^{+}$ is obtained by applying the underlying \textcolor{smBlue}{non-orthogonal central-moment multiple-relaxation-time (NOCM-MRT)}~\cite{Rosis2017,Rosis2019} collision step \textcolor{smBlue}{(see~\autoref{app:collision_models} for more details)} to $\bm{f}$:
\begin{equation}
\bm{f}^{+}
=
\bm{f} + \bm{\Omega}^{\mathrm{NOCM}}(\bm{f})
=
\bm{f} - \bm{M}^{-1}\bm{R}\bm{M}\bigl(\bm{f}-\bm{f}^{\mathrm{eq}}\bigr),
\label{eq:home_postcollision}
\end{equation}

\textcolor{smOrange}{where $\bm{M} = \bm{M}^{\mathrm{NOCM}}(\bm{u})$ is the moment transformation matrix of the NOCM-MRT collision model}, and $\bm{R}$ is a diagonal matrix of relaxation rates $\{r_i\}_{i=1,\dots,q}$~\cite{Wei2020}. 

This ``moment representation + reconstruction'' procedure first converts the post-collision distribution into a set of moments and then reconstructs the distribution by projecting it onto a truncated third-order Hermite subspace. Although this reconstruction is algebraically equivalent at the macroscopic level, it alters the effective linearized update operator, thereby affecting numerical stability and error amplification.

\paragraph{Linearization of Composite Mapping.}
Following the standard perturbation and linearization procedure introduced in \autoref{app:standard_lbm_stability_analysis}, we linearize the composite mapping $\bm{g}^{\mathrm{HOME}}$ around a uniform equilibrium and analyze its Jacobian with respect to the distribution function. Since $\bm{g}^{\mathrm{HOME}}$ is a composition of moment extraction, collision, and reconstruction operators, the chain rule gives its Jacobian directly.

To obtain a compact closed-form expression for the Jacobian, we exploit the Hermite-based decomposition of the distribution function under the Chapman--Enskog expansion assumption~\cite{Malaspinas2015} and write
\begin{equation}
\bm{f} = \bm{f}^{\mathrm{eq}} + \bm{f}^{\mathrm{neq}}.
\label{eq:decompose}
\end{equation}
Following prior analysis~\cite{Malaspinas2015}, the zeroth- and first-order non-equilibrium Hermite coefficients vanish, i.e., $\bm{a}^{[0],\mathrm{neq}} = 0$ and $\bm{a}^{[1],\mathrm{neq}} = 0$, so the non-equilibrium distribution can be expressed using only the second- and third-order Hermite components:
\begin{equation}
\bm{f}^{\mathrm{neq}} = \bm{A}_2\,\bm{a}^{[2],\mathrm{neq}} + \bm{A}_3\,\bm{a}^{[3],\mathrm{neq}},
\label{eq:non-equilibrium}
\end{equation}
where $\bm{A}_2$ and $\bm{A}_3$ collect the discrete Hermite basis contributions. Exploiting the recursive structure in~\cite{Malaspinas2015}, the third-order non-equilibrium coefficient can be written as $\bm{a}^{[3],\mathrm{neq}} = \bm{R}_{\bm{u}}\,\bm{a}^{[2],\mathrm{neq}}$, allowing \autoref{eq:non-equilibrium} to be rewritten as
\begin{equation}
\bm{f}^{\mathrm{neq}} = \bm{Q}_{\bm{u}} \bm{a}^{[2],\mathrm{neq}},
\label{eq:non-equilibrium_compact}
\end{equation}
where $\bm{Q}_{\bm{u}} = \bm{A}_2 + \bm{A}_3\,\bm{R}_{\bm{u}}$, \textcolor{smOrange}{and the recursive matrix $\bm{R}_{\bm{u}}$ has size $\mathbb{R}^{4\times 3}$ in $2$D and $\mathbb{R}^{10\times 6}$ in $3$D}.

As noted by~\citet{li2023high}, the second-order Hermite coefficient satisfies $\bm{a}^{[2]} = \rho \bm{S} = \bm{B}\bm{f}$ and $\bm{a}^{[2],\mathrm{eq}} = \rho \bm{S}^{\mathrm{eq}} = \bm{B}\bm{f}^{\mathrm{eq}}$, which gives
\begin{equation}
\bm{a}^{[2],\mathrm{neq}}
=
\rho\bigl(\bm{S} - \bm{S}^{\mathrm{eq}}\bigr)
=
\bm{B}\,(\bm{f}^{+} - \bm{f}^{\mathrm{eq}}),
\label{eq:second-non-equilibrium-distribution}
\end{equation}
where $\bm{B}$ is a constant projection matrix that maps distribution functions to second-order moment components; its entries are defined symbolically in terms of the lattice velocities $\bm{c}$ and the lattice sound speed $c_s$.


Substituting \autoref{eq:second-non-equilibrium-distribution} into \autoref{eq:non-equilibrium_compact}, and then replacing the non-equilibrium component in \autoref{eq:decompose}, yields the first-order variation of the reconstructed distribution:
\begin{equation}
g_i
=
\bar{f}_i
+
\left[
\bm{Q}_{\bm{u}}\bm{B}\,(\bm{f}^{+} - \bm{f}^{\mathrm{eq}})
\right]_i.
\label{eq:decompose_distribution}
\end{equation}
Finally, substituting $\bm{f}^{+}$ from \autoref{eq:home_postcollision} into \autoref{eq:decompose_distribution} gives
\begin{equation}
g_i
=
\bar{f}_i
+
\left[
\bm{Q}_{\bm{u}}\bm{B}
\left(
\bm{f}
-
\bm{M}^{-1}\bm{R}\bm{M}(\bm{f}-\bm{f}^{\mathrm{eq}})
-
\bm{f}^{\mathrm{eq}}
\right)
\right]_i.
\end{equation}
This expression explicitly defines the Jacobian operator of the HOME-LBM composite mapping evaluated at equilibrium. Accordingly, the amplification matrix $\bm{G}$ of HOME-LBM can be written as
\begin{equation}
\bm{G}_{ij}
=
e^{-\mathrm{i}\bm{k}\cdot\bm{c}_i}
\left[
\bm{\Lambda}
+
\bm{Q}_{\bm{u}}\bm{B}
\left(
\bm{I}
-
\bm{M}^{-1}\bm{R}\bm{M}(\bm{I}-\bm{\Lambda})
-
\bm{\Lambda}
\right)
\right]_{ij}.
\label{eq:Mo-NOCM}
\end{equation}
Here $\bm{k}$ is the wave vector of the plane-wave perturbation and $\bm{\Lambda}$ is the Jacobian of the equilibrium distribution. The explicit forms and dimensions of $\bm{A}_2$, $\bm{A}_3$, $\bm{R}_{\bm{u}}$, and $\bm{B}$ are provided in \autoref{app:matrix_definition_2d} and \autoref{app:matrix_definition_3d}. \textcolor{smBlue}{The corresponding lattice definitions, moment transformation matrices, and relaxation rates of the collision models for D2Q9 and D3Q27 are also given in Appendices~\ref{app:d2q9_details} and~\ref{app:d3q27_details}, respectively. The derivation does not depend on the lattice and applies to D3Q19 as well.}

\subsection{Quantization}
\label{sec:quantization}

\begin{figure}[ht]
    \centering
    \includegraphics[width=0.6\columnwidth]{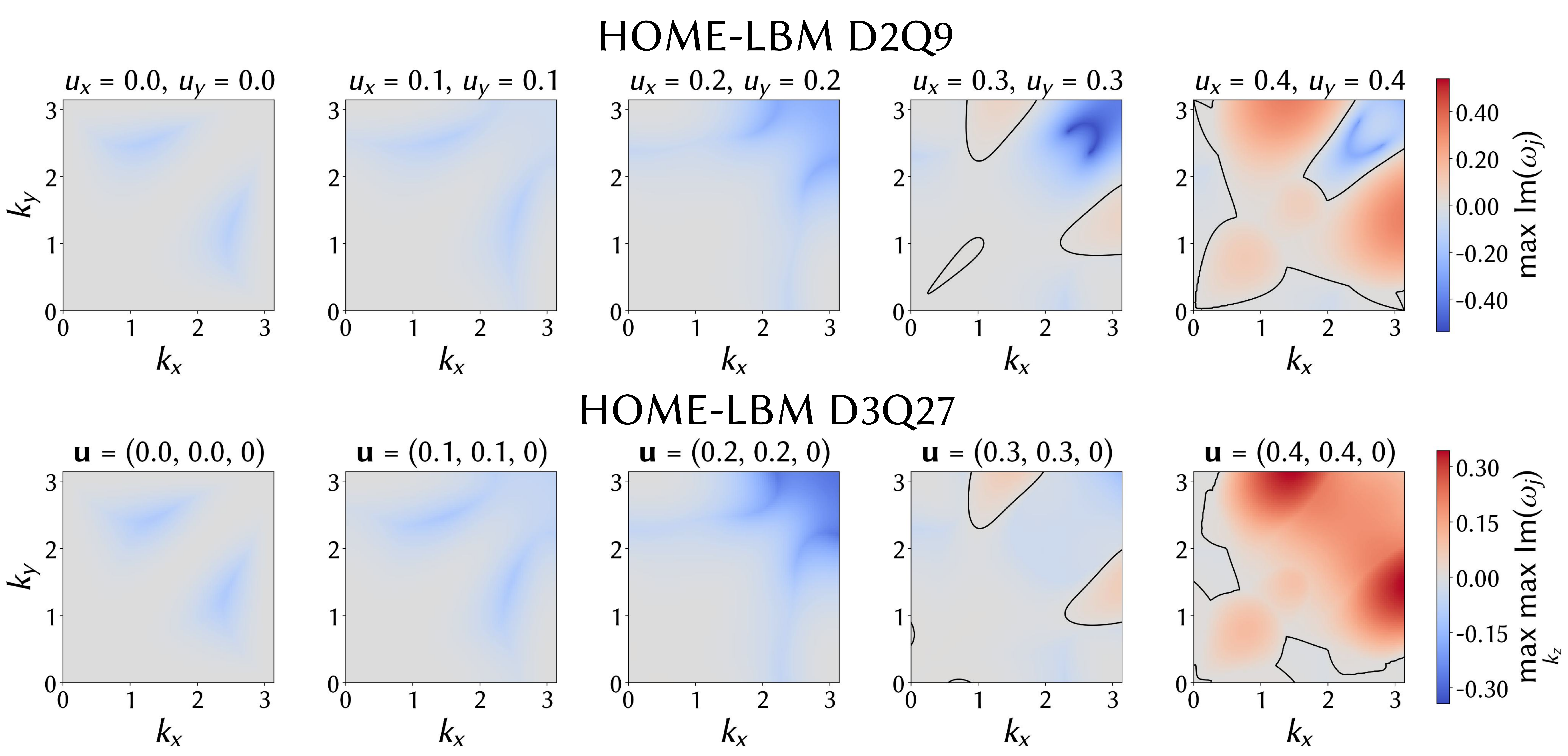}
    \caption{The maximal imaginary part of the eigenvalues, $\max_j \operatorname{Im}(\omega_j)$, derived from the eigenvalues of the HOME-LBM amplification matrix, is visualized over the Fourier wavenumber space $(k_x,k_y)$. From left to right, the panels correspond to increasing background velocities $(u_x,u_y)=(a,a)$ for 2D and $\bm{u}=(a,a,0)$ for 3D, with $a\in\{0,0.1,0.2,0.3,0.4\}$.
    Warmer colors indicate larger positive growth rates and therefore \textcolor{smBlue}{linearly unstable modes}, whereas cooler colors indicate
    \textcolor{smBlue}{negative growth rates (damping) and grey indicates a maximal growth rate of zero}. Black contours denote the neutral stability boundary ($\max_j \operatorname{Im}(\omega_j)=0$).
    \textcolor{smBlue}{For D3Q27, $u_z=0$, and the plotted growth rate is additionally maximized over $k_z\in[0,\pi]$. \textcolor{smBlue}{The D3Q27 maps show the same transition between $a = 0.3$ and $a = 0.4$ as D2Q9, with a slightly lower maximal growth rate at $a = 0.3$.}}
    }
    \Description{}
    \label{fig:velocity_analysis}
\end{figure}

\paragraph{Conservative Range for Velocity.}
Using the derived amplification matrix $\bm{G}$ for HOME-LBM (\autoref{eq:Mo-NOCM}), \textcolor{smOrange}{we evaluate its eigenvalues $\lambda_j$ and the corresponding growth rates $\mathrm{Im}(\omega_j)=\ln|\lambda_j|$ in lattice-time units. In \autoref{fig:velocity_analysis}, we visualize the maximum growth rate $\max_j \mathrm{Im}(\omega_j)$ over the sampled wavenumber domain $(k_x,k_y)\in[0,\pi]^2$}, which quantifies the strongest dissipation or growth rate among all modes. Regions where $\max \mathrm{Im}(\omega) > 0$ correspond to linearly unstable modes and are outlined by solid contours. \textcolor{smOrange}{We evaluate five background velocities, $\bm{u}=(a,a)$ for D2Q9 and $\bm{u}=(a,a,0)$ for D3Q27, with $a\in\{0,0.1,0.2,0.3,0.4\}$. For the 3D lattice, the plotted growth rate is additionally maximized over $k_z\in[0,\pi]$.} These maps characterize the dependence of linear error amplification on the background velocity and reveal increasingly pronounced growing modes at higher advection velocities. More details and comparisons can be found in~\autoref{app:vn_analysis_results}. 

\begin{figure}[ht!]
  \centering
    \includegraphics[width=\columnwidth]{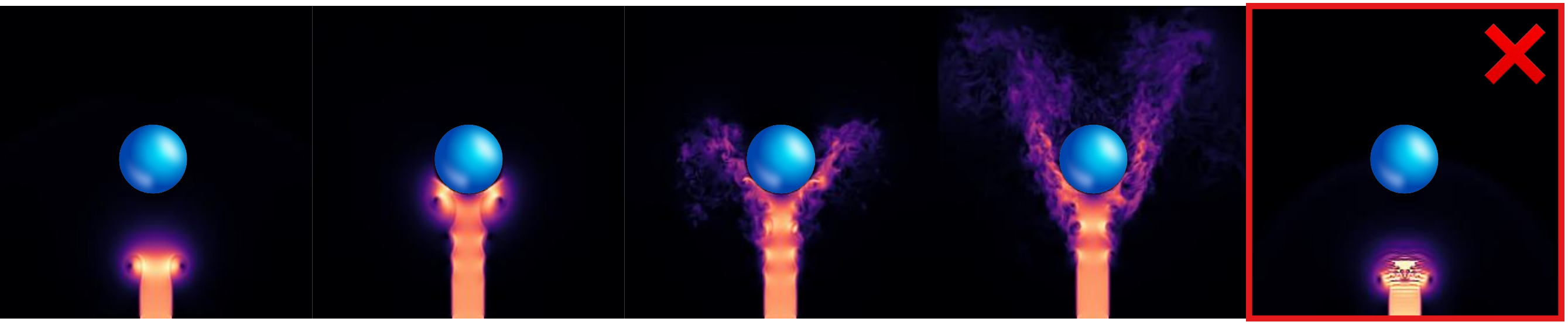}  
     \caption{A D3Q27 plume flow past a sphere is simulated under varying inlet velocities. \textcolor{smOrange}{With other parameters fixed}, the simulation remains stable at an inlet speed of $u=0.4$ in lattice units, whereas increasing the speed to $u=0.45$ leads to rapid numerical failure.} 
  \label{fig:3d_sphere_vel_tol}
  \Description{}
\end{figure}

\textcolor{smOrange}{We further examine the practical behavior using a D3Q27 simulation of turbulent plume flow past a 3D spherical obstacle, as illustrated in \autoref{fig:3d_sphere_vel_tol}, which includes curved solid boundaries and unsteady wake dynamics.}
By increasing the inlet velocity while keeping all other parameters fixed, we find the simulation remains stable at an inlet velocity magnitude of $0.4$ but rapidly becomes unstable when it increases to $0.45$. This behavior qualitatively matches the growing modes observed at higher background velocities in our provided VN analysis, \textcolor{smBlue}{although the observed transition is specific to the tested configuration.} 

For a fixed bit width, using a tighter quantization interval improves the effective precision, \textcolor{smOrange}{but leaves less headroom against clipping.} Guided by the linear analysis and the empirical test, we adopt a conservative velocity quantization range of $u \in [-0.4,0.4]$. \textcolor{smBlue}{Since the actual stored first moment is momentum $\rho\bm{u}$, we quantize each component over the interval $[-0.4\rho_{\max},\,0.4\rho_{\max}]$, where $\rho_{\max}$ is the prescribed density upper bound specified below. For $\rho_{\max} = 1.5$, the interval is $[-0.6, 0.6]$, which bounds the stored momentum rather than the velocity itself (\autoref{tab:clamping}).} 

\textcolor{smBlue}{We note what this analysis does not establish: it is linear about a uniform equilibrium, it does not include the rounding and clamping perturbations introduced by quantization, and it does not determine a minimal bit width.}

\begin{figure}[t!]
  \centering
\includegraphics[width=0.5\columnwidth]{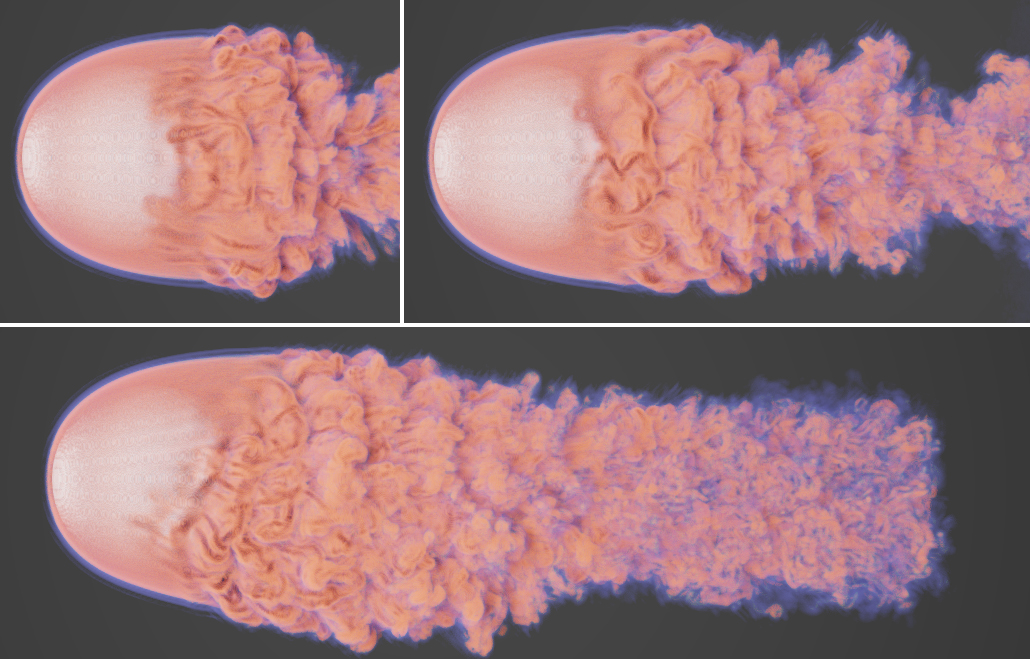}  
 \caption{\textcolor{smOrange}{3D experimental setup for determining the storage ranges of the moment variables.}}
  \label{fig:3d_sphere}
  \Description{}
\end{figure}

\paragraph{\textcolor{smOrange}{Experimental Range for Density.}}
In weakly compressible LBM, density fluctuations are expected to remain small around the reference value $\rho = 1$. Nevertheless, because the method is not strictly incompressible, we explicitly account for density variation. In an extreme-turbulence $3$D flow-over-sphere experiment (\autoref{fig:3d_sphere}) with inlet velocity $0.256$, viscosity $\nu = 0$, and grid resolution $512\times256\times256$, we observe density values within the range $\rho \in [0.8,1.5]$, even in highly vortical regions. \textcolor{smBlue}{We empirically select our fixed quantization ranges based on representative simulations shown in this paper and we clamp out-of-range values to the nearest endpoint.}

\paragraph{Range Reduction of Second-Order Moments.}
In the same flow-over-sphere experiment (\autoref{fig:3d_sphere}), we observe
density-normalized second-order moment components $\bm{S}$ within the range $[-0.1, 0.6]$. We further reduce the quantization range by decomposing the stress tensor into equilibrium and non-equilibrium components. Specifically, \textcolor{smOrange}{following the definition in~\autoref{eq:rho_u_stress}, the isotropic contribution has already been subtracted,} so that the second-order moment tensor can be written as
\begin{equation}
    \textcolor{smOrange}{\rho S_{\alpha\alpha} = \rho S^{\mathrm{neq}}_{\alpha\alpha} + \rho u_\alpha^2},
    \quad
    \rho S_{\alpha\beta} = \rho S^{\mathrm{neq}}_{\alpha\beta} + \rho u_\alpha u_\beta
    \quad (\alpha \neq \beta),
\end{equation}
where the equilibrium contribution is determined analytically from the density and velocity fields. We store and quantize only the high-frequency non-equilibrium components, whose amplitudes are progressively smaller under multiscale perturbation analysis~\cite{harris2004introduction} and are therefore particularly amenable to quantization:
\begin{equation}
    \rho S^{\mathrm{neq}}_{\alpha\alpha}
    =
    \textcolor{smOrange}{\rho S_{\alpha\alpha} - \rho u_\alpha^2},
    \quad
    \rho S^{\mathrm{neq}}_{\alpha\beta}
    =
    \rho S_{\alpha\beta} - \rho u_\alpha u_\beta
    \quad (\alpha \neq \beta).
\end{equation}
This decomposition reduces the range of the second-order moment from $[-0.1,0.6]$ to $[-0.1,0.1]$, \textcolor{smOrange}{decreasing the uniform
quantization step size by a factor of $3.5$ at the same bit width. Since the actual stored quantity is the stress density $\rho S^{\mathrm{neq}}$ rather than $S^{\mathrm{neq}}$ itself, so each of its components is quantized over $[-0.1\rho_{\max}, 0.1\rho_{\max}] = [-0.15, 0.15]$ for $\rho_{\max} = 1.5$ (\autoref{tab:clamping}). The factor 3.5 is unchanged because both intervals scale with $\rho_{\max}$}. This design is also consistent with the preceding stability analysis, which indicates that the dissipation and stability properties of HOME-LBM are governed primarily by non-equilibrium moments, whereas the equilibrium components mainly contribute to macroscopic transport and can therefore be reconstructed analytically without compromising stability.

\paragraph{Normalization and Fixed-point Quantization.}
Each moment component $m$ is normalized to the unit interval as
\begin{equation}
    m' = \frac{m - m_{\min}}{m_{\max} - m_{\min}},
\end{equation}
where $m_{\min}$ and $m_{\max}$ denote the predefined bounds obtained from the range analysis above. This normalization maps all stored moments to $[0,1]$. Each normalized component is then quantized to a $b$-bit unsigned integer using uniform quantization:
\begin{equation}
    q(m) = \left\lfloor m' \left(2^b - 1\right) + \tfrac{1}{2} \right\rfloor.
\end{equation}
Dequantization is performed using the inverse mapping
\begin{equation}
    \hat{m}
    =
    m_{\min} + q(m)\left(\frac{m_{\max} - m_{\min}}{2^b - 1}\right),
\end{equation}
which reconstructs an approximation of the original value $m$ up to quantization error. In our implementation, most moment components are stored as $16$-bit fixed-point integers ($b=16$), allowing two moments to be packed into a single $32$-bit word. Consequently, the moment buffer requires only five \texttt{uint32} values per lattice node, rather than $10$ \texttt{float32} values, reducing both the memory footprint and the global memory write bandwidth by approximately $50\%$.

\paragraph{Clamping Statistics.}
\textcolor{smBlue}{\autoref{tab:clamping} reports how often clamping actually occurred in the two largest coupled scenes (\autoref{fig:teaser} and \autoref{fig:highres_f1}) over their full 9000-iteration runs, counted at the fluid-kernel write-back for every stored quantity and both bounds: saturation occurs only for the density, almost exclusively at the lower bound, and only during the first 900 iterations
of the start-up transient, affecting $3.1 \times 10^{-8}$ (F1) and $5.3 \times 10^{-8}$ (Ducati) of all stored values. No momentum or stress component saturates, and no event is recorded after iteration 900. For a new configuration, the velocity bound carries over, since it is a property of the scheme rather than of the scene; the density and non-equilibrium stress bounds are read from a short FP32 run of a representative configuration, with headroom added before quantization.}

\begin{table}[t]
\centering
\caption{
Clamping fractions (in units of $10^{-8}$) over 9000 iterations of the Ducati (\autoref{fig:teaser}) and F1 (\autoref{fig:highres_f1}) simulations. Fractions are normalized by the total number of moment writes, $10N_{\mathrm{updated}}\times9000$, where $N_{\mathrm{updated}}=158{,}087{,}192$ is the number of lattice nodes updated per iteration. Saturation events are recorded during fixed-point quantization at fluid-kernel write-back. Storage ranges for $\rho u_\alpha$ and $\rho S^{\mathrm{neq}}_{\alpha\beta}$ are scaled by $\rho_{\max}=1.5$ from \autoref{sec:quantization}. (a) Clamping fractions by quantity and bound. (b) Contributions from individual iteration windows to the full-run fraction. Zeros indicate no saturation events.
}
\label{tab:clamping}

\scriptsize
\setlength{\tabcolsep}{2pt}
\renewcommand{\arraystretch}{1.05}

\begin{tabular}{@{}llrrrr@{}}
\toprule

\multicolumn{6}{@{}l}{
\textbf{(a) Clamping by stored quantity}
} \\
\addlinespace[3pt]

& & \multicolumn{2}{c}{Ducati}
  & \multicolumn{2}{c}{F1} \\
\cmidrule(lr){3-4}
\cmidrule(l){5-6}

Quantity & Storage range
& Lower & Upper
& Lower & Upper \\
\midrule

$\rho$
& $[0.8,1.5]$
& 5.329 & 0.004189
& 3.072 & 0.03345 \\

$\rho u_\alpha$
& $[-0.6,0.6]$
& 0 & 0 & 0 & 0 \\

$\rho S^{\mathrm{neq}}_{\alpha\beta}$
& $[-0.15,0.15]$
& 0 & 0 & 0 & 0 \\

\addlinespace[3pt]

\multicolumn{2}{@{}l}{Total (lower + upper)}
& \multicolumn{2}{c}{5.333}
& \multicolumn{2}{c}{3.106} \\

\midrule

\multicolumn{6}{@{}l}{
\textbf{(b) Temporal contributions}
} \\
\addlinespace[3pt]

\multicolumn{2}{@{}l}{Iterations}
& \multicolumn{2}{c}{Ducati}
& \multicolumn{2}{c}{F1} \\
\midrule

\multicolumn{2}{@{}l}{1--900}
& \multicolumn{2}{c}{5.333}
& \multicolumn{2}{c}{3.106} \\

\multicolumn{2}{@{}l}{901--9000}
& \multicolumn{2}{c}{0}
& \multicolumn{2}{c}{0} \\

\bottomrule
\end{tabular}
\end{table}

\section{Results}
\label{sec:results}

\begin{table*}[hbt]
\setlength{\tabcolsep}{1pt}
\caption{{Statistics.} Per-iteration timings are reported in milliseconds, while total time is reported in seconds and corresponds to the accumulated time of the entire simulation (all iterations). \textcolor{smBlue}{All experiments except~\autoref{fig:plume_3d} are conducted on a workstation equipped with an AMD Ryzen Threadripper 3970X 32-core CPU and a single GPU with 16,384 cores and 24~GB of GPU memory.} $*$ denotes the large-scale fluid-only simulation, executed on a single GPU with $6{,}912$ cores and $80\,\mathrm{GB}$ of memory. \textcolor{smBlue}{$\nu$ is the kinematic viscosity in lattice units $(\times 10^{-3})$, and $\nu = 0$ marks inviscid runs. Speedup is relative to the original FP32 HOME-LBM implementation measured on the same 24GB GPU}.}
\begin{tabular}{lccccccccc|cc}
			\toprule
			 & \makecell{Grid \\ resolution } &
             \textcolor{smBlue}{\makecell{$\nu$ \\ $(\times 10^{-3})$}} & \makecell{Face \\ \#} & \makecell{Fluid \\ update \\ (ms)} & \makecell{Buffer \\ copy \\ (ms)} & \makecell{Solid correction \\ (+ voxelization) \\ (ms)} & \makecell{ Iter \\ \# }& \makecell{Total \\ time \\ (s)} & \makecell{Total \\ memory \\ (GB)} 
             & \textcolor{smBlue}{\makecell{HOME-LBM \\ time per iter \\ (ms)}} & \textcolor{smBlue}{ \makecell{Speedup}}\\
			\midrule
\autoref{fig:teaser} & 1000$\times$400$\times$400 & $4.7$ & 2,268k & \hspace{0.45em}17.7 & 10.8 & 7.6\hspace{0.45em} & 9000 & \hspace{0.45em}325 & \hspace{0.5em}9.8 & 324.7 & 9.0$\times$\\		
\autoref{fig:highres_f1} & 1000$\times$400$\times$400 & $4.7$ & 5,329k & \hspace{0.45em}17.7 & 10.8 & 12.4\hspace{0.9em}  & 9000 & \hspace{0.45em}368 & \hspace{0.5em}9.8 & 237.3 & 5.8$\times$\\            
\autoref{fig:plume_3d}$^*$ & 1024$\times$1024$\times$1024 & \makecell{$5.0$ \\ $0.1$} & -- & 151.3 & -- & -- & 8100 & 1225 & 42.9 & -- & --\\	
\autoref{fig:hilbert_3d} LoD~0  & 256$\times$512$\times$256 & $1.2$ & \hspace{1.25em}15k & \hspace{0.9em}3.1 & \hspace{0.45em}2.2 & 0.2\hspace{0.45em} & 6750 & \hspace{0.9em}37 & \hspace{0.5em}2.0 & 33.8 & 6.2$\times$\\
\autoref{fig:hilbert_3d} LoD~1 & 256$\times$512$\times$256 & $1.2$ & \hspace{1.25em}38k & \hspace{0.9em}3.1 & \hspace{0.45em}2.2 & 0.3\hspace{0.45em} & 6750 & \hspace{0.9em}38 & \hspace{0.5em}2.0 & 53.4 & 9.5$\times$\\
\autoref{fig:hilbert_3d} LoD~2 & 256$\times$512$\times$256 & $1.2$ & \hspace{1.25em}71k & \hspace{0.9em}3.1 & \hspace{0.45em}2.2 & 0.6\hspace{0.45em} & 6750 & \hspace{0.9em}40 & \hspace{0.5em}2.0 & 105.4 & 17.9$\times$\\
\autoref{fig:hilbert_3d} LoD~3 & 256$\times$512$\times$256 & $1.2$ & \hspace{0.85em}145k & \hspace{0.9em}3.1 & \hspace{0.45em}2.2 & 1.3\hspace{0.45em} & 6750 & \hspace{0.9em}44 & \hspace{0.5em}2.0 & 214.6 & 32.5$\times$\\
\autoref{fig:3d_sphere} & 512$\times$256$\times$256 & 0 & \hspace{1.85em}1k & \hspace{0.9em}3.2 & \hspace{0.45em}2.2 & 0.05 & 6000 & \hspace{0.9em}33 & \hspace{0.5em}2.0 & 28.8 & 5.3$\times$\\
\autoref{fig:yeahright_3d} Col~1 & 720$\times$360$\times$360 & $3.4$ & \hspace{0.85em}188k & \hspace{0.45em}10.1 & \hspace{0.45em}6.3 & 0.8\hspace{0.45em} & 7500 & \hspace{0.45em}129 & \hspace{0.5em}5.7 & 132.7 & 7.7$\times$\\
\autoref{fig:yeahright_3d} Col~2 & 720$\times$360$\times$360 & $3.4$ & \hspace{0.85em}754k & \hspace{0.45em}10.1 & \hspace{0.45em}6.3 & 2.1\hspace{0.45em} & 7500 & \hspace{0.45em}139 & \hspace{0.5em}5.7 & 139.3 & 7.5$\times$\\
\autoref{fig:yeahright_3d} Col~3 & 720$\times$360$\times$360 & $3.4$ & \hspace{0.85em}754k & \hspace{0.9em}9.1 & \hspace{0.45em}6.1 & 2.6\hspace{0.45em} & 7500 & \hspace{0.45em}133 & \hspace{0.5em}5.7 & -- & --\\
\autoref{fig:3d_trefoil} & 256$\times$128$\times$128 & 0 & \hspace{0.85em}-- & \hspace{0.9em}0.4 & \hspace{0.45em}-- & --\hspace{0.45em} & 5400 & \hspace{1.35em}2 & \hspace{0.5em}0.2 & 2.4 & 6.1$\times$\\
\autoref{fig:deltawing_3d} & 660$\times$250$\times$330 & $3.3$ & \hspace{0.85em}240k & \hspace{0.9em}5.3 & \hspace{0.45em}3.4 & 0.4\hspace{0.45em} & 9000 & \hspace{0.9em}82 & \hspace{0.5em}3.1 & 37.3 & 4.1$\times$\\
\autoref{fig:deltawing_comparison} & 256$\times$128$\times$128 & 0 & \hspace{2.25em}8 & \hspace{0.9em}0.4 & \hspace{0.45em}0.3 & 0.01 & 4500 & \hspace{1.35em}3 & \hspace{0.5em}0.3 & 3.2 & 4.5$\times$\\
\autoref{fig:turbine_comparison} & 256$\times$256$\times$256 & 0 & \hspace{1.85em}5k & \hspace{0.9em}1.4 & \hspace{0.45em}1.1 & 0.56 (+ 2.1) & 3000 & \hspace{0.9em}16 & \hspace{0.5em}1.1 & -- & --\\
\autoref{fig:porous}   & 256$\times$128$\times$128 & $1.2$ & \hspace{0.85em}128k & \hspace{0.9em}0.4 & \hspace{0.45em}0.3 & 0.4 & 4500 & \hspace{1.35em}5 & \hspace{0.5em}0.3 & 20.0 & 18.2$\times$\\	            
\bottomrule
\end{tabular}
\label{table:timing_memory}
\end{table*}

All experiments are conducted on a workstation with an AMD Ryzen Threadripper 3970X 32-core CPU and a single GPU (16,384 cores, 24 GB memory), except for the $1024^3$ fluid-only simulation (\autoref{fig:plume_3d}), which uses a GPU with 6,912 cores and 80 GB memory.
All algorithms are implemented in CUDA and compiled on both Windows and Ubuntu~22.04.
We summarize per-scene statistics in \autoref{table:timing_memory}, which include only simulation costs unless otherwise specified. When solids are present, we additionally report mesh complexity and the corresponding computation costs. Unless stated otherwise, we report all per-iteration timings in milliseconds, measured with NVIDIA Nsight profiling after a short warm-up phase to exclude initialization overhead. Total execution time is reported in seconds.

\subsection{Implementation Details}
\label{sec:implementation_details}
For real-time visualization, we employ the Vulkan-based ray-marching volume renderer of~\citet{Sun2025Leapfrog}, which enables direct rendering from GPU simulation buffers. For high-quality offline rendering, we export particle data, convert it to volumetric representations using OpenVDB~\cite{museth2013openvdb}, and render it in Cinema~4D with Redshift~\cite{maxon}.

\paragraph{Data Layout.}
To improve memory efficiency and access coherence, we store fluid nodes in a structure-of-arrays (SoA) layout~\cite{Chen2022} and organize them into fixed-size $8\times8\times8$ tiles. This layout improves spatial locality when accessing boundary-adjacent fluid nodes and enables coalesced memory access during solid correction.

\paragraph{Dithering.}
To reduce bias from deterministic rounding, we apply spatial dithering during quantization. Specifically, a small zero-mean noise, uniformly distributed in $[-\tfrac{1}{2}, \tfrac{1}{2}]$ of one quantization unit~\cite{Liu2022autoquantization}, is added to the normalized value before rounding, drawn independently for every node and every time step, and the result is clamped to $[0, 2^b - 1]$. This converts deterministic rounding error into zero-mean stochastic noise, thereby preventing systematic bias accumulation over long time integration.

\paragraph{Atomic Accumulation with Mixed Precision.}
When solid coupling is present, multiple triangle faces may contribute moment corrections to the same lattice node, which requires atomic accumulation of moment updates in the solid-correction kernel. Performing atomic operations directly on quantized fixed-point values would introduce order-dependent rounding errors and may lead to noticeable bias accumulation over long simulations. To mitigate this issue, we adopt a mixed-precision strategy.

Specifically, we maintain two moment buffers: \textcolor{smOrange}{a full-precision buffer and a compact 16-bit quantized buffer. The fluid update kernel (\autoref{alg:homelbm_fluid}) reads the compact buffer and writes post-streaming moments to the FP32 buffer (the output buffer of~\autoref{alg:homelbm_fluid}). The buffer copy kernel (\autoref{alg:moment_buffer_copy}) then encodes these post-streaming moments into the compact buffer, overwriting the previous time step's data. During solid correction (\autoref{alg:homelbm_solid}), the compact buffer remains read-only: the solid-correction kernel decodes its contents and accumulates
the resulting moment corrections into the FP32 buffer using \texttt{atomicAdd}. Once all solid corrections are complete, the corrected moments are encoded back into the compact buffer for the next fluid update.} Since only one of the moment buffers is quantized, the overall memory reduction is about $25\%$ compared with the FP32 implementation, which we find is a favorable trade-off between accuracy and bandwidth savings in solid-coupled simulations.

\subsection{Robustness and Accuracy of Quantization}
\label{sec:results_quantization}

We evaluate the robustness of the proposed reduced-precision representation using long-horizon vortex benchmarks and bit-width ablation studies. In particular, we focus on whether the quantized moment representation preserves the stability and visual fidelity of the full-precision HOME-LBM baseline over long simulations.

\begin{figure*}[t]
\centering
\includegraphics[width=\linewidth]{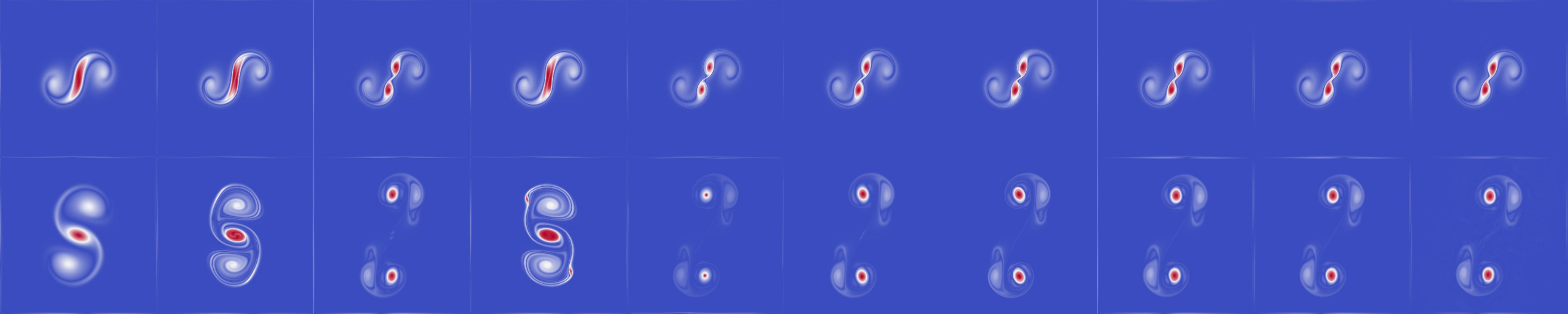}
\vspace{-1.5em}
\caption{Comparison of vorticity advection results on the 2D Taylor vortex, using
    Semi-Lagrangian~\cite{Stam1999}, MacCormack~\cite{Selle2008},
    MC+R~\cite{Zehnder2018}, PolyPIC~\cite{Fu2017},
    BiMocq$^2$~\cite{Qu2019}, CF~\cite{nabizadeh2022covector},
    and the Eulerian vortex method on flow maps~\cite{wang2024eulerian}. We additionally include
    HOME-LBM~\cite{li2023high} baselines using full-precision double (64-bit) and single-precision float (32-bit) variants, as well as our 16-bit quantized solver. All methods use the same $256 \times 256$ grid resolution.}
    \Description{}
    \label{fig:taylor_vortex}
\end{figure*}

\paragraph{2D Taylor Vortex}
We first evaluate the long-term robustness of vortex evolution using the 2D Taylor vortex benchmark, following the setup in prior work~\cite{Qu2019, nabizadeh2022covector}. This benchmark is sensitive to numerical dissipation and perturbation growth, making it well suited for assessing the stability of our proposed quantized simulations. We compare against representative grid-based and particle-based fluid solvers, including Semi-Lagrangian~\cite{Stam1999}, MacCormack~\cite{Selle2008}, MC+R~\cite{Zehnder2018}, PolyPIC~\cite{Fu2017}, BiMocq$^2$~\cite{Qu2019}, Covector Fluids (CF)~\cite{nabizadeh2022covector}, and the Eulerian vortex method on flow maps~\cite{wang2024eulerian}, as well as HOME-LBM~\cite{li2023high} baselines using both double precision (64-bit) and single precision (32-bit). All methods use the same grid resolution, $256\times256$. 
As shown in \autoref{fig:taylor_vortex}, classical semi-Lagrangian schemes show progressive vortex-core diffusion, while particle-based methods preserve sharper structures but introduce more high-frequency artifacts. In contrast, HOME-LBM preserves coherent vortex structures with limited dissipation. Our proposed 16-bit quantized solver remains visually consistent with the 64-bit and 32-bit HOME-LBM baselines and closely matches other high-quality advection methods, indicating that moment-space quantization preserves visual fidelity and numerical behavior while reducing memory usage by 50$\%$.

\begin{figure}[ht!]
	\centering
    \includegraphics[width=0.6\columnwidth]{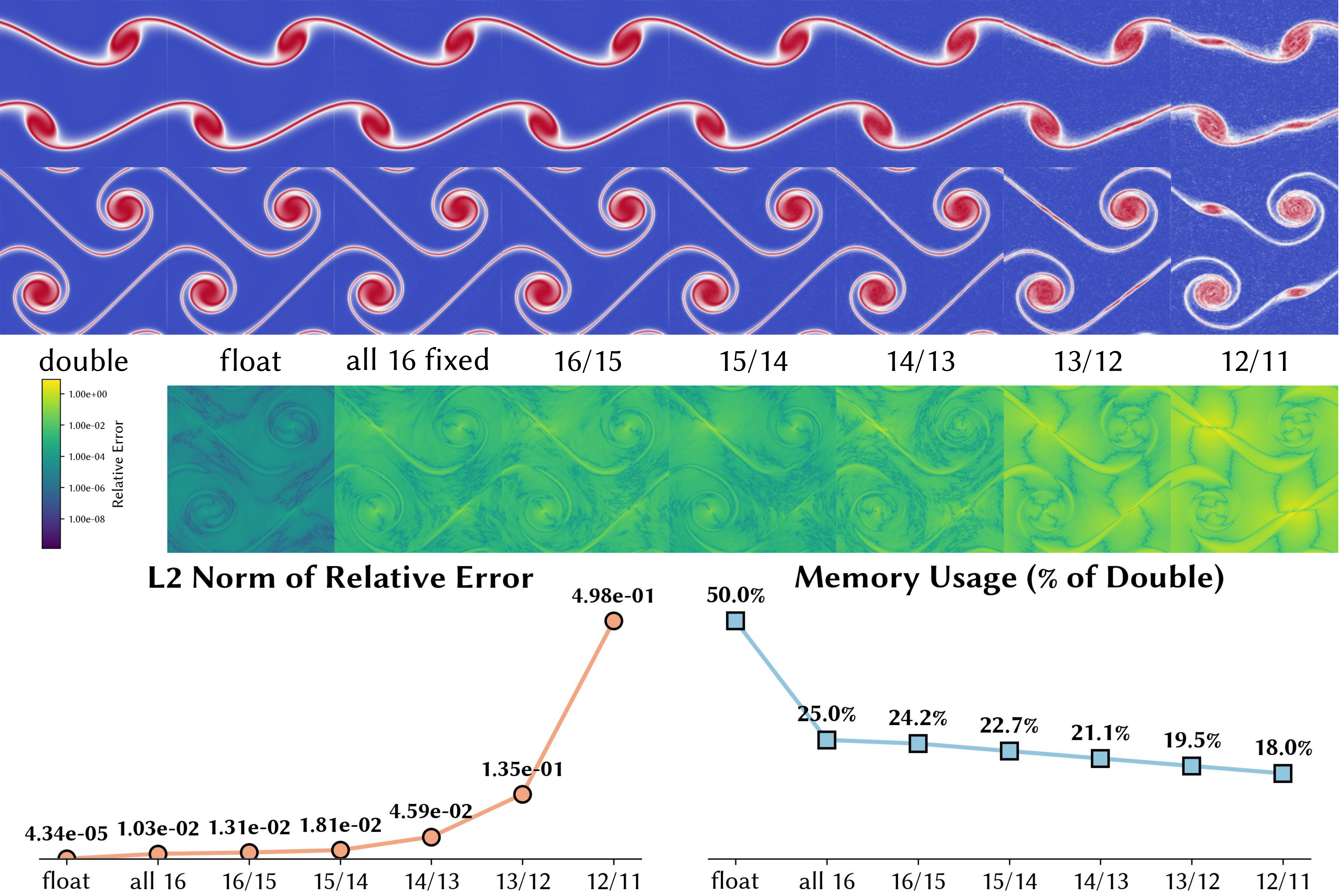}
	\caption{{2D double-layer vortex under different quantization configurations.} Top: vorticity snapshots for the 64-bit double-precision reference, 32-bit float, uniform fixed 16-bit quantization, and adaptive bit-allocation configurations $b_{\rho, u} / b_S \in \{16/15, 15/14, 14/13, 13/12, 12/11\}$. Middle: spatial distribution of the pointwise relative error with respect to the double-precision reference. Bottom: quantitative comparison of the accuracy--memory trade-off, measured by the L2 norm of the relative error and the memory usage \textcolor{smOrange}{evaluated relative to the 64-bit double-precision baseline}.} \vspace{-1em}
	\label{fig:double_layer_vortex}
    \Description{}   
\end{figure}

\paragraph{2D Double-layer Vortex}
We next evaluate the proposed quantized representation's sensitivity to different bit allocations using the 2D double-layer vortex benchmark. We adopt an adaptive bit-allocation strategy to balance quantization error across different moment components. Since density and velocity variables exhibit similar normalized ranges and directly affect macroscopic transport, we assign the same bit width to $\rho$ and $\rho\bm{u}$. In contrast, after the range reduction described in~\autoref{sec:quantization}, the non-equilibrium stress moments $\rho\bm{S}^{\mathrm{neq}}$ have a much smaller dynamic range and can therefore be represented with fewer bits. 
We denote each configuration as $b_{\rho, u} / b_S$, where $b_{\rho, u}$ and $b_S$ are the bit widths for density–velocity and stress moments, respectively. For example, the \texttt{16/15} configuration stores $\rho$ and $\rho\bm{u}$ using 16 bits and stores $\rho\bm{S}^{\mathrm{neq}}$ using 15 bits. 
As shown in~\autoref{fig:double_layer_vortex}, \textcolor{smOrange}{where both the L2 relative error and the memory usage are measured against the 64-bit double-precision run}, this allocation yields a smooth accuracy–memory trade-off. Reducing stress precision to 14–15 bits introduces only minor error increases, whereas more aggressive reduction rapidly degrades both numerical accuracy and visual quality.

\paragraph{3D Trefoil Knot.}
\textcolor{smBlue}{To assess the ability of our 16-bit moment quantization to preserve 3D vortex structures, we adopt the trefoil-knot benchmark of~\citet{nabizadeh2022covector}. 
We compare our 16-bit quantized solver against Semi-Lagrangian~\cite{Stam1999}, MacCormack~\cite{Selle2008}, MC+R~\cite{Zehnder2018}, BiMocq$^2$~\cite{Qu2019}, Covector Fluids (CF) and its extension with the method of characteristic mapping (MCM)~\cite{nabizadeh2022covector}, as well as the FP32 HOME-LBM baseline~\cite{li2023high}. 
As shown in \autoref{fig:3d_trefoil}, the quantized solver produces vortex structures visually consistent with the FP32 baseline, with both retaining more kinetic energy than CF but less than CF+MCM. 
Their kinetic-energy and enstrophy histories closely agree (\autoref{fig:3d_trefoil} (a,b)), including the transient enstrophy increase during vortex reconnection. 
After 5.5~s, the relative mass drifts remain small, at $3.5\times10^{-5}$ for FP32 and $4\times10^{-5}$ for the quantized solver (\autoref{fig:3d_trefoil} (c)). 
These results suggest that 16-bit moment quantization largely preserves vortex dynamics and mass conservation behavior while reducing storage requirements.} 

\begin{figure*}[t]
\centering
\includegraphics[width=0.333\linewidth,clip,trim=3cm 2cm 2cm 3cm]{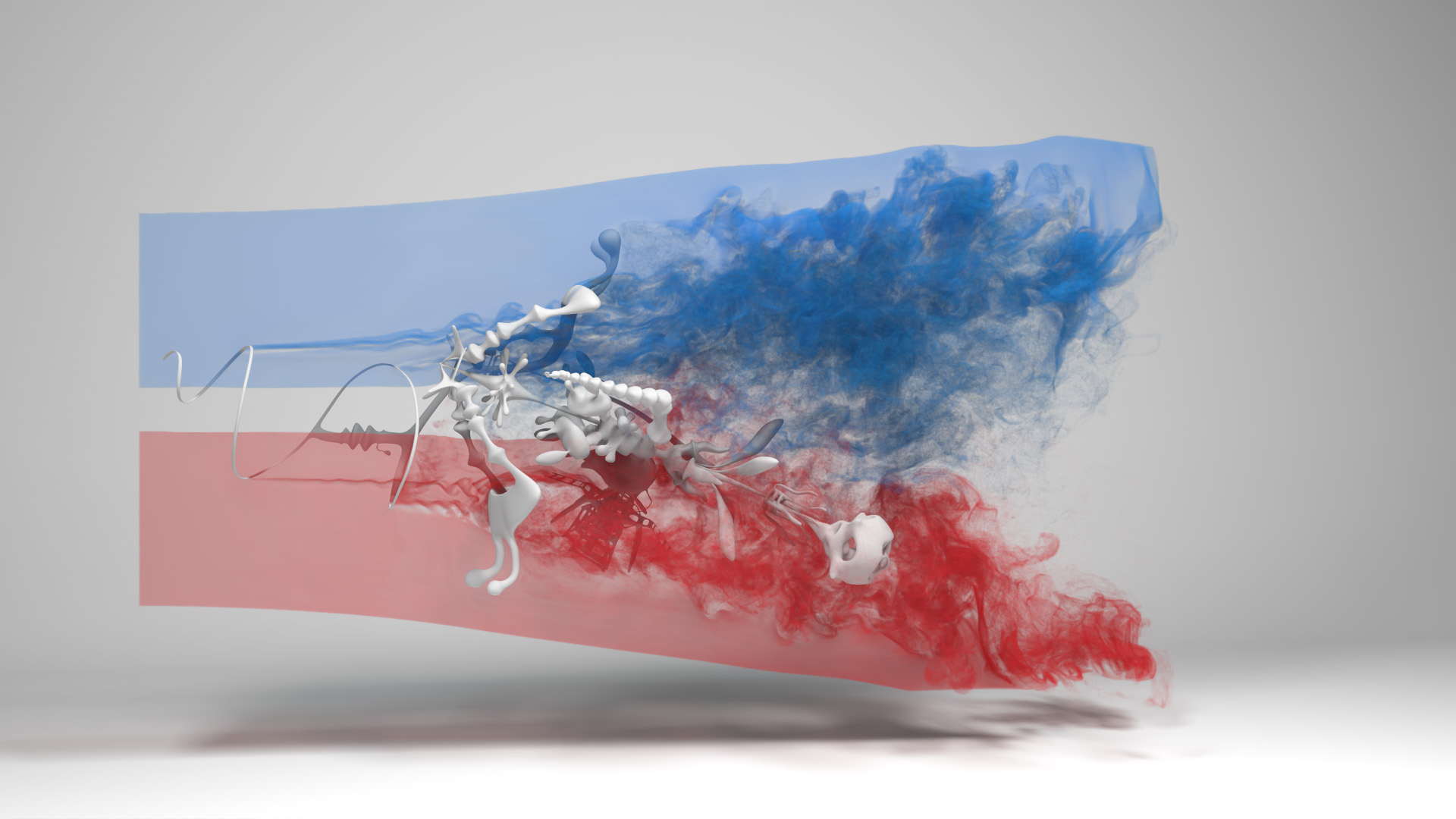}\hfill
\includegraphics[width=0.333\linewidth,clip,trim=3cm 2cm 2cm 3cm]{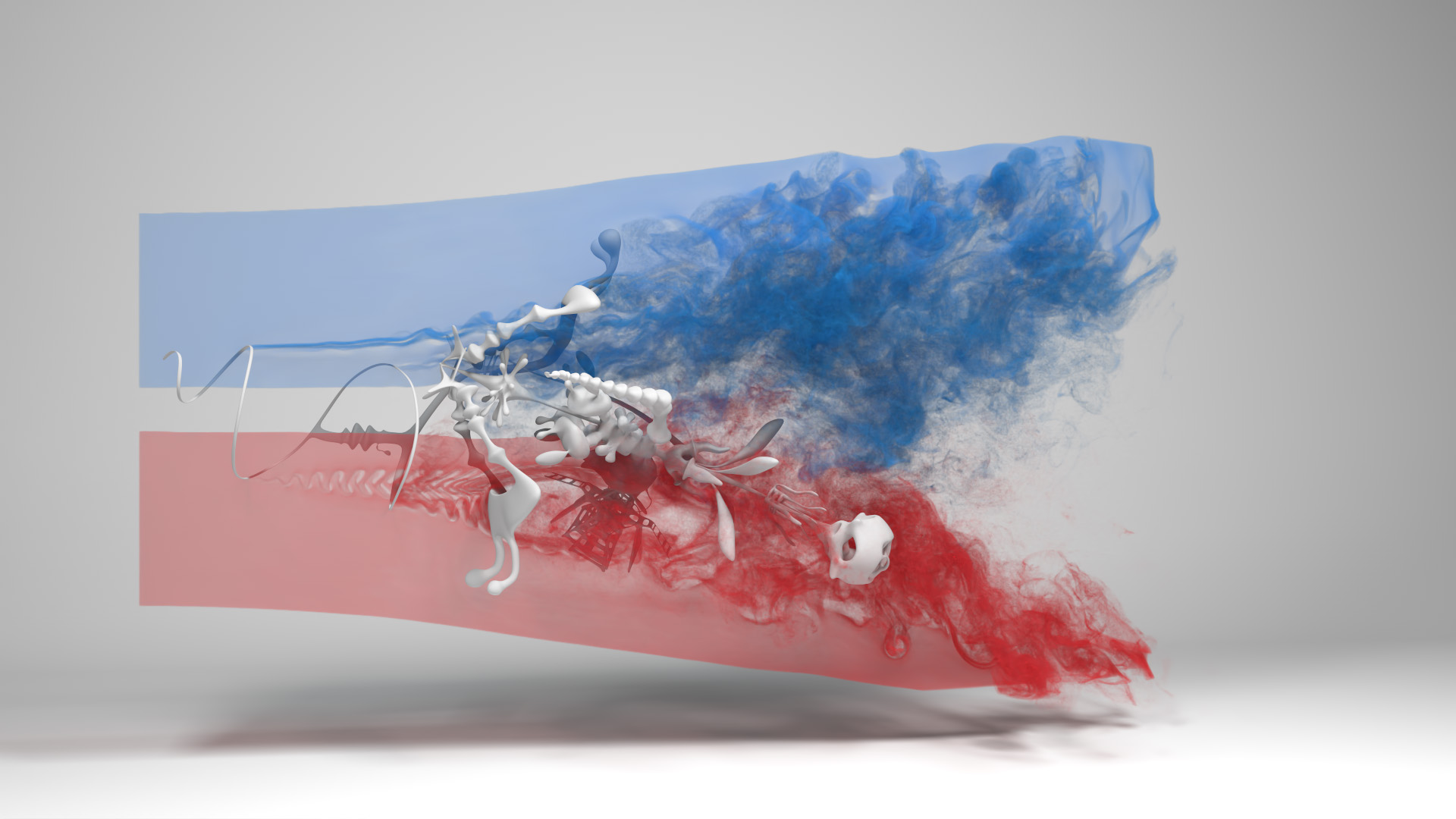}\hfill
\includegraphics[width=0.333\linewidth,clip,trim=3cm 2cm 2cm 3cm]{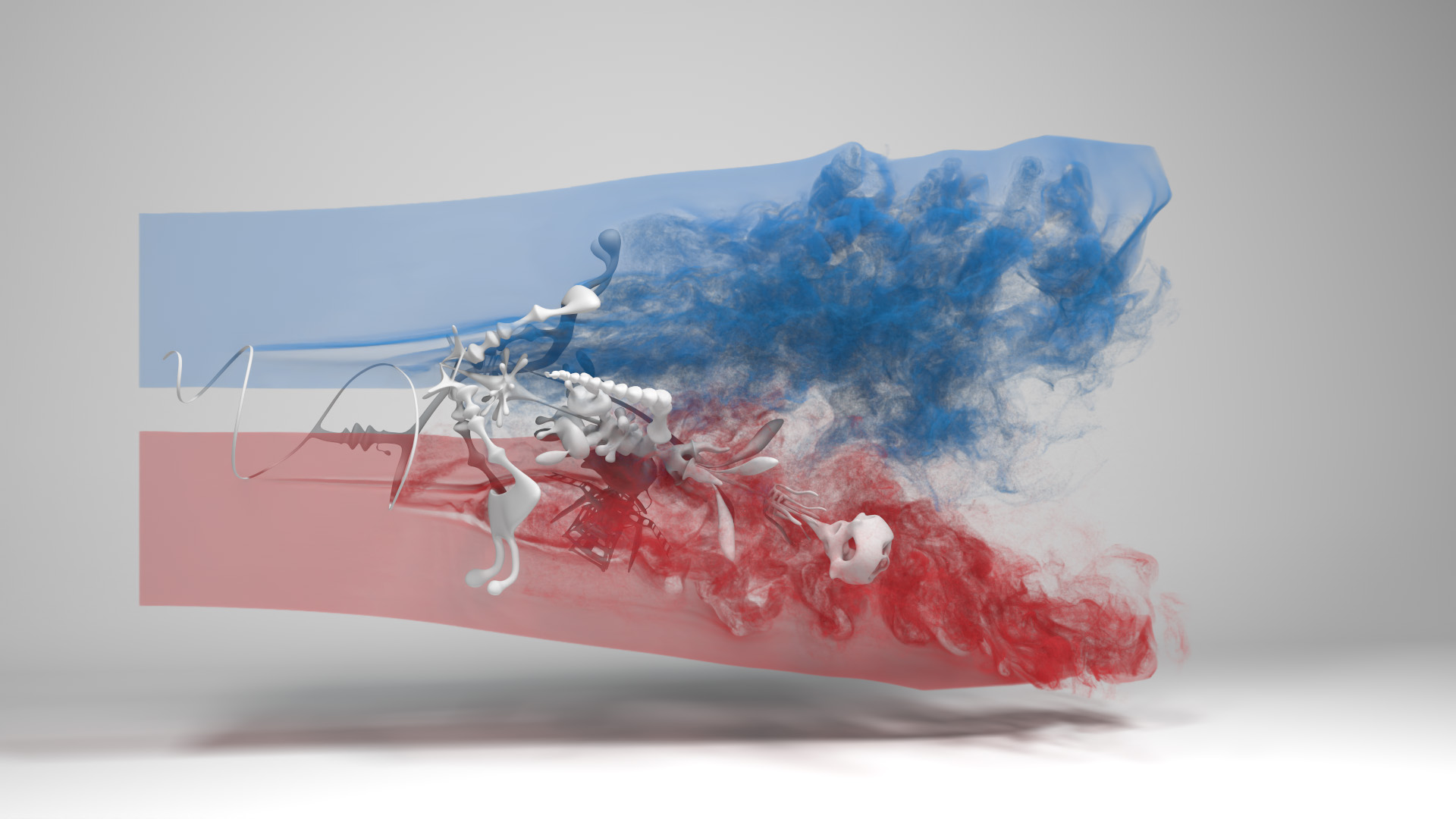}\\
\includegraphics[width=0.333\linewidth,clip,trim=3cm 2.54cm 3.92cm 3.54cm]{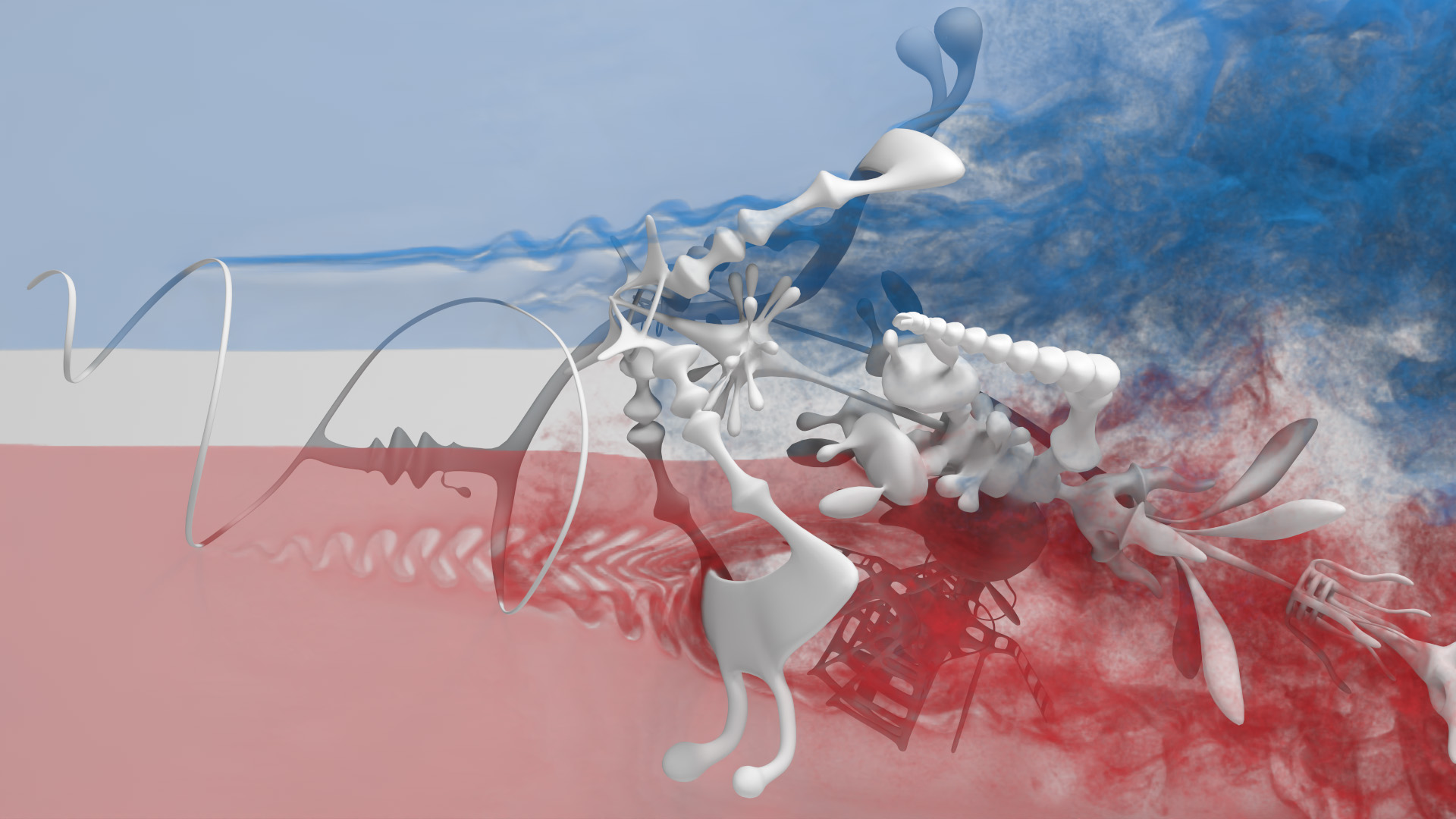}\hfill
\includegraphics[width=0.333\linewidth,clip,trim=3cm 2.54cm 3.92cm 3.54cm]{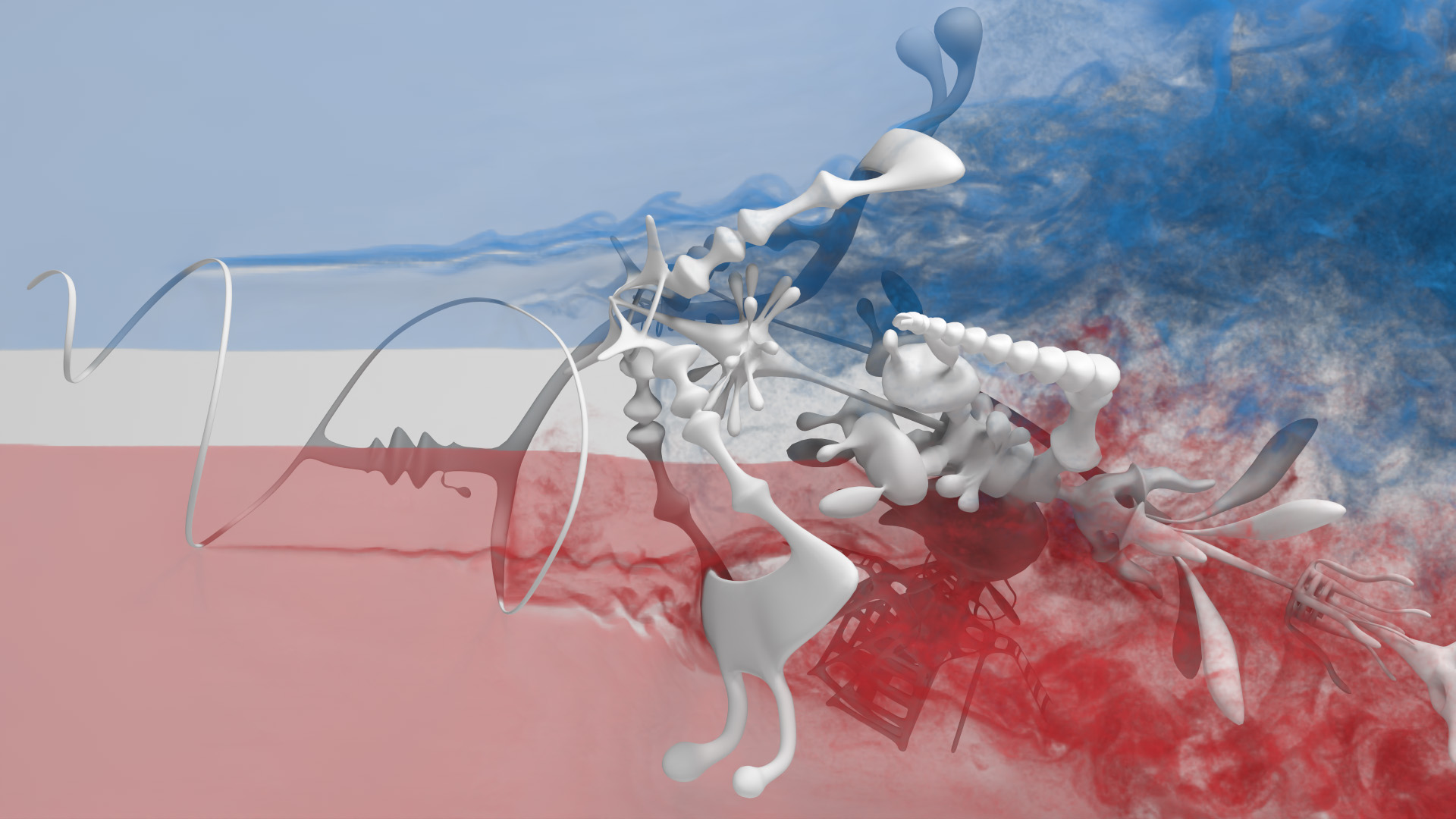}\hfill
\includegraphics[width=0.333\linewidth,clip,trim=3cm 2.54cm 3.92cm 3.54cm]{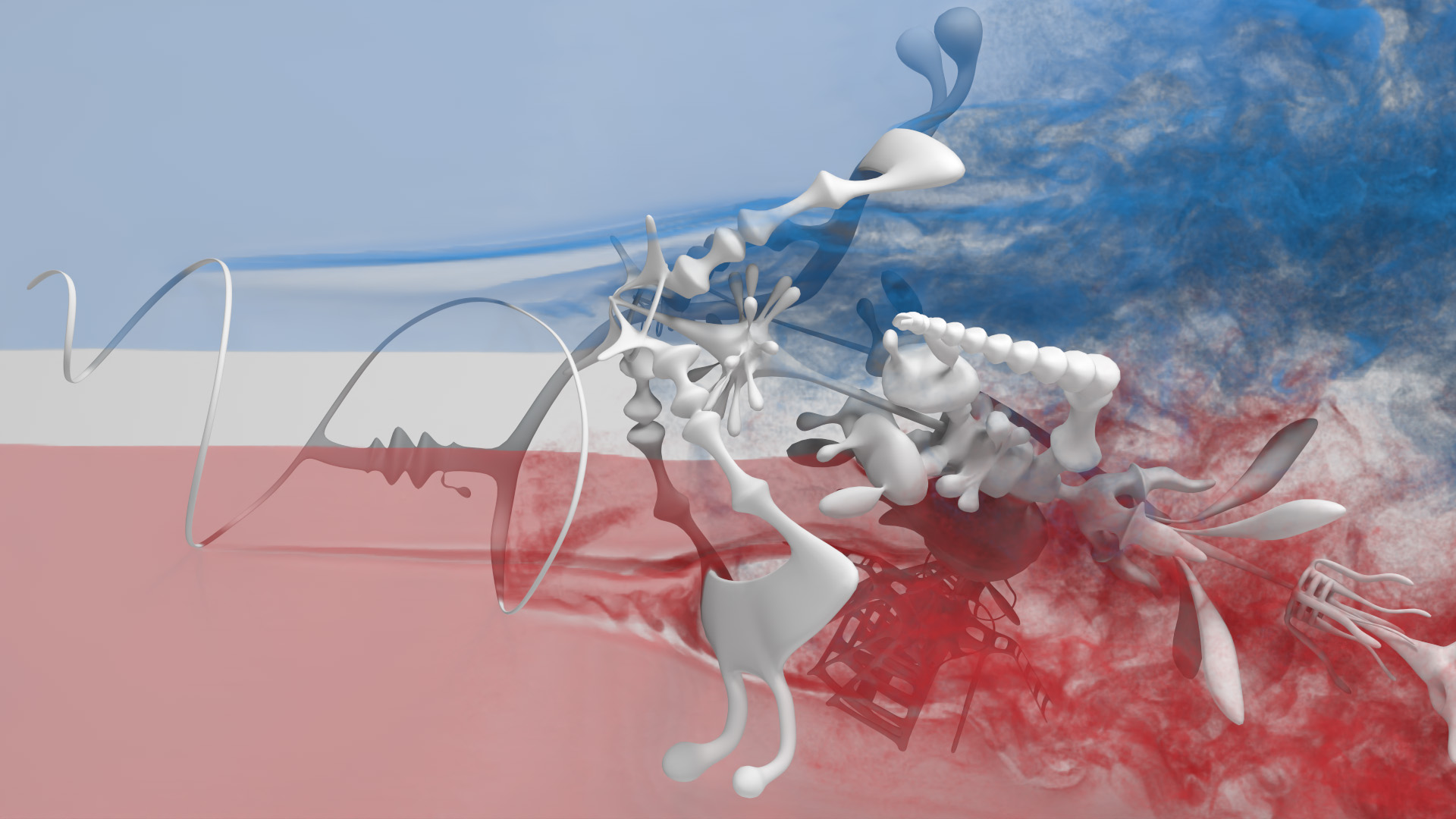}
\caption{Comparison of different lattice velocity sets on the 3D yeahright mesh at different mesh resolutions, all on the same $720 \times 360 \times 360$ grid. Left: D3Q27 with a low-resolution mesh (188k triangles), middle: D3Q27 with a high-resolution mesh (754k triangles), and 
right: D3Q19 with the same high-resolution mesh (754k triangles).}
\label{fig:yeahright_3d}
\Description{}
\end{figure*}

\subsection{Accuracy of the Geometry-Decoupled Formulation}
\label{sec:boundary_accuracy}
\begin{figure}[ht]
\centering
\includegraphics[width=0.6\columnwidth]{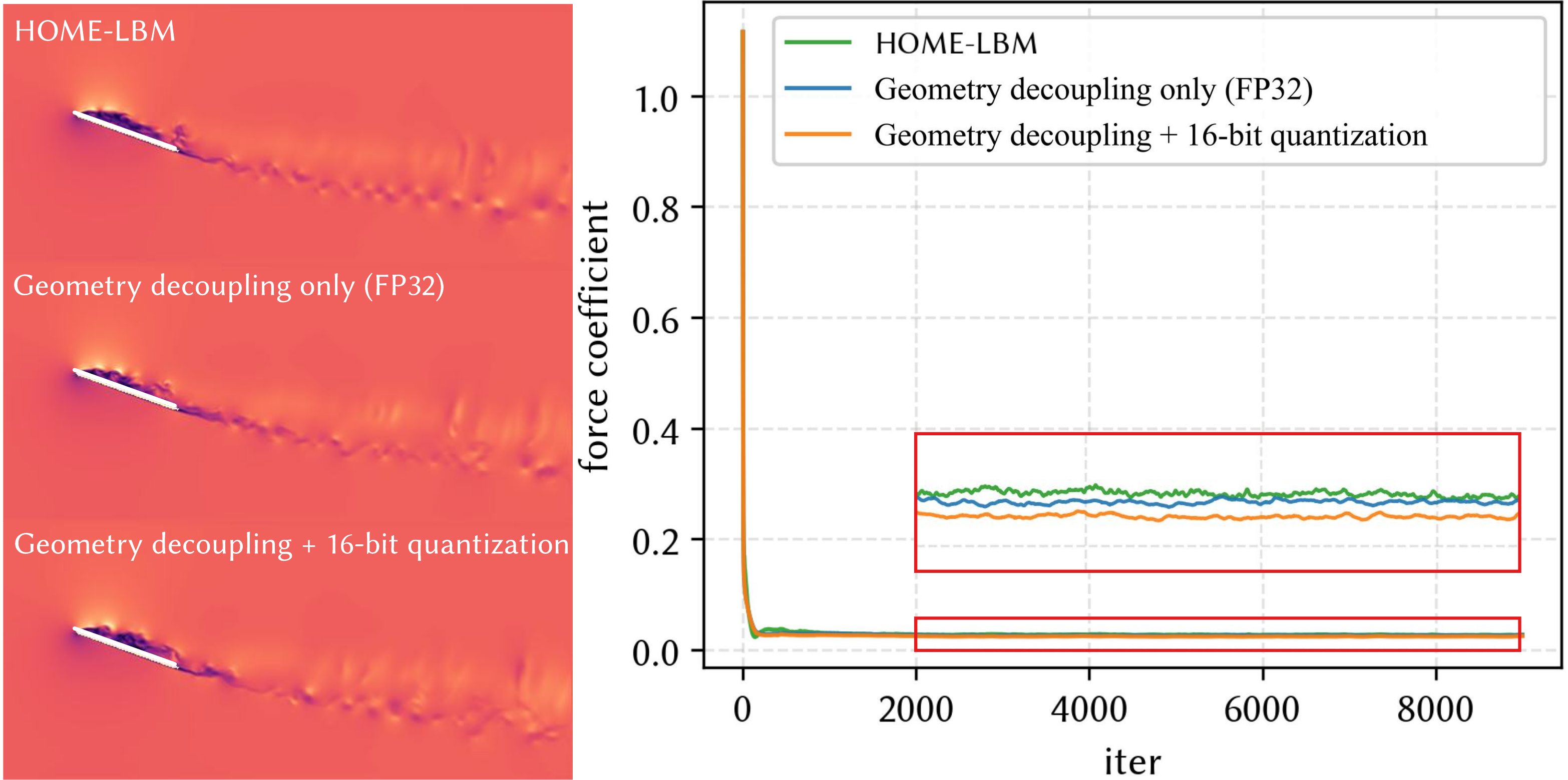}
\caption{\textcolor{smOrange}{Accuracy of the geometry-decoupled formulation on airflow over a thin-shell delta wing (the scene of~\autoref{fig:deltawing_3d}). Left: velocity-magnitude cross-sections for HOME-LBM~\cite{li2023high} (reference), geometry decoupling only (FP32), and geometry decoupling with 16-bit quantization. Right: drag-coefficient histories of the three solvers over 9000 iterations, where the lower red box marks the window of iterations 2000--9000 used for~\autoref{tab:boundary_processing}.
}}
\Description{}
\label{fig:boundary_processing}
\end{figure}

In~\autoref{fig:boundary_processing}, we evaluate the accuracy of our geometry-decoupled boundary formulation. \autoref{fig:boundary_processing} (left) visualizes cross-sections of the velocity magnitude for airflow over a thin-shell delta wing, comparing HOME-LBM~\cite{li2023high}, our geometry-decoupled formulation, and our geometry-decoupled formulation with quantization. The results show that both methods produce tail-flow structures that differ negligibly from the reference generated by HOME-LBM~\cite{li2023high}. All three methods produce consistent vortical structures near the leading edge and exhibit similar long-term force-coefficient behavior after the initial transient phase.

To quantify the accuracy, we compare the evolution of the drag force coefficient from $2000$ to $9000$ iterations after the initial transient phase in~\autoref{fig:boundary_processing} (right). \autoref{tab:boundary_processing} reports the corresponding performance improvement of each optimization strategy and the quantitative error statistics for the drag-coefficient comparison. Our decoupled solver without quantization achieves a relative error of $2.9\%$, while providing a $2.8\times$ speedup over the original HOME-LBM solver. Enabling $16$-bit quantization further increases the speedup to $4.1\times$ while increasing the relative error to $3.6\%$. 

\textcolor{smBlue}{As discussed in~\autoref{sec:gpu_optimization}, the discrepancy has two sources, both present in 32-bit precision: the boundary correction is reconstructed from the post-streaming moments at $t + 1$, whereas the distribution it replaces was streamed from the post-collision moments at $t$; and the per-triangle correction cannot resolve multi-layer geometry, so links that cross both faces of the thin shell within one lattice spacing receive more than one correction, whereas the original kernel replaces them once at their nearest intersection. Quantization adds $0.7\%$ points on top of these two effects.}

\begin{table}[ht]
\setlength{\tabcolsep}{3pt}
\caption{\textcolor{smOrange}{Performance and the relative drag error of our solver with geometry decoupling only (FP32) and with geometry decoupling plus 16-bit quantization}, compared with HOME-LBM~\cite{li2023high}, for the example shown in \autoref{fig:boundary_processing}.
}
\label{tab:boundary_processing}
\footnotesize
\begin{tabular}{l c c c}
\toprule
Method & {\thead{Time \\ (ms/iter)}} & Speedup & {\thead{Relative \\ drag error}} \\
\midrule
HOME-LBM (baseline) & 37.3 & - & -\\
Geometry decoupling only (FP32) & 13.1 & $2.8\times$ & $2.9\%$\\
Geometry decoupling + 16-bit quantization & \hspace{0.5em}9.1 & $4.1\times$ & $3.6\%$\\
\bottomrule
\end{tabular}
\end{table}

\begin{figure*}[t]
\centering
\includegraphics[width=\linewidth]{images/trefoil_3d.pdf}
\caption{\textcolor{smBlue}{Comparison of vorticity evolution results on the 3D trefoil-knot
benchmark, using Semi-Lagrangian~\cite{Stam1999}, MacCormack~\cite{Selle2008}, MC+R~\cite{Zehnder2018}, 
BiMocq$^2$~\cite{Qu2019}, CF and CF+MCM~\cite{nabizadeh2022covector}, the HOME-LBM~\cite{li2023high} FP32 baseline, and our 16-bit quantized solver. We show vortex evolution at three successive times (top) and the histories of (a) kinetic energy, (b) enstrophy, and (c) relative mass drift $\Delta m/m_0$ of the two LBM solvers (bottom). All methods use the same grid resolution of $256 \times 128 \times 128$.}}
    \Description{}
    \label{fig:3d_trefoil}
\end{figure*}

\subsection{Performance and Scalability}
\label{sec:results_gpu_optimization}


\paragraph{Yeahright}
\autoref{fig:yeahright_3d} shows a coupled fluid--solid simulation in which smoke flows through a humanoid skeleton under different lattice configurations and mesh resolutions while keeping the fluid resolution fixed. The three columns correspond to D3Q27 with a low-resolution mesh (left), D3Q27 with a high-resolution mesh (middle), and D3Q19 with a high-resolution mesh (right). The overall flow structure remains visually consistent across all configurations; however, noticeable differences emerge primarily near fine geometric features, such as the slender rod elements of the skeleton. In these regions, higher mesh resolution with the D3Q27 lattice produces more detailed, coherent vortical structures, whereas coarser discretizations tend to smooth or attenuate small-scale vorticity.
\textcolor{smOrange}{From a performance perspective, \autoref{table:timing_memory} reports both solvers on the two meshes: from 188k to 754k triangles, the original HOME-LBM grows from 132.7 to 139.3~ms per iteration, because BVH traversal and link--triangle intersection tests inside its fused update kernel become more expensive, and our quantized solver grows from 17.2 to 18.5~ms, with the entire increase in the solid-correction stage (0.8 to 2.1~ms) while the fluid update and the buffer copy are unchanged. On this mesh, refinement subdivides a surface whose footprint on the lattice barely changes, so both costs grow only mildly. The Hilbert-curve sequence below isolates the effect of increasing obstacle complexity.}

\paragraph{Hilbert Space-filling Curve}
To further evaluate our solver under highly intricate boundaries, we simulate flow past obstacles generated by 3D Hilbert curves with increasing iteration levels, as shown in \autoref{fig:hilbert_3d}. As the curve density increases, the resulting geometric complexity creates increasingly tortuous flow paths and stronger local flow confinement. Our solver captures the resulting multi-scale vortical structures and turbulent cascades. We also observe a clear morphological change in the downstream wake: at higher curve densities, the wake contracts inward because the densely packed obstacles induce a stronger pressure gradient, as illustrated in \autoref{fig:hilbert_3d} (LoD 3).

\textcolor{smBlue}{Across LoD~0--3, the original HOME-LBM runtime increases from $33.8$ to $214.6\,\mathrm{ms}$ per iteration, while our quantized solver increases from $5.5$ to $6.6\,\mathrm{ms}$. The FP32 variant with geometry decoupling only also shows lower runtime growth, reaching a $27.9\times$ speedup at LoD~3; quantization further increases this to $32.5\times$. This comparison distinguishes the performance benefit of geometry decoupling from the additional benefit of quantization. In the quantized variant, fluid update and buffer copy remain at $3.1$ and $2.2\,\mathrm{ms}$, while solid correction increases from $0.2$ to $1.3\,\mathrm{ms}$. The lower runtime growth thus holds across this tested refinement sequence, in which both triangle count and obstacle structure change.}

\subsection{Comparisons with SOTAs}

\subsubsection{Compared to~\citet{li2023high}}
\begin{figure}[t!]
\centering
\newcommand{\figcap}[1]{\begin{minipage}{0.33\linewidth}\centering#1\end{minipage}}
\includegraphics[width=\linewidth]{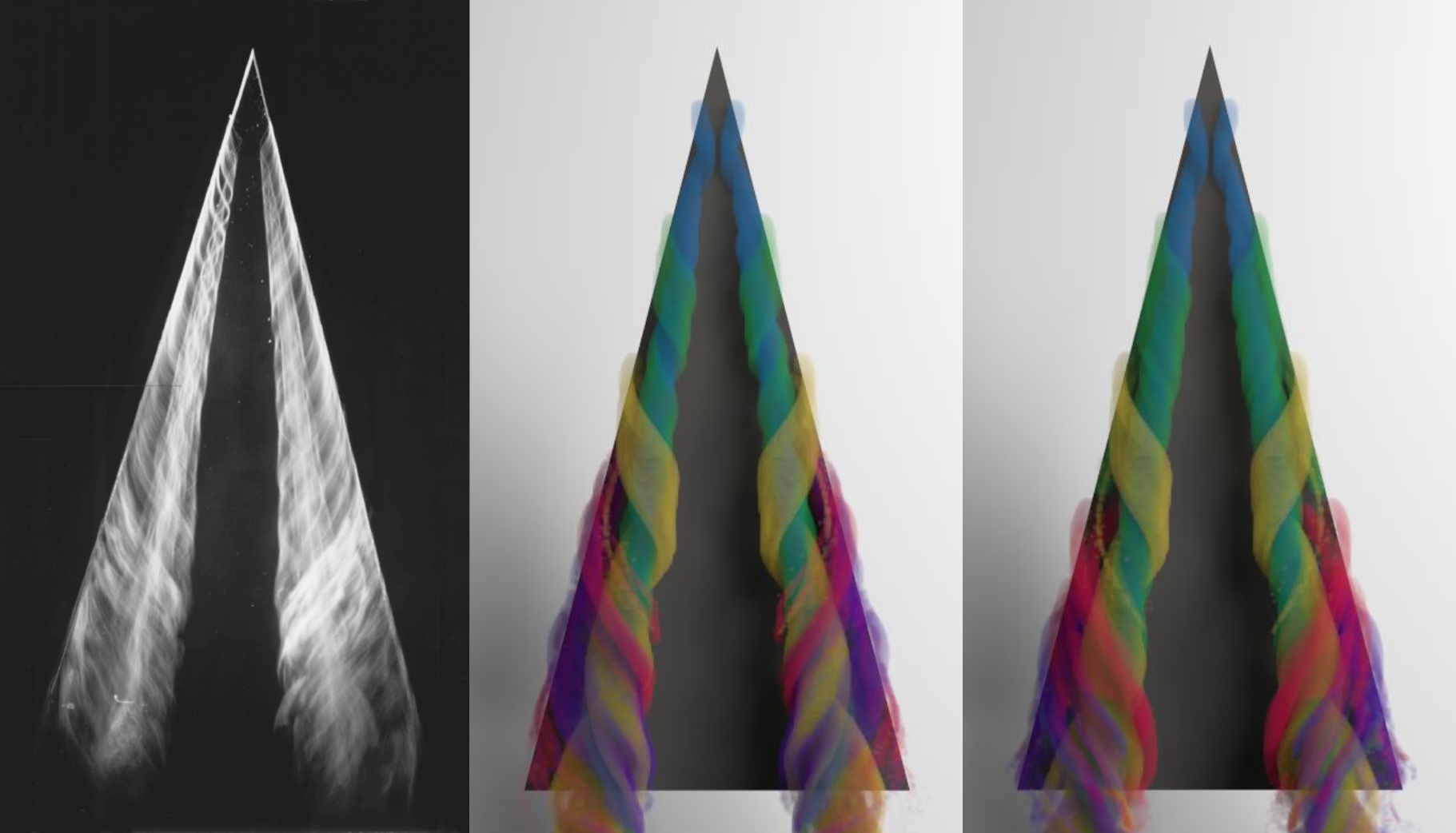}
\figcap{\small Real experiment}\hfill
\figcap{\small HOME-LBM}\hfill
\figcap{\small Ours}
\caption{Comparison with real experimental flow visualization reported by~\citet{delery2001robert}, HOME-LBM~\cite{li2023high}, and our quantized kinetic solver on an example of airflow over a thin-shell delta wing simulated at a grid resolution of $660\times 250 \times 330$, achieving a $4.1\times$ speedup over HOME-LBM, with $2.8\times$ contributed by the geometry decoupling alone and additional gains from 16-bit quantization. The simulation reproduces the characteristic spiral vortex structures that form near the wing's leading edges, closely matching the real experiment. }
\Description{}
\label{fig:deltawing_3d}
\end{figure}

We compare the end-to-end simulation performance of our GPU-oriented geometry-decoupled implementation with the original single-kernel HOME-LBM algorithm~\cite{li2023high}. Both implementations use identical collision models, boundary conditions, and time step sizes.
In terms of memory footprint, the proposed solver reduces storage by approximately $50\%$ in fluid-only simulations and $25\%$ in fluid--solid coupling scenarios. 
\textcolor{smBlue}{In terms of runtime, it is $4.1\times$ to $32.5\times$ faster than HOME-LBM on the thirteen configurations the original implementation supports (\autoref{table:timing_memory})}. The speedup comes from regular memory access patterns and reduced warp divergence in the fluid update, and from the sparse solid-correction stage, which restricts geometric execution to boundary-intersecting cells and significantly reduces unnecessary thread activity. For airflow over the thin-shell delta wing shown in~\autoref{fig:deltawing_3d}, our method achieves a $4.1\times$ speedup, $2.8\times$ of which comes from geometry decoupling alone (\autoref{tab:boundary_processing}), and reproduces the spiral vortices near the wing edges seen in the experiment of~\citet{delery2001robert}.

\subsubsection{Compared to~\citet{Sun2025Leapfrog}}

We compare our method with the state-of-the-art real-time fluid--solid coupling approach~\cite{Sun2025Leapfrog} using both static and moving-boundary examples. 

\begin{figure}[t]
\centering
\begin{tabular}{
    @{}c@{}
    c@{}c@{}
}
\multicolumn{3}{c}{\small Angle of attack = $0^\circ$}
\\

\includegraphics[
    trim=150 150 50 0, clip, width=0.35\linewidth
]{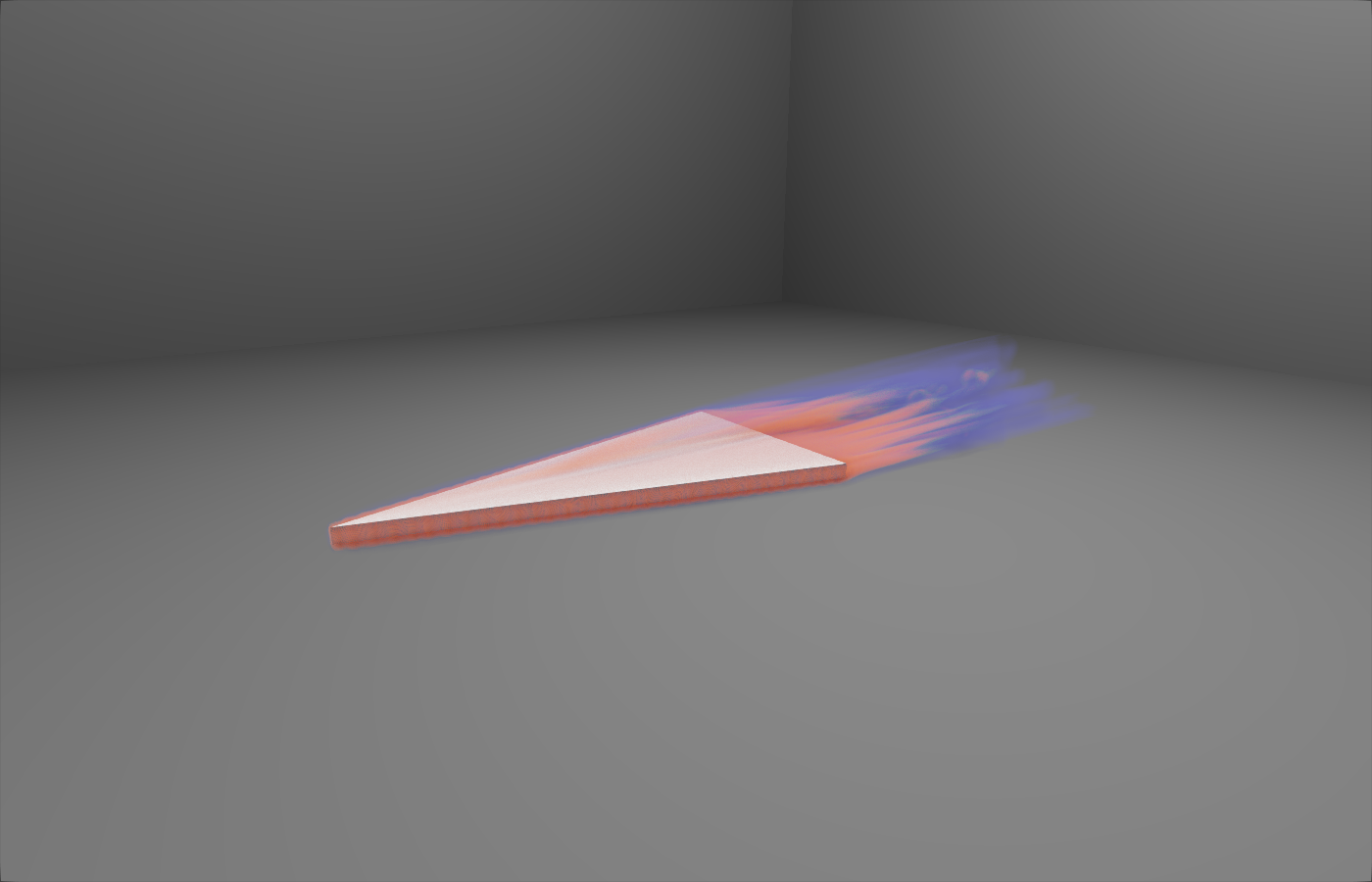}
&
\includegraphics[
    trim=150 150 50 0, clip, width=0.35\linewidth
]{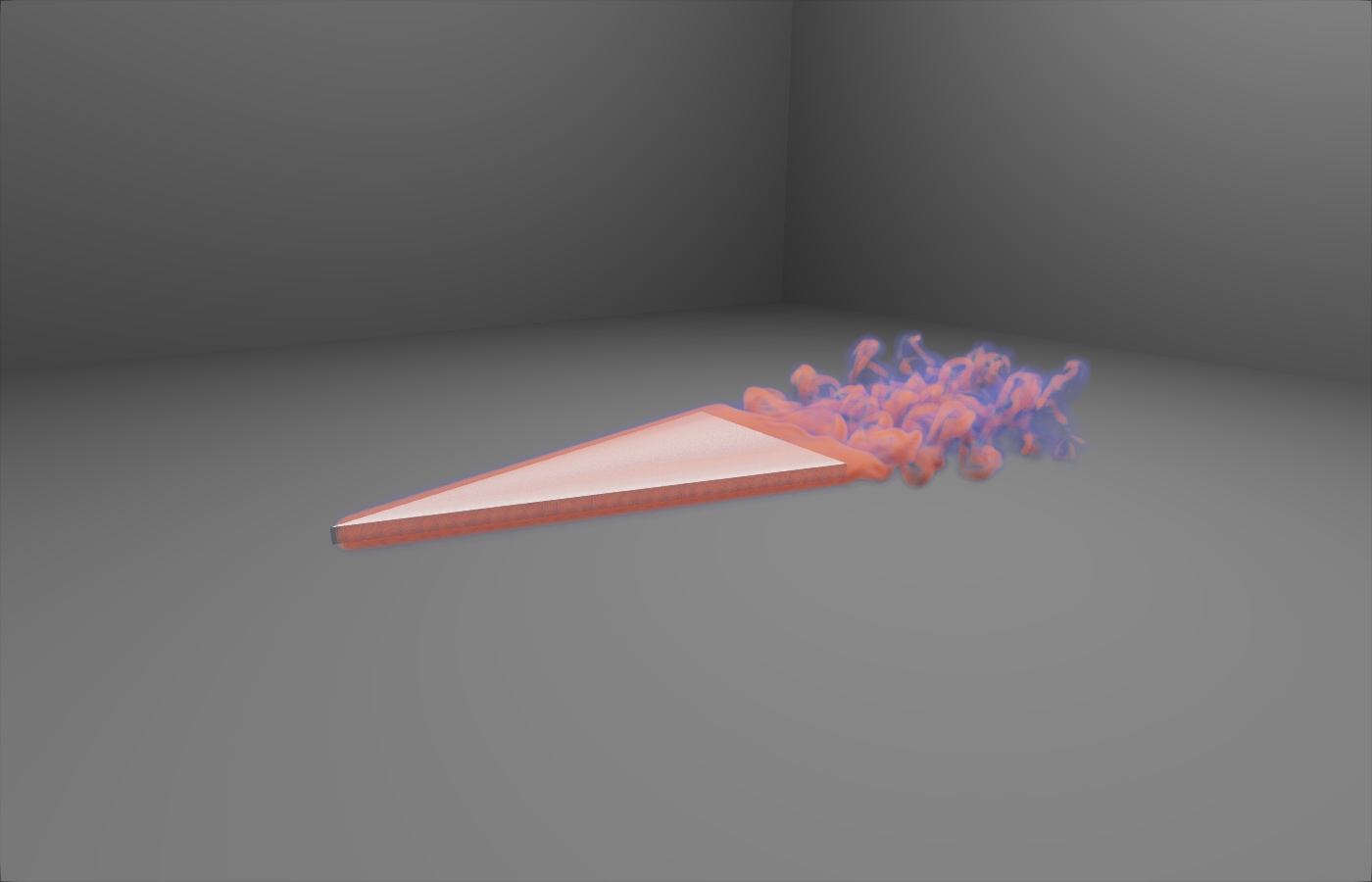}
&
\includegraphics[
    trim=150 0 150 200, clip, width=0.28\linewidth
]{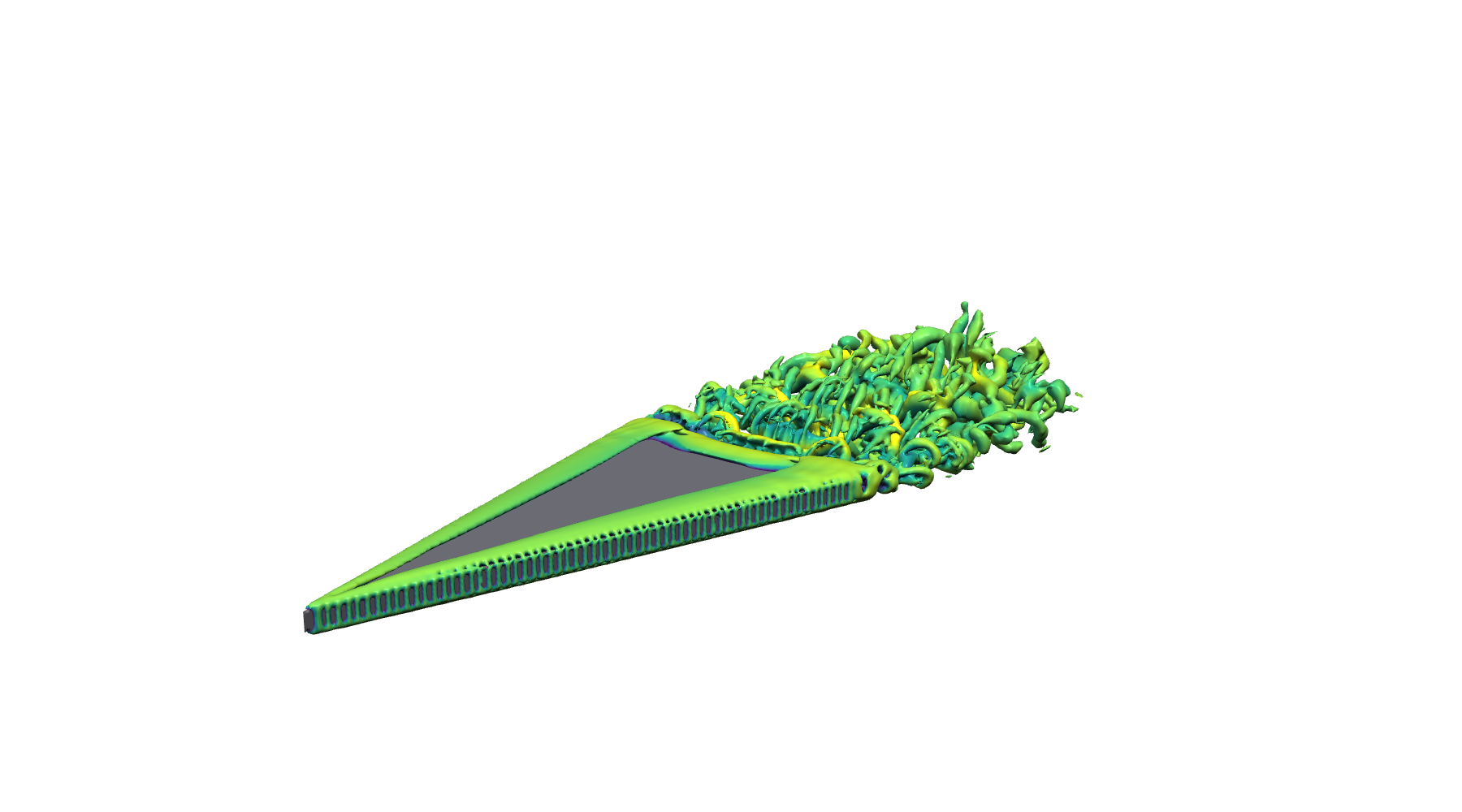}
\\

\multicolumn{3}{c}{\small Angle of attack = $20^\circ$}
\\

\includegraphics[
    trim=150 150 50 0, clip, width=0.35\linewidth
]{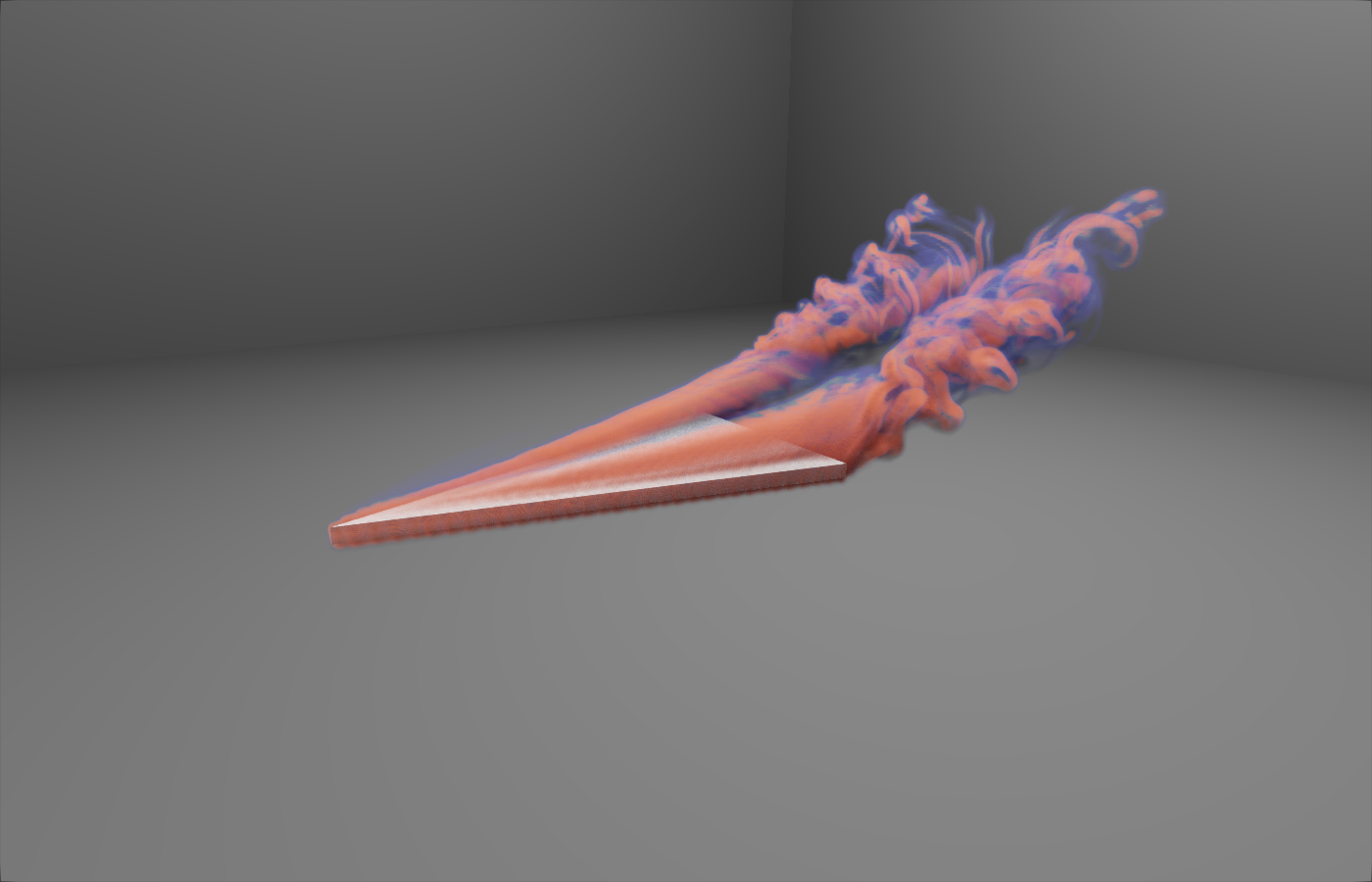}
&
\includegraphics[
    trim=150 150 50 0, clip, width=0.35\linewidth
]{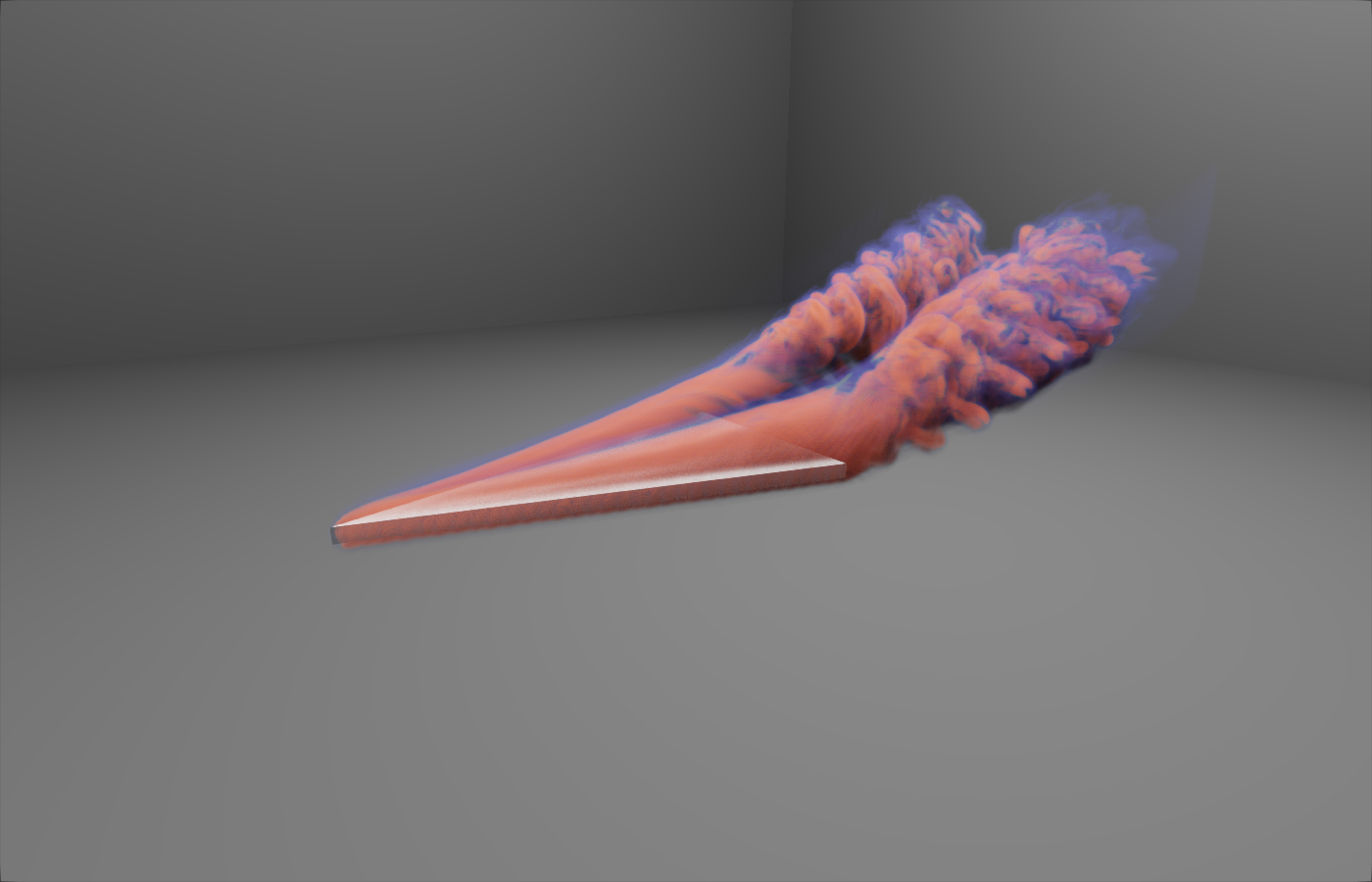}
&
\includegraphics[
    trim=150 0 150 200, clip, width=0.28\linewidth
]{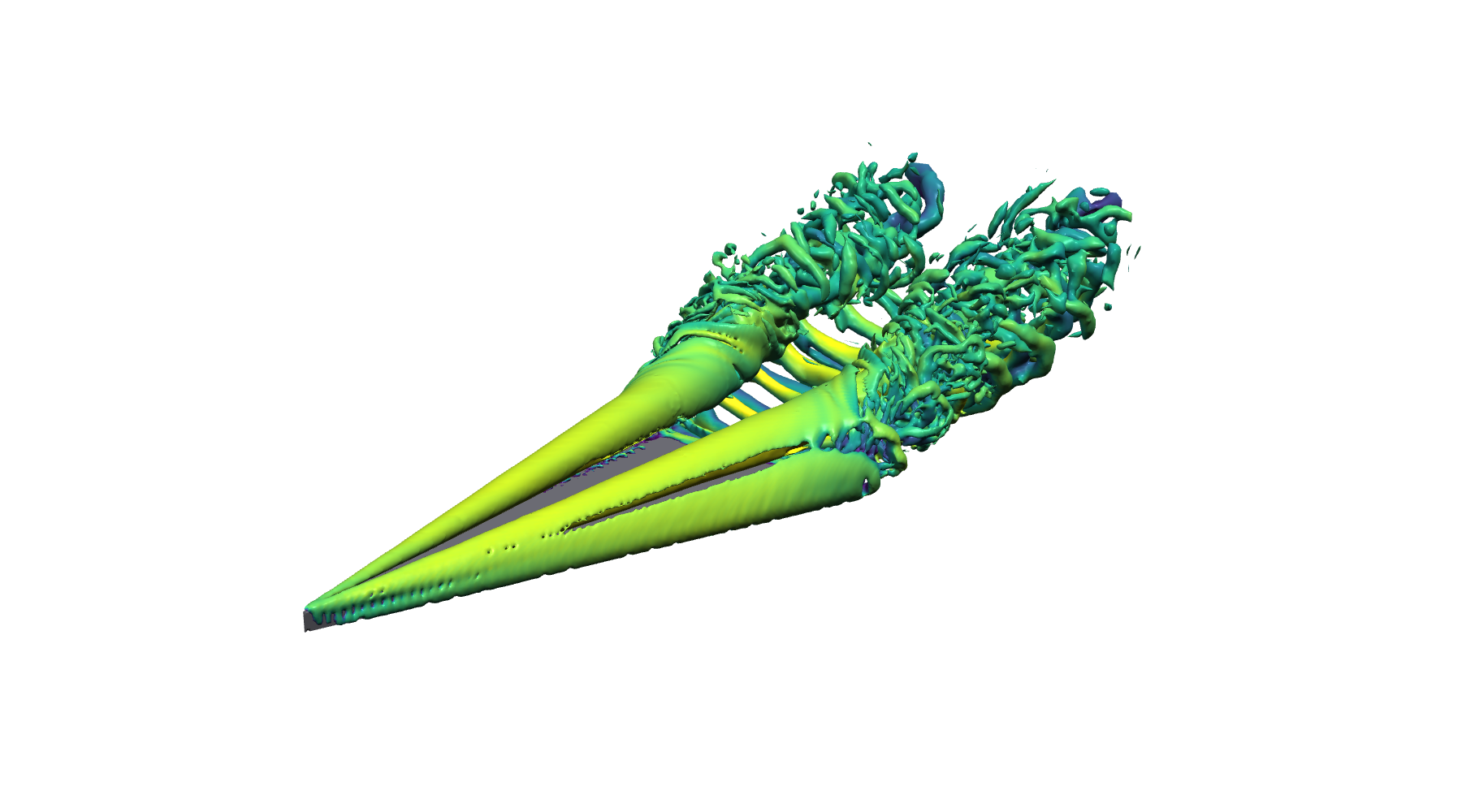}
\\

{\small LFM}
&
{\small Ours}
&
{\small OpenFOAM}
\end{tabular}

\caption{Comparison of our real-time 3D simulation and Leapfrog Flow Maps (LFM) by~\citet{Sun2025Leapfrog} on a static delta wing at a grid resolution of $256 \times 128 \times 128$. Two angles of attack are shown: $0^\circ$ and $20^\circ$. For each angle, LFM with its default reinitialization interval ($n=5$) is shown on the left, followed by our method in the middle. LFM and our method apply the same visualization settings. \textcolor{smBlue}{An additional OpenFOAM~\cite{jasak2009openfoam} result, obtained using the same input surface mesh, is shown on the right, visualized by $Q$-criterion isosurfaces. It serves as a qualitative reference only.}}
\label{fig:deltawing_comparison}
\Description{}
\end{figure}

\paragraph{Static 3D Delta Wing}
We first evaluate the static delta-wing benchmark at two angles of attack, $0^\circ$ and $20^\circ$ (\autoref{fig:deltawing_comparison}). LFM and our solver use a $256\times128\times128$ grid with identical boundary conditions, inflow settings, and visualization parameters. \textcolor{smBlue}{The wing is a closed mesh of finite thickness, so the flow can separate at its edges and shed vortices even at $0^\circ$ angle of attack. We additionally perform an OpenFOAM reference simulation~\cite{jasak2009openfoam} using the same finite-thickness surface mesh, with \texttt{pimpleFoam} and the WALE LES model. The locally refined OpenFOAM mesh has a finest cell spacing matching that of a $512\times256\times256$ uniform grid over the same domain. The OpenFOAM result provides a qualitative reference for wake comparison. At $0^\circ$, OpenFOAM captures distinct vortical structures behind the wing, with a wake extent closer to ours than to LFM's; at $20^\circ$, all methods produce the characteristic leading-edge vortex pair. Note that OpenFOAM is visualized using $Q$-criterion isosurfaces, whereas LFM and our solver are rendered as smoke.}

\textcolor{smOrange}{Our solver requires approximately $0.71$~ms per LBM iteration. Since three LBM iterations advance the same physical time as one LFM step, the corresponding cost is $2.13$~ms, compared with $11.1$~ms per LFM step measured under the same setup, resulting in a $5.2\times$ speedup}. The LFM cost and accuracy depend on the reinitialization interval $n$ and time step; we use the published $n=5$ configuration as a practical setting. Therefore, the reported speedup reflects this configuration and is not intended as an accuracy-matched comparison.

\begin{figure}[ht!]
\centering
\newcommand{\figcap}[1]{%
    \vspace*{1mm} 
    \begin{minipage}{0.498\linewidth}
        \centering
        #1
    \end{minipage}
}
\newcommand{\figcapp}[1]{\begin{minipage}{0.5\linewidth}\centering#1\end{minipage}}
\includegraphics[trim = 360 190 360 190, clip, width=0.498\linewidth]{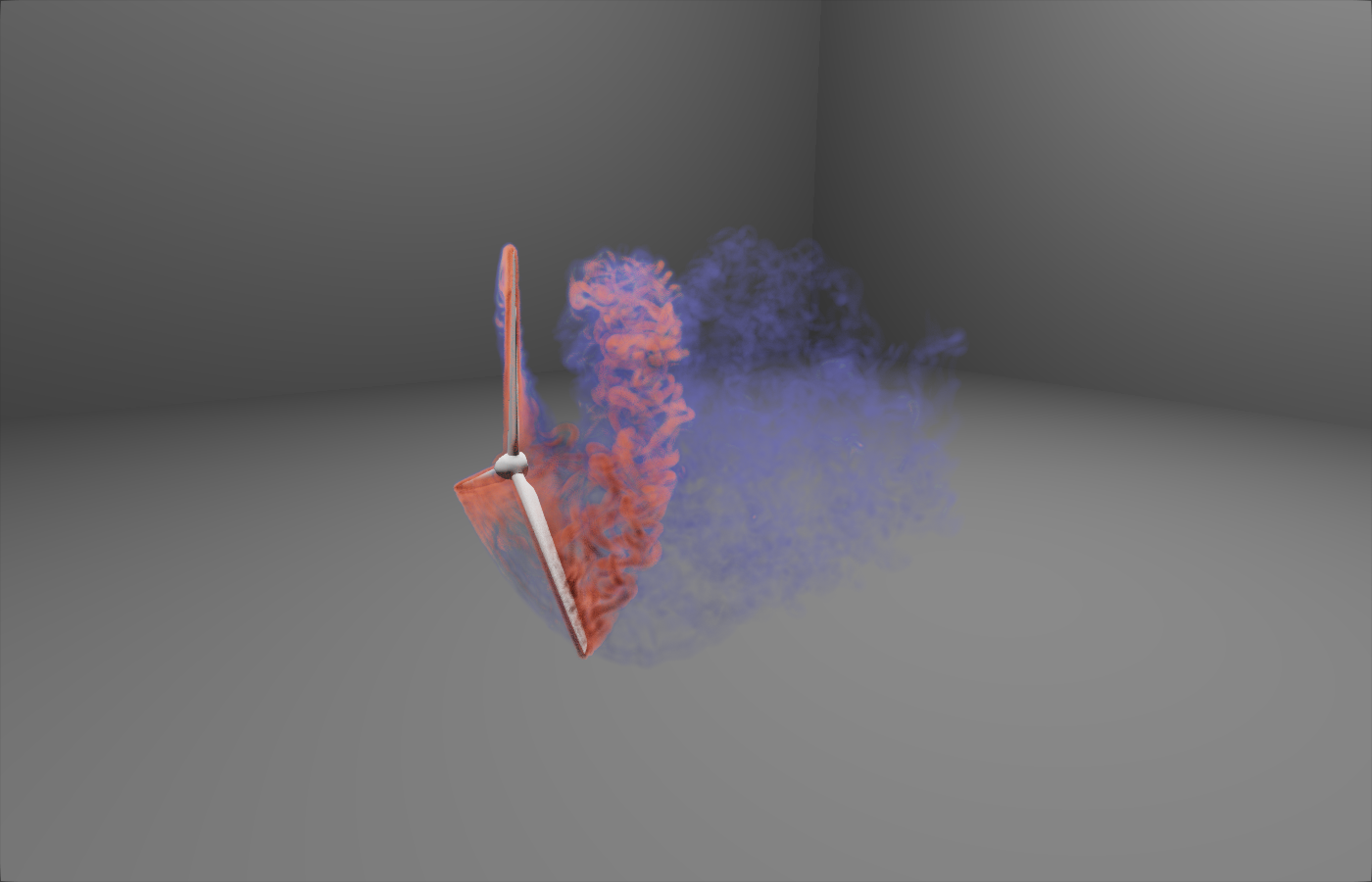}\hfill
\includegraphics[trim = 360 190 360 190, clip, width=0.498\linewidth]{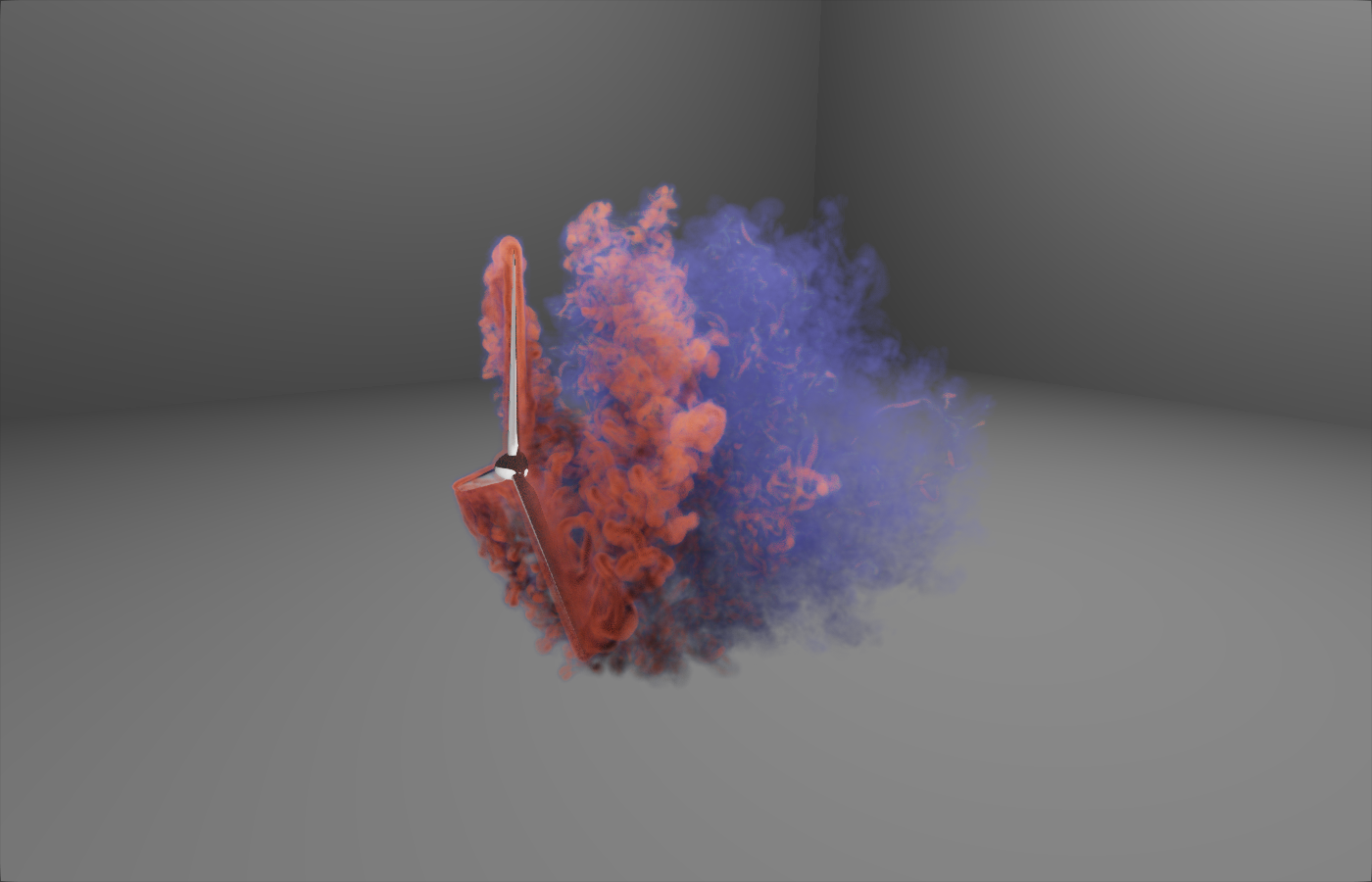}\\
\figcapp{\small LFM}\hfill
\figcapp{\small Ours}
\caption{Ours and the Leapfrog Flow Maps (LFM) method \cite{Sun2025Leapfrog} with a rotating propeller, \textcolor{smOrange}{both at frame 200 (the same physical time and rotation phase)}, at a grid resolution of $256 \times 256 \times 256$. All simulations are rendered using the same visualization settings.}
\label{fig:turbine_comparison}
\Description{}
\end{figure}

\paragraph{Rotating 3D Propeller}
We further compare our method with~\citet{Sun2025Leapfrog} on a rotating-propeller benchmark with dynamic solid boundaries whose motion is prescribed, as shown in \autoref{fig:turbine_comparison}. In this setting, the solid geometry must be voxelized dynamically at each time step, introducing an average overhead of approximately $2.1\,\mathrm{ms}$ per \textcolor{smOrange}{LBM iteration} in our implementation. Including this cost, our solver takes \textcolor{smOrange}{$5.16\,\mathrm{ms}$ per LBM iteration (\autoref{table:timing_memory}) and two LBM iterations advance the same physical time as one LFM step, taking approximately $10.3\,\mathrm{ms}$ in total.} Compared to the $40.7\,\mathrm{ms}$ per step of~\citet{Sun2025Leapfrog}, our solver achieves nearly a $4\times$ speedup.

\section{Applications}


\paragraph{$1024^3$ Fluid Simulation.} 
To demonstrate the scalability and fidelity of our solver at extreme resolutions, we simulate two cases of a plume formed by obliquely impacting jets on a $1024^3$ grid at different Reynolds numbers, as shown in~\autoref{fig:plume_3d}. \textcolor{smBlue}{The Reynolds number is defined as $\mathrm{Re} = UL/\nu$ with the inlet speed $U = 0.2$ and the domain edge length $L = 1024$ lattice units.} At lower Reynolds numbers, the flow exhibits stable, coherent vortex structures, while at higher Reynolds numbers, the solver resolves fine-scale turbulent filaments and intermittent eddies across a wide range of spatial scales. 
Our quantized formulation reduces the memory footprint by $50\%$ compared to FP32 baselines, enabling large-scale simulations that would otherwise exceed the GPU memory capacity of a single GPU. This demonstrates that our framework can efficiently support high-resolution fluid simulations while preserving rich small-scale flow details.

\paragraph{Large-scale Fluid--solid Coupling.} 
We further demonstrate large-scale fluid--solid interaction using turbulent smoke simulations around a Ducati motorcycle and a Ferrari F1 racing car at a grid resolution of $1000\times400\times400$, as shown in~\autoref{fig:teaser} and~\autoref{fig:highres_f1}. \textcolor{smBlue}{Both simulations use a closed computational domain with stationary no-slip walls.} The simulations use 16-bit quantization and incorporate highly detailed solid geometry with intricate boundary features. Despite the large simulation size and geometric complexity, the solver develops complex shear layers and coherent vortex structures while maintaining stable temporal evolution and consistent visual quality over long sequences, achieving an average performance of approximately 1 second per frame. This shows that the proposed framework efficiently supports high-resolution, geometry-rich fluid--solid simulations while preserving fine-scale flow details.

\begin{figure}[ht!]
    \centering
    \includegraphics[trim = 0 1090 1920 0, clip, width=\linewidth]{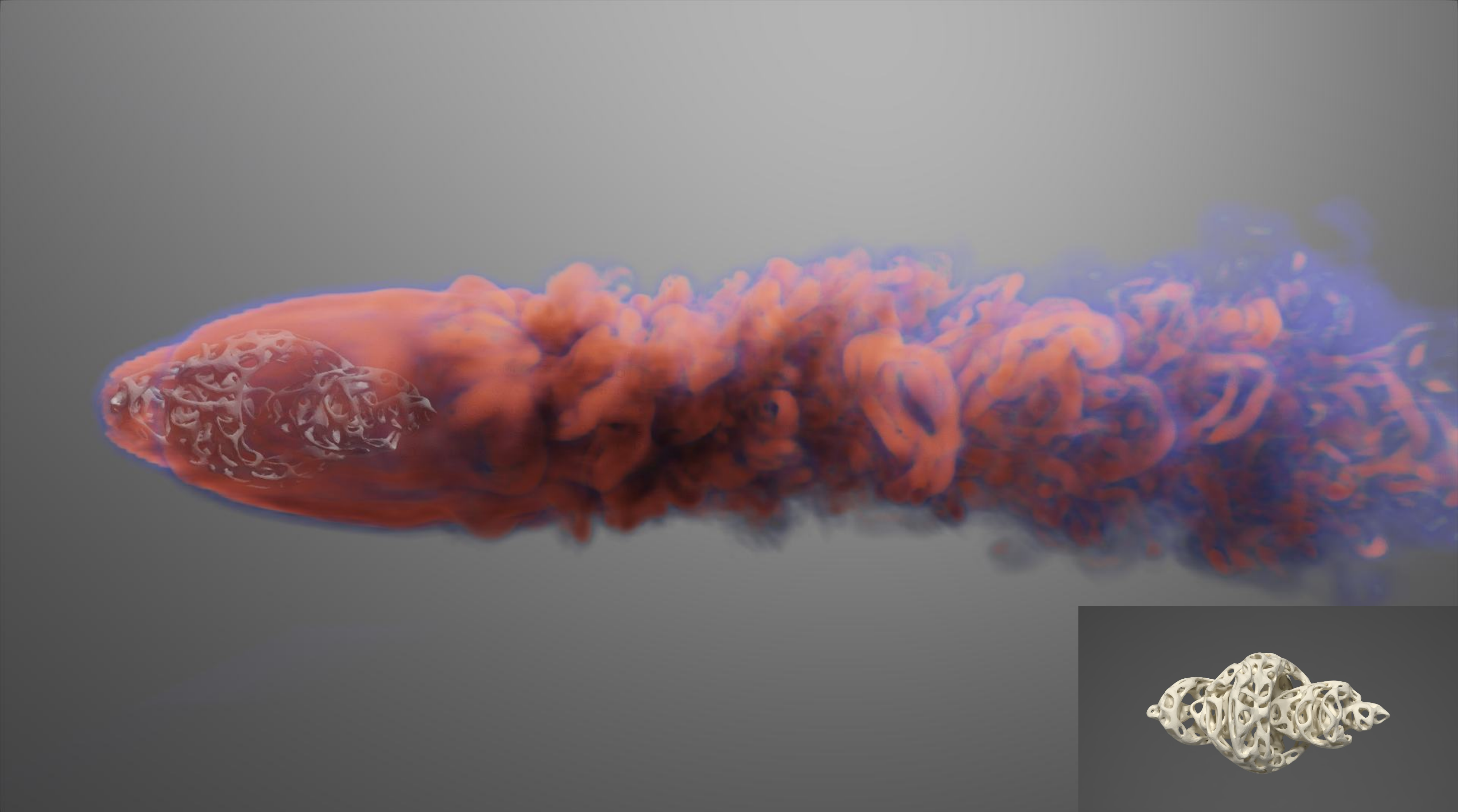}
    \caption{Real-time 3D smoke simulation through a porous inlet with intricate microgeometry (128k triangle faces) on a $256 \times 128 \times 128$ grid.
    }
    \Description{}
    \label{fig:porous}
\end{figure}

\paragraph{Real-time Simulation}
We demonstrate real-time smoke simulations interacting with solids that exhibit a highly complex porous microgeometry, as shown in~\autoref{fig:porous}. The dense, irregular fluid--solid interfaces generate intense multi-scale turbulence while simultaneously imposing stringent stability and performance demands on the solver. Despite these challenges, our method remains numerically stable and consistently sustains real-time performance across all tested porous configurations. As reported in \autoref{table:timing_memory}, the example run at a grid resolution of $256\times128\times128$ took $1.1\,\mathrm{ms}$ per iteration.

\section{Conclusion}
\label{sec:conclusion}
We presented a stability-constrained GPU framework for memory-efficient, high-performance moment-based lattice Boltzmann simulation with complex fluid--solid interaction. The proposed geometry-decoupled computation scheme decouples dense fluid evolution from sparse boundary correction, substantially reducing warp divergence and improving GPU utilization in complex geometries. In addition, our von Neumann stability analysis provides moment-wise stability constraints that inform the design of a robust reduced-precision representation for HOME-LBM. Together, these techniques significantly reduce memory footprint without sacrificing numerical stability. \textcolor{smOrange}{Our experiments demonstrate moment-storage reductions of $50\%$ in fluid-only and $25\%$ in fluid--solid simulations, speedups of $4.1\times$ to $32.5\times$ over the original HOME-LBM across the configurations it supports, and a $5.2\times$ speedup over LFM on the static delta-wing benchmark under the tested configuration, enabling higher-resolution simulations within the constraints of a single GPU at approximately one second per frame for our largest fluid--solid scenes.} More broadly, this work illustrates how theoretical stability analysis can directly inform practical GPU system design for scalable LBM simulation.

\paragraph{Limitations and Future Work}
Our current implementation assumes single-layer surface meshes and does not yet support layered geometry or complex self-intersections, which would require more advanced inside–outside classification and intersection handling. \textcolor{smBlue}{In the current per-triangle correction, links that cross more than one surface layer within one lattice spacing are corrected more than once, which contributes to the boundary discrepancy measured in~\autoref{sec:boundary_accuracy}.} Moreover, while our boundary treatment is optimized for robustness and throughput, we do not explicitly evaluate sub-cell geometric accuracy or boundary error convergence; incorporating wall-function models~\cite{liu2025hybrid} is a natural next step.
%
%
Several promising directions remain, including extending the framework to fully coupled two-way interactions and integrating multi-resolution or adaptive discretization to further improve scalability for large-scale scenes. \textcolor{smBlue}{Our implementation retains two moment buffers, and the buffer copy between the fluid and solid-correction stages accounts for a substantial share of the per-step time in coupled scenes; adapting in-place streaming schemes~\cite{lehmann2022esoteric}, which remove the second distribution buffer in standard LBM, to moment-encoded storage provides a way to reduce this cost.}


\bibliographystyle{ACM-Reference-Format}
\bibliography{bibliography}



\appendix
\section{Standard Stability Analysis of LBM}
\label{app:standard_lbm_stability_analysis}
The standard von Neumann stability analysis for LBM~\cite{sterling1996stability} is a linear Fourier-mode analysis that studies whether small perturbations in the distribution functions grow or decay under the discrete streaming–collision update. By linearizing the dynamics about a uniform equilibrium state and introducing plane-wave perturbations, the method yields an amplification matrix $\GG$ whose eigenvalues characterize linear stability, numerical dissipation, and dispersion across wavenumbers. Here, we briefly summarize the key elements of classical formulations required for extending the analysis to HOME-LBM in \autoref{sec:home-lbm-analysis}. A more detailed review of the classical VN formulation of LBM is provided in~\cite{chavez2018improving}.

\subsection{General Collision–Streaming Update}
Starting from the lattice Boltzmann equation (\autoref{eq:lbe_normalized}) and following~\citet{sterling1996stability} to neglect the external force term for clarity, the discrete kinetic update can be written as:
\begin{equation}
f_i(\bm{x}+\bm{c}_i, t+1) = f_i(\bm{x}, t) + \Omega_i(\bm{f}),
\label{eq:lbe_without_force}
\end{equation}
where $\Omega_i$ is the discrete collision term. We define the post-collision mapping 
\begin{equation}
g_i(\bm{f}) = f_i + \Omega_i(\bm{f}),
\label{eq:g_mapping}
\end{equation}
so that the update becomes
\begin{equation}
f_i(\bm{x}+\bm{c}_i , t+1) = g_i(\bm{f}(\bm{x}, t)).
\label{eq:lbe_without_force_update}
\end{equation}

\subsection{Linearization via Perturbation and Jacobians}
We introduce a small perturbation around a spatially uniform equilibrium distribution:
\begin{equation} \label{eq:fbar}
f_i(\bm{x}, t) = \bar{f}_i + f_i'(\bm{x}, t), \quad \text{with} \quad |\bar{f}_i| \gg |f_i'|,
\end{equation}
where $\bar{f}_i$ denotes the globally uniform equilibrium distribution that does not vary with time and space, and $f_i' (\bm{x}, t)$ at node $\bm{x}$ and time $t$ accounts for a small perturbation from $\bar{f}_i$.
Performing the first-order Taylor expansion of $g_i$ around $\bar{f}_i$ with a small perturbation $f'_i$ yields:
\begin{equation} \label{eq:linearCollision}
    g_i(\bar{\bm{f}} + \bm{f}') \approx g_i(\bar{\bm{f}}) + \sum_{j} \left. \frac{\partial g_i}{\partial f_j} \right|_{\bar{\bm{f}}} f'_j.
\end{equation}
To proceed, we note that most practical LBM collision models can be interpreted as linear relaxation of moments toward their equilibrium values~\cite{higuera1989boltzmann}. When expressed in distribution-function space, this relaxation admits the unified matrix representation:
\begin{equation} \label{eq:collision_models}
   \Omega(\bm{f}) = \bm{K}(\bm{f}^{eq} - \bm{f}),
\end{equation}
where the collision matrix $\bm{K}$ characterizes how the distribution function relaxes toward its equilibrium state. Therefore, the corresponding linearized evolution becomes:
\begin{equation}\label{eq:linearCollision_K}
\bm{g}(\bar{\bm{f}} + \bm{f}')
= \bar{\bm{f}} + \bm{f}'(\bm{x},t) + \bm{K}\bigl(\bm{\Lambda} - \bm{I}\bigr)\,\bm{f}'(\bm{x},t),
\end{equation}
where $\bm{\Lambda}$ is the Jacobian of the equilibrium distribution with elements 
\begin{equation}\label{eq:Lambda_chainrule}
\Lambda_{ij}
=
\frac{\partial f^{eq}_i}{\partial \rho}\frac{\partial \rho}{\partial f_j}
+
\sum_{\alpha}\frac{\partial f^{eq}_i}{\partial u_\alpha}\frac{\partial u_\alpha}{\partial f_j}.
\end{equation}
\textcolor{smBlue}{Without forcing, \autoref{eq:rho_u_stress} gives
$\partial\rho/\partial f_j = 1$ and
$\partial u_\alpha/\partial f_j = (c_{j\alpha}-u_\alpha)/\rho$, and every equilibrium used in this paper is $\rho$ times a polynomial in $\bm{u}$, so that $\partial\bm{f}^{\text{eq}}/\partial\rho = \bm{f}^{\text{eq}}/\rho$. Hence}
{\color{smBlue}
\begin{equation}\label{eq:Lambda_general}
\bm{\Lambda} = \frac{1}{\rho}\Bigl[\bm{f}^{\text{eq}}\,\bm{1}^{\!\top} + \sum_{\alpha}\frac{\partial\bm{f}^{\text{eq}}}{\partial u_\alpha}\,\bigl(\bm{c}_{\alpha} - u_\alpha\bm{1}\bigr)^{\!\top}\Bigr],
\end{equation}
}%
\textcolor{smBlue}{where $\bm{1}\in\mathbb{R}^{q}$ is the vector of ones and $(\bm{c}_\alpha)_j = c_{j\alpha}$. The equilibrium $\bm{f}^{\text{eq}}$ and its derivatives $\partial\bm{f}^{\text{eq}}/\partial u_\alpha$ are given in Appendix~\ref{app:eq2d} for D2Q9 and in Appendix~\ref{app:eq3d} for D3Q27.}

Combining ~\autoref{eq:lbe_without_force_update} and~\autoref{eq:linearCollision_K}, the evolution of the perturbation is given by:
\begin{equation}\label{eq:linearized_fprime}
 f'_i(\bm{x}+\bm{c}_i ,\, t+1) 
=
\Bigl[\bm{I}+\bm{K}\bigl(\bm{\Lambda}-\bm{I}\bigr)\Bigr]_{ij} f'_j(\bm{x},t).
\end{equation}
This equation defines a linear dynamical system and serves as the starting point for the subsequent Fourier (von Neumann) stability analysis.

\subsection{Plane-Wave Substitution and Amplification Matrix}
Following the standard VN procedure, we substitute into~\autoref{eq:linearized_fprime} a plane-wave solution of the form~\cite{wissocq2020linear}:
\begin{equation} \label{eq:solution}
f_i'(\bm{x}, t) =
\hat{f}_i e^{\mathrm{i}(\bm{k}\cdot\bm{x} - \omega t)},
\end{equation}
where $\hat{f}_i$ denotes the complex amplitude, and the exponential term represents a monochromatic plane wave with wave vector $\bm{k}$ and angular frequency $\omega$. Combining the streaming phase factor with the Jacobian operator yields the amplification matrix $\bm{G}(\bm{k})$:
\begin{equation}\label{eq:G_general}
    \bm{G}(\bm{k}) = \bm{D}(\bm{k}) \left[ \bm{I} + \bm{K}(\bm{\Lambda} - \bm{I}) \right],
\end{equation}
where $\bm{D}(\bm{k}) = \text{diag}(e^{-\mathrm{i}\bm{k}\cdot\bm{c}_0}, \dots, e^{-\mathrm{i}\bm{k}\cdot\bm{c}_{q-1}})$ represents the advection operator in Fourier space. $\bm{K}$ and $\bm{\Lambda}$ denote the collision and Jacobian matrices, respectively. This matrix $\bm{G}(\bm{k})$ compactly encodes the linearized evolution of Fourier modes and provides a convenient basis for analyzing the stability, dispersion, and numerical properties of the scheme. 
\textcolor{smBlue}{\autoref{eq:G_general} does not depend on the lattice: the dimension and the velocity set enter only through the $q$ phases of $\bm{D}(\bm{k})$, the collision matrix $\bm{K}$ and the Jacobian $\bm{\Lambda}$. \autoref{app:collision_models} specifies $\bm{K}$ for each collision model. \autoref{app:d2q9_details} and~\autoref{app:d3q27_details} provide the lattice data, equilibria and reconstruction matrices of D2Q9 and D3Q27, respectively. The numerical results are reported in \autoref{app:vn_analysis_results}.}

\section{Collision Models}
\label{app:collision_models}
Collision modeling plays a central role in determining both the accuracy and the stability of lattice Boltzmann solvers. \textcolor{smBlue}{Each collision model considered in this work is specified by its collision matrix $\bm{K}$ in \autoref{eq:collision_models}. This section gives $\bm{K}$ for the three reference models and refers to the original papers for their derivation. We list the moment bases and relaxation rates per lattice in~\autoref{app:d2q9_details} and~\autoref{app:d3q27_details}, and we assemble the amplification matrices for all four models, including HOME-LBM. All moment-matrix inverses are computed numerically (\texttt{np.linalg.inv}), and the matrices are at most $27\times27$.}

\subsection{Single Relaxation Time BGK model}
\label{app:bgk_collision_operator}
The classical BGK formulation~\textcolor{smBlue}{\cite{bhathnagor1954model}} applies a single relaxation time to relax each lattice population toward its local
equilibrium. The collision operator is written as
\begin{equation}
    \Omega_i^{\mathrm{BGK}}
    = \frac{1}{\tau}\left(f_i^{\mathrm{eq}} - f_i \right),
    \qquad
    \textcolor{smBlue}{\bm{K}^{\mathrm{BGK}} = \frac{1}{\tau}\,\bm{I},}
    \label{eq:bgk_collision}
\end{equation}
where $\tau$ is the relaxation time determining the rate at which the distribution function approaches equilibrium. The BGK model yields the following relation between $\tau$ and the kinematic viscosity $\nu$: $\tau = 3\nu + \tfrac{1}{2}$ (equivalently $\nu = c_s^2(\tau - \tfrac{1}{2})$).

\subsection{Raw Moment Multiple Relaxation Time (RM-MRT) Model}
\label{app:rm_collision_operator}
To overcome the stability limitations of the BGK model, the MRT framework introduces a more stable collision operator that acts in moment space rather than directly on the distribution functions~\textcolor{smOrange}{\cite{d1992generalized, lallemand2000theory, chavez2018improving}}. In the RM-MRT formulation, the collision model is performed on polynomial moments constructed from the lattice velocities\textcolor{smOrange}{, $\bm{m} = \bm{M}^{\mathrm{RM}}\bm{f}$ and $\bm{m}^{\mathrm{eq}} = \bm{M}^{\mathrm{RM}}\bm{f}^{\mathrm{eq}}$, where the rows of the raw-moment transformation matrix $\bm{M}^{\mathrm{RM}}$ are polynomials in the lattice velocities and do not depend on the local fluid velocity} \textcolor{smBlue}{(Appendices~\ref{app:collision_models_RM} and~\ref{app:collision_models_RM_3d})}. \textcolor{smOrange}{Each moment is relaxed toward its equilibrium at its own rate, $\bm{\Omega}^{\mathrm{RM}} =(\bm{M}^{\mathrm{RM}})^{-1}\bm{R}\,(\bm{m}^{\mathrm{eq}} - \bm{m})$, so that the RM-MRT collision operator is}
\begin{equation}
\begin{aligned}
    \bm{\Omega}^{\mathrm{RM}} &= \bm{K}^{\mathrm{RM}}\left(\bm{f}^{\mathrm{eq}} - \bm{f}\right),\\
    \textcolor{smBlue}{\bm{K}^{\mathrm{RM}}} &\textcolor{smBlue}{{}= (\bm{M}^{\mathrm{RM}})^{-1}\bm{R}\,\bm{M}^{\mathrm{RM}},}
\end{aligned}
\label{eq:mrt_rm_collision}
\end{equation}
where $\mathbf{R} = \mathrm{diag}(\bm{r}^{\mathrm{RM}})$ is a diagonal matrix of relaxation rates for each moment \textcolor{smBlue}{(Appendices~\ref{app:relaxation_rates_2d} and~\ref{app:relaxation_rates_3d})}.

\subsection{Non-Orthogonal Central Moment Multiple Relaxation Time (NOCM-MRT) Model}
\label{app:nocm_collision_operator}
While RM-MRT improves stability over BGK, its use of raw moments lacks Galilean invariance, which can degrade accuracy and cause instabilities at low viscosities. The non-orthogonal central moment multiple relaxation time model (NOCM-MRT) addresses this issue by defining moments relative to the local fluid velocity, thereby achieving significantly improved invariance properties and enhanced robustness and accuracy for turbulent flows~\textcolor{smOrange}{\cite{Rosis2017, Rosis2019}}.
Let $\bm{u}$ denote the local macroscopic velocity, and let $\mathbf{M}^{\mathrm{NOCM}}(\bm{u})$ be the central-moment transformation matrix, which shifts the velocity basis so that the resulting moments\textcolor{smOrange}{, $\bm{m} = \mathbf{M}^{\mathrm{NOCM}}(\bm{u})\,\bm{f}$ and $\bm{m}^{\mathrm{eq}} = \mathbf{M}^{\mathrm{NOCM}}(\bm{u})\,\bm{f}^{\mathrm{eq}}$,} are computed with respect to $\bm{c}_i - \bm{u}$ \textcolor{smBlue}{(Appendices~\ref{app:collision_models_NOCM} and~\ref{app:collision_models_NOCM_3d})}; $\bm{m}^{\mathrm{eq}}$ represents the central-moment equilibrium obtained from the Maxwellian equilibrium distribution. The NOCM-MRT collision operator is then expressed as
\begin{equation}
\begin{aligned}
    \bm{\Omega}^{\mathrm{NOCM}} &= \mathbf{K}^{\mathrm{NOCM}}\left(\bm{f}^{\mathrm{eq}} - \bm{f} \right),\\
    \textcolor{smBlue}{\mathbf{K}^{\mathrm{NOCM}}} &\textcolor{smBlue}{{}= \mathbf{M}^{\mathrm{NOCM}}(\bm{u})^{-1}\mathbf{R}\,\mathbf{M}^{\mathrm{NOCM}}(\bm{u}),}
\end{aligned}
\label{eq:cm_collision}
\end{equation}
where $\mathbf{R} = \mathrm{diag}(\bm{r}^{\mathrm{NOCM}})$ is a diagonal relaxation matrix specifying independent relaxation rates for each central moment. Relaxing higher-order moments more aggressively provides strong damping
of non-hydrodynamic modes while preserving the correct recovery of Navier--Stokes dynamics~\cite{Wei2020}.

\section{2D Implementation Details}
\label{app:d2q9_details}
\label{app:matrix_definition_2d}
\subsection{D2Q9 Lattice Definitions}
For completeness, we list the discrete velocities, lattice weights, and the lattice speed of sound used throughout this work. The D2Q9 lattice consists of nine discrete velocity directions
\[
\bm{c}_i = (c_{x,i}, c_{y,i}), \qquad i = 0,\ldots,8,
\]
with components: 
\[
c_{x} =
\begin{pmatrix}
0 & 1 & 0 & -1 & 0 & 1 & -1 & -1 & 1
\end{pmatrix}
\]
\[
c_{y} =
\begin{pmatrix}
0 & 0 & 1 & 0 & -1 & 1 & 1 & -1 & -1
\end{pmatrix}.
\]
The corresponding lattice weights are
\[
w =
\begin{pmatrix}
\frac{4}{9},\; 
\frac{1}{9},\; \frac{1}{9},\; \frac{1}{9},\; \frac{1}{9},\;
\frac{1}{36},\; \frac{1}{36},\; \frac{1}{36},\; \frac{1}{36}
\end{pmatrix}.
\]
The lattice speed of sound is $c_s= \sqrt{\frac{1}{3}}$ .
The above definitions are used for constructing the RM-MRT matrix $\mathbf{M}^\mathrm{RM}$, the NOCM-MRT transformation matrix $\mathbf{M}^\mathrm{NOCM}(\bm{u})$, and all Hermite-based reconstructions described in this appendix.

\subsection{RM-MRT Transformation Matrix}
\label{app:collision_models_RM}
For the D2Q9 lattice with discrete velocities $(c_{x,i}, c_{y,i})_{i=0}^{8}$, the transformation matrix $\mathbf{M}^{\mathrm{RM}}$ used in \autoref{eq:mrt_rm_collision} \textcolor{smBlue}{has the rows $(\mathbf{M}^{\mathrm{RM}})_{ki} = p_k(\bm{c}_i)$, $k = 0,\ldots,8$, with the raw-moment polynomials}
{\color{smBlue}
\[
\begin{aligned}
(p_0,\ldots,p_{8}) = \bigl(&1,\; c_x,\; c_y,\; c_x^2 + c_y^2,\\
&c_x^2 - c_y^2,\; c_xc_y,\; c_x(c_x^2 + c_y^2),\\
&c_y(c_x^2 + c_y^2),\; c_x^2c_y^2\bigr).
\end{aligned}
\]
}

\subsection{NOCM-MRT Transformation Matrix}
\label{app:collision_models_NOCM}
Let the shifted velocities be $X_i = c_{x,i} - u_x$ and $Y_i = c_{y,i} - u_y$. \textcolor{smBlue}{The rows of the transformation matrix $\mathbf{M}^{\mathrm{NOCM}}(\bm{u})$ are the central-moment polynomials $(\mathbf{M}^{\mathrm{NOCM}}(\bm{u}))_{ki} = p_k(X_i, Y_i)$ with}
{\color{smBlue}
\[
\begin{aligned}
(p_0,\ldots,p_{8}) = \bigl(&1,\; X_i,\; Y_i,\; X_i^2 + Y_i^2,\\
&X_i^2 - Y_i^2,\; X_iY_i,\; X_i^2Y_i,\; X_iY_i^2,\; X_i^2Y_i^2\bigr).
\end{aligned}
\]
}

\subsection{Relaxation Rates}
\label{app:relaxation_rates_2d}
The relaxation rates for the RM-MRT model are collected in the vector
\[
\bm{r}^{\mathrm{RM}}
=
\begin{pmatrix}
0,\, 0,\, 0,\,
1.64,\,
r_\nu,\,
r_\nu,\,
1.9,\, 1.9,\,
1.54
\end{pmatrix}^{\!\top},
\]
where $r_\nu = \frac{1}{0.5 + \nu/c_s^2} = \frac{1}{3\nu + 0.5}$, \textcolor{smOrange}{according to the values of \cite{lallemand2000theory}}. Similarly, the relaxation rate vector for the NOCM-MRT model 
\[
\bm{r}^{\mathrm{NOCM}}
=
\begin{pmatrix}
0,\, 0,\, 0,\,
1.0,\,
s_\nu,\,
s_\nu,\,
1.0,\, 1.0,\,
1.0
\end{pmatrix}^{\!\top},
\]
with $s_\nu = r_\nu$.
The diagonal relaxation matrix is then constructed as
\[
\mathbf{R} = \mathrm{diag}(\bm{r}),
\]
where the operator $\mathrm{diag}(\cdot)$ places the vector components on the main diagonal of the matrix.

\subsection{\textcolor{smBlue}{Equilibrium Distribution}}
\label{app:eq2d}
\textcolor{smOrange}{The fourth-order equilibrium distribution $f_i^{\text{eq}}$, restricted to the Hermite components supported by the D2Q9 lattice, is defined as}:
\begin{align}
f_i^{\text{eq}}
= \rho w_i\bigg[&
1 + \tfrac{\bm{c}_i\!\cdot\!\bm{u}}{c_s^2}
  + \tfrac{(\bm{c}_i\!\cdot\!\bm{u})^2}{2c_s^4}
  - \tfrac{\bm{u}^2}{2c_s^2}
  + \tfrac{c_{xi}c_{yi}\,u_xu_y\,(\bm{c}_i\!\cdot\!\bm{u})}{2c_s^6} \\
  - &\tfrac{u_xu_y\,(c_{yi}u_x + c_{xi}u_y)}{2c_s^4} 
  + \tfrac{c_{xi}^2c_{yi}^2\,u_x^2u_y^2}{4c_s^8}
  - \tfrac{(c_{xi}^2+c_{yi}^2)\,u_x^2u_y^2}{4c_s^6}
  + \tfrac{u_x^2u_y^2}{4c_s^4}\bigg] \nonumber
\end{align}
\textcolor{smOrange}{and the derivatives} \textcolor{smBlue}{entering \autoref{eq:Lambda_general}} \textcolor{smOrange}{are}
\begin{align}
\tfrac{\partial f_i^{\text{eq}}}{\partial u_x}
=~ \rho w_i\bigg[ &
\tfrac{c_{xi}}{c_s^2}
+ \tfrac{(c_{xi}u_x+c_{yi}u_y)\,c_{xi}}{c_s^4}
- \tfrac{u_x}{c_s^2} \\
& + \tfrac{c_{xi}c_{yi}\,u_y\,(2c_{xi}u_x + c_{yi}u_y)}{2\,c_s^6}
- \tfrac{u_y\,(2c_{yi}u_x + c_{xi}u_y)}{2\,c_s^4} \nonumber \\
& + \tfrac{c_{xi}^2c_{yi}^2\,u_xu_y^2}{2\,c_s^8}
- \tfrac{(c_{xi}^2 + c_{yi}^2)\,u_xu_y^2}{2\,c_s^6}
+ \tfrac{u_xu_y^2}{2\,c_s^4}
\bigg] \nonumber \\
\tfrac{\partial f_i^{\text{eq}}}{\partial u_y}
=~ \rho w_i\bigg[ &
\tfrac{c_{yi}}{c_s^2}
+ \tfrac{(c_{xi}u_x+c_{yi}u_y)\,c_{yi}}{c_s^4}
- \tfrac{u_y}{c_s^2} \\
& + \tfrac{c_{xi}c_{yi}\,u_x\,(c_{xi}u_x + 2c_{yi}u_y)}{2\,c_s^6}
- \tfrac{u_x\,(c_{yi}u_x + 2c_{xi}u_y)}{2\,c_s^4} \nonumber \\
& + \tfrac{c_{xi}^2c_{yi}^2\,u_x^2u_y}{2\,c_s^8}
- \tfrac{(c_{xi}^2 + c_{yi}^2)\,u_x^2u_y}{2\,c_s^6}
+ \tfrac{u_x^2u_y}{2\,c_s^4}
\bigg] \nonumber.
\end{align}
The same equilibrium is used for the Jacobian $\bm{\Lambda}$ of all four collision models.

\subsection{Recursive Matrices}
\label{app:recursive_2d}
\textcolor{smOrange}{Following the recursive regularization of \cite{Malaspinas2015}, the non-equilibrium part of the
distribution is expanded in Hermite polynomials, with the second-order coefficients given by the non-equilibrium stress in the convention of \autoref{eq:rho_u_stress}, i.e.\ $S^{\text{eq}}_{\alpha\beta} = u_\alpha u_\beta$.} In component form:
\begin{align}
a_{\alpha\alpha}^{[2],neq} = \rho (S_{\alpha\alpha} - u_\alpha^2), \quad
a_{\alpha\beta}^{[2],neq} = \rho (S_{\alpha\beta} - u_\alpha u_\beta). 
\end{align}
In the recursive regularization procedure, the non-equilibrium third-order coefficients $a^{[3],neq}$ are obtained recursively from non-equilibrium
$a^{[2],neq}$ and the local velocity $u_{\alpha}$.
The recursion is given by:
\begin{align}
a_{\alpha\alpha\alpha}^{[3],neq} = 3\,u_{\alpha}\,a_{\alpha\alpha}^{[2],neq}, \quad
a_{\alpha\alpha\beta}^{[3],neq} = 2\,u_{\alpha}\,a_{\alpha\beta}^{[2],neq}
+ u_{\beta}\,a_{\alpha\alpha}^{[2],neq},
\label{eq:a3_neq}
\end{align}
where repeated indices are not summed unless explicitly stated. 

The non-equilibrium distribution function  $f^{\mathrm{neq}}_i$
is reconstructed from these quantities in Hermite space as
\begin{equation}
f^{\mathrm{neq}}_i
= w_i \left[
\tfrac{1}{2 c_s^4}\,
\bm{H}^{(2)}_i : a^{[2],neq}
+ \tfrac{1}{2 c_s^6}
\left(
\bm{H}^{(3)}_{i,xxy}\, a^{[3],neq}_{xxy}
+ \bm{H}^{(3)}_{i,xyy}\, a^{[3],neq}_{xyy}
\right)
\right],
\label{eq:fneq_reconstruction}
\end{equation}
where for the D2Q9 lattice, the second- and third-order
Hermite tensors are
\begin{align}
\bm{H}^{(2)}_{i,\alpha\beta}
&= c_{i\alpha} c_{i\beta} - c_s^2 \delta_{\alpha\beta}, \nonumber \\
\bm{H}^{(3)}_{i,xxy} &= c_{iy}\,(c_{ix}^2 - c_s^2), \\
\bm{H}^{(3)}_{i,xyy} &= c_{ix}\,(c_{iy}^2 - c_s^2). \nonumber
\end{align}
Collecting all populations into the vector $\bm{f}^{neq} \in \mathbb{R}^{9 \times 1}$
and grouping the Hermite coefficients into
\(
\bm{a}^{[2],neq} \in \mathbb{R}^{3 \times 1}
\)
and
\(
\bm{a}^{[3],neq} \in \mathbb{R}^{4 \times 1}
\)
\textcolor{smBlue}{defined as}
\begin{equation}
\bm{a}^{[2],neq}
=
\begin{bmatrix}
a_{xx}^{[2],neq}\\
2a_{xy}^{[2],neq}\\
a_{yy}^{[2],neq}
\end{bmatrix}
\in\mathbb{R}^{3\times1},
\qquad
\bm{a}^{[3],neq}
=
\begin{bmatrix}
a_{xxx}^{[3],neq}\\
a_{xxy}^{[3],neq}\\
a_{xyy}^{[3],neq}\\
a_{yyy}^{[3],neq}
\end{bmatrix}
\in\mathbb{R}^{4\times1},
\end{equation}
the above expression can be written as
\begin{equation}
\bm{f}^{\mathrm{neq}} = \mathbf{A}_2\,\bm{a}^{[2],neq} + \mathbf{A}_3\,\bm{a}^{[3],neq},
\label{eq:fneq_base}
\end{equation}
where $\mathbf{A}_2 \in \mathbb{R}^{9\times3}$ and $\mathbf{A}_3 \in \mathbb{R}^{9\times4}$ 
collect the discrete Hermite basis contributions:
\begin{align}
(\mathbf{A}_2)_{i,:} &= 
\frac{w_i}{2c_s^{4}}
\begin{bmatrix}
c_{ix}^2 - c_s^2 & c_{ix}c_{iy} & c_{iy}^2 - c_s^2
\end{bmatrix},\\
(\mathbf{A}_3)_{i,:} &=
\frac{w_i}{2c_s^{6}}
\begin{bmatrix}
c_{ix}^3 - 3c_s^2 c_{ix} &
c_{ix}^2 c_{iy} - c_s^2 c_{iy} &
c_{ix} c_{iy}^2 - c_s^2 c_{ix} &
c_{iy}^3 - 3c_s^2 c_{iy}
\end{bmatrix}.
\end{align}
Using the recursive relation
\(
\bm{a}^{[3],neq} = \mathbf{R}_{\bm{u}}\,\bm{a}^{[2],neq},
\)
with
\[
\mathbf{R}_{\bm{u}} =
\begin{bmatrix}
3u_x & 0 & 0\\[4pt]
u_y & u_x & 0\\[4pt]
0 & u_y & u_x\\[4pt]
0 & 0 & 3u_y
\end{bmatrix},
\]
\autoref{eq:fneq_base} reduces to $\bm{f}^{\text{neq}} = (\mathbf{A}_2 + \mathbf{A}_3\mathbf{R}_{\bm{u}})\,\bm{a}^{[2],neq}
= \mathbf{Q}_{\bm{u}}\,\bm{a}^{[2],neq}$, which is \autoref{eq:non-equilibrium_compact} of the main text.

\subsection{Explicit Form of the Matrix $\mathbf{B}$ for D2Q9}
\label{app:Bneq_d2q9}

We define the second-order moment vector in Voigt form as
\begin{equation}
\mathbf{S}^{\ast}
=
\begin{bmatrix}
S_{xx}^{\ast} \\
2 S_{xy}^{\ast} \\
S_{yy}^{\ast}
\end{bmatrix},
\qquad
S_{\alpha\beta}^{\ast}
=
\sum_i f_i \left( c_{i\alpha} c_{i\beta} - c_s^2 \delta_{\alpha\beta} \right),
\end{equation}
where $c_s$ denotes the lattice sound speed. The linear mapping from the distribution function vector $\mathbf{f}$ to
$\mathbf{S}^{\ast}$ is written as
\begin{equation}
\mathbf{S}^{\ast} = \mathbf{B}\, \mathbf{f},
\end{equation}
with $\mathbf{B} \in \mathbb{R}^{3 \times 9}$.
For the standard D2Q9 velocity set
\[
\mathbf{c}_i \in \{(0,0),(\pm1,0),(0,\pm1),(\pm1,\pm1)\},
\]
the explicit entries of $\mathbf{B}$ are listed in
Table~\ref{tab:Bneq_d2q9}.

\begin{table}[t]
\centering
\caption{Explicit form of $\mathbf{B}$ for the D2Q9 lattice, using the Voigt ordering
$\mathbf{S}^{\ast} = [S_{xx}^{\ast},\, 2S_{xy}^{\ast},\, S_{yy}^{\ast}]^{\mathsf{T}}$.}
\label{tab:Bneq_d2q9}
\scriptsize

\resizebox{\linewidth}{!}{
\begin{tabular}{c|ccccccccc}
$i$ & 0 & 1 & 2 & 3 & 4 & 5 & 6 & 7 & 8 \\ \hline

$(c_{ix},c_{iy})$
& (0,0) & (1,0) & (0,1) & (-1,0) & (0,-1)
& (1,1) & (-1,1) & (-1,-1) & (1,-1) \\

$B_{1i}=c_{ix}^2-c_s^2$
& $-c_s^2$ & $1-c_s^2$ & $-c_s^2$ & $1-c_s^2$ & $-c_s^2$
& $1-c_s^2$ & $1-c_s^2$ & $1-c_s^2$ & $1-c_s^2$ \\

$B_{2i}=2c_{ix}c_{iy}$
& 0 & 0 & 0 & 0 & 0 & 2 & $-2$ & 2 & $-2$ \\

$B_{3i}=c_{iy}^2-c_s^2$
& $-c_s^2$ & $-c_s^2$ & $1-c_s^2$ & $-c_s^2$ & $1-c_s^2$
& $1-c_s^2$ & $1-c_s^2$ & $1-c_s^2$ & $1-c_s^2$
\end{tabular}
}
\end{table}

\section{3D Implementation Details}
\label{app:d3q27_details}
\label{app:matrix_definition_3d}
\textcolor{smBlue}{The 3D analysis follows the same construction as \autoref{app:standard_lbm_stability_analysis}; only the lattice, the moment bases, the relaxation-rate vectors, the equilibrium and the Hermite matrices change. We list them here in the same order as in \autoref{app:d2q9_details}.}

\subsection{\textcolor{smBlue}{D3Q27 Lattice Definitions}}
\textcolor{smBlue}{The D3Q27 lattice consists of the 27 discrete velocities $\bm{c}_i = (c_{x,i}, c_{y,i}, c_{z,i})\in\{-1,0,1\}^3$, $i = 0,\ldots,26$, ordered as rest velocity, six face velocities, twelve edge velocities and eight corner velocities:}
{\color{smBlue}
\[
\begin{aligned}
c_x = \bigl(&0,\,1,\,-1,\,0,\,0,\,0,\,0,\,1,\,-1,\,1,\,-1,\,1,\,-1,\,0,\\
            & 0,\,0,\,0,\,1,\,-1,\,1,\,-1,\,1,\,-1,\,1,\,-1,\,1,\,-1\bigr).
\end{aligned}
\]

\[
\begin{aligned}
c_y = \bigl(&0,\,0,\,0,\,1,\,-1,\,0,\,0,\,1,\,1,\,-1,\,-1,\,0,\,0,\,1,\\
            &{-1},\,1,\,-1,\,0,\,0,\,1,\,1,\,-1,\,-1,\,1,\,1,\,-1,\,-1\bigr).
\end{aligned}
\]

\[
\begin{aligned}
c_z = \bigl(&0,\,0,\,0,\,0,\,0,\,1,\,-1,\,0,\,0,\,0,\,0,\,1,\,1,\,1,\\               &1,\,-1,\,-1,\,-1,\,-1,\,1,\,1,\,1,\,1,\,-1,\,-1,\,-1,\,-1\bigr).
\end{aligned}
\]
}%
\textcolor{smBlue}{The lattice weights depend only on $|\bm{c}_i|^2$,}
{\color{smBlue}
\[
w_i =
\begin{cases}
\tfrac{8}{27},  & |\bm{c}_i|^2 = 0 \quad (i = 0),\\[2pt]
\tfrac{2}{27},  & |\bm{c}_i|^2 = 1 \quad (i = 1,\ldots,6),\\[2pt]
\tfrac{1}{54},  & |\bm{c}_i|^2 = 2 \quad (i = 7,\ldots,18),\\[2pt]
\tfrac{1}{216}, & |\bm{c}_i|^2 = 3 \quad (i = 19,\ldots,26),
\end{cases}
\]
and the lattice speed of sound is again $c_s = \sqrt{1/3}$. These definitions are used for $\mathbf{M}^{\mathrm{RM}}$, $\mathbf{M}^{\mathrm{NOCM}}(\bm{u})$ and the Hermite matrices.}

\subsection{\textcolor{smBlue}{RM-MRT Transformation Matrix}}
\label{app:collision_models_RM_3d}
\textcolor{smBlue}{Row $k$ of $\mathbf{M}^{\mathrm{RM}}\in\mathbb{R}^{27\times27}$ is the polynomial $p_k(\bm{c}_i)$ evaluated at the 27 velocities, $(\mathbf{M}^{\mathrm{RM}})_{ki} = p_k(\bm{c}_i)$, with}
{\color{smBlue}
\[
\begin{aligned}
(p_0,\ldots,p_{26}) = \bigl(&1,\; c_x,\; c_y,\; c_z,\; |\bm{c}|^2,\\
&c_x^2c_y^2 + c_x^2c_z^2 + c_y^2c_z^2,\; c_x|\bm{c}|^2,\; c_y|\bm{c}|^2,\\
&c_z|\bm{c}|^2,\; 3c_x^2 - |\bm{c}|^2,\; c_y^2 - c_z^2,\; c_xc_y,\; c_yc_z,\\
&c_xc_z,\; c_x^2c_y^2 + c_x^2c_z^2 - 2c_y^2c_z^2,\\
&c_x^2c_y^2 - c_x^2c_z^2,\; c_x(c_y^2 - c_z^2),\\
&c_y(c_z^2 - c_x^2),\; c_z(c_x^2 - c_y^2),\; c_xc_yc_z,\\
&c_x^2c_yc_z,\; c_xc_y^2c_z,\; c_xc_yc_z^2,\; c_x^2c_y^2c_z,\\
&c_x^2c_yc_z^2,\; c_xc_y^2c_z^2,\; c_x^2c_y^2c_z^2\bigr).
\end{aligned}
\]
}

\subsection{\textcolor{smBlue}{NOCM-MRT Transformation Matrix}}
\label{app:collision_models_NOCM_3d}
\textcolor{smBlue}{With the shifted velocities $X_i = c_{x,i} - u_x$, $Y_i = c_{y,i} - u_y$, $Z_i = c_{z,i} - u_z$, the rows of $\mathbf{M}^{\mathrm{NOCM}}(\bm{u})\in\mathbb{R}^{27\times27}$ are the central-moment polynomials $(\mathbf{M}^{\mathrm{NOCM}}(\bm{u}))_{ki} = p_k(X_i, Y_i, Z_i)$ with}
{\color{smBlue}
\[
\begin{aligned}
(p_0,\ldots,p_{26}) = \bigl(&1,\; X_i,\; Y_i,\; Z_i,\\
&X_i^2 + Y_i^2 + Z_i^2,\; X_i^2 - Y_i^2,\; X_i^2 - Z_i^2,\; X_iY_i,\\
&X_iZ_i,\; Y_iZ_i,\; X_i^2Y_i,\; X_i^2Z_i,\; Y_i^2Z_i,\; X_iY_i^2,\\
&X_iZ_i^2,\; Y_iZ_i^2,\; X_i^2Y_i^2,\; X_i^2Z_i^2,\; Y_i^2Z_i^2,\\
&X_iY_iZ_i,\; X_i^2Y_iZ_i,\; X_iY_i^2Z_i,\; X_iY_iZ_i^2,\\
&X_i^2Y_i^2Z_i,\; X_i^2Y_iZ_i^2,\; X_iY_i^2Z_i^2,\\
&X_i^2Y_i^2Z_i^2\bigr).
\end{aligned}
\]
}

\subsection{\textcolor{smBlue}{Relaxation Rates}}
\label{app:relaxation_rates_3d}
\textcolor{smBlue}{In the row ordering of $\mathbf{M}^{\mathrm{RM}}$ above, the RM-MRT rates are taken from~\cite{d2002multiple} for the 19 moments defined there, with the eight additional moments assigned by order:}
{\color{smBlue}
\[
\begin{aligned}
\bm{r}^{\mathrm{RM}} = \bigl(
&0,\,0,\,0,\,0,\,
1.19,\,1.4,\,
1.2,\,1.2,\,1.2,\\
&s_\nu,\,s_\nu,\,s_\nu,\,s_\nu,\,s_\nu,\,
1.4,\,1.4,\,
1.98,\,1.98,\,1.98,\\
&1.98,\,
1.4,\,1.4,\,1.4,\,1.4,\,1.4,\,1.4,\,1.4
\bigr)^{\!\top},
\end{aligned}
\]
}%
\textcolor{smBlue}{with $s_\nu = 1/(3\nu + \tfrac12)$ on the five deviatoric second-order moments, as in two dimensions. For NOCM-MRT, in the row ordering of Appendix~\ref{app:collision_models_NOCM_3d},}
{\color{smBlue}
\[
\begin{aligned}
\bm{r}^{\mathrm{NOCM}} = \bigl(
&0,\,0,\,0,\,0,\,1.0,\,
s_\nu,\,s_\nu,\,s_\nu,\,s_\nu,\\
&s_\nu,\,
1.0,\,1.0,\,1.0,\,1.0,\,1.0,\,1.0,\,1.0,\,1.0,\\
&1.0,\,1.0,\,1.0,\,1.0,\,1.0,\,1.0,\,1.0,\,1.0,\,1.0
\bigr)^{\!\top},
\end{aligned}
\]
}%
\textcolor{smBlue}{i.e.\ only the five deviatoric second-order central moments carry the viscosity-dependent rate; the trace and all higher-order central moments are relaxed to their Maxwellian values with unit rate. 
The diagonal relaxation matrices are $\mathbf{R} = \mathrm{diag}(\bm{r})$ as before.}

\subsection{\textcolor{smBlue}{Equilibrium Distribution}}
\label{app:eq3d}
\textcolor{smBlue}{On D3Q27 all four collision models use the product-form equilibrium~\cite{Rosis2017, Rosis2019}}
{\color{smBlue}
\begin{align}
f_i^{\text{eq}} &= \rho\,\varphi(c_{ix},u_x)\,\varphi(c_{iy},u_y)\,\varphi(c_{iz},u_z), \nonumber\\
\varphi(c,u) &= w(c)\Bigl[1 + \tfrac{c\,u}{c_s^2} + \tfrac{(c^2 - c_s^2)\,u^2}{2c_s^4}\Bigr],
\label{eq:feq_3d}
\end{align}
}%
\textcolor{smBlue}{with $w(0) = 2/3$, $w(\pm1) = 1/6$, $w_i = w(c_{ix})\,w(c_{iy})\,w(c_{iz})$. Expanded in the Hermite components supported by D3Q27 (no index more than twice), with $H^{(2)}_{i,\alpha\alpha} = c_{i\alpha}^2 - c_s^2$ and, in the fifth-order term, $\gamma\notin\{\alpha,\beta\}$:}
{\color{smBlue}
\begin{align}
f_i^{\text{eq}} = \rho w_i\bigg[&
1 + \tfrac{\bm{c}_i\!\cdot\!\bm{u}}{c_s^2}
+ \tfrac{(\bm{c}_i\!\cdot\!\bm{u})^2 - c_s^2\bm{u}^2}{2c_s^4} \nonumber\\
&+ \tfrac{1}{2c_s^6}\sum_{\alpha\neq\beta} c_{i\beta}H^{(2)}_{i,\alpha\alpha}\,u_\alpha^2u_\beta
+ \tfrac{c_{ix}c_{iy}c_{iz}}{c_s^6}\,u_xu_yu_z \nonumber\\
&+ \tfrac{1}{4c_s^8}\sum_{\alpha<\beta} H^{(2)}_{i,\alpha\alpha}H^{(2)}_{i,\beta\beta}\,u_\alpha^2u_\beta^2 \nonumber\\
&+ \tfrac{1}{2c_s^8}\sum_{\alpha}\sum_{\substack{\beta<\gamma\\ \beta,\gamma\neq\alpha}} c_{i\beta}c_{i\gamma}H^{(2)}_{i,\alpha\alpha}\,u_\alpha^2u_\beta u_\gamma \nonumber\\
&+ \tfrac{1}{4c_s^{10}}\sum_{\alpha<\beta} H^{(2)}_{i,\alpha\alpha}H^{(2)}_{i,\beta\beta}\,c_{i\gamma}\,u_\alpha^2u_\beta^2u_\gamma \nonumber\\
&+ \tfrac{1}{8c_s^{12}}\,H^{(2)}_{i,xx}H^{(2)}_{i,yy}H^{(2)}_{i,zz}\,u_x^2u_y^2u_z^2
\bigg], \nonumber
\end{align}
}%
\textcolor{smBlue}{which reduces to the D2Q9 equilibrium of Appendix~\ref{app:eq2d} without the $z$ factor. All central moments of Appendix~\ref{app:collision_models_NOCM_3d} are Maxwellian,}
{\color{smBlue}
\[
\begin{aligned}
\mathbf{M}^{\mathrm{NOCM}}(\bm{u})\,\bm{f}^{\text{eq}} &= \rho\,\bm{m}^{\text{eq}},\\
\bm{m}^{\text{eq}} &= \bigl(1,0,0,0,3c_s^2,\underbrace{0,\ldots,0}_{11},c_s^4,c_s^4,c_s^4,\underbrace{0,\ldots,0}_{7},c_s^6\bigr)^{\!\top},
\end{aligned}
\]
}%
\textcolor{smBlue}{and the derivatives entering \autoref{eq:Lambda_general} are $\partial f_i^{\text{eq}}/\partial\rho = f_i^{\text{eq}}/\rho$ and}
{\color{smBlue}
\begin{equation}
\frac{\partial f_i^{\text{eq}}}{\partial u_\alpha}
= \rho\,w(c_{i\alpha})\Bigl[\tfrac{c_{i\alpha}}{c_s^2} + \tfrac{(c_{i\alpha}^2 - c_s^2)\,u_\alpha}{c_s^4}\Bigr]\prod_{\beta\neq\alpha}\varphi(c_{i\beta},u_\beta).
\label{eq:dfeq_3d}
\end{equation}
}

\subsection{\textcolor{smBlue}{Recursive Matrices}}
\label{app:recursive_3d}
\textcolor{smBlue}{The second-order non-equilibrium coefficients are $a^{[2],neq}_{\alpha\beta} = \rho\,(S_{\alpha\beta} - u_\alpha u_\beta)$ with $\alpha,\beta\in\{x,y,z\}$, and the recursive closure for the third-order coefficients reads}
{\color{smBlue}
\begin{align}
a^{[3],neq}_{\alpha\alpha\alpha} &= 3\,u_\alpha\,a^{[2],neq}_{\alpha\alpha}, \nonumber\\
a^{[3],neq}_{\alpha\alpha\beta} &= 2\,u_\alpha\,a^{[2],neq}_{\alpha\beta} + u_\beta\,a^{[2],neq}_{\alpha\alpha},\label{eq:a3_neq_3d} \\
a^{[3],neq}_{xyz} &= u_x\,a^{[2],neq}_{yz} + u_y\,a^{[2],neq}_{xz} + u_z\,a^{[2],neq}_{xy}, \nonumber
\end{align}
}%
\textcolor{smBlue}{where repeated indices are not summed. The non-equilibrium distribution is reconstructed in Hermite space as}
{\color{smBlue}
\begin{align}
f^{\mathrm{neq}}_i
= w_i \Bigl[
\frac{1}{2c_s^4}\,\bm{H}^{(2)}_i : a^{[2],neq}
+ \frac{1}{2c_s^6}\!\sum_{\alpha\neq\beta}\bm{H}^{(3)}_{i,\alpha\alpha\beta}\,a^{[3],neq}_{\alpha\alpha\beta} \nonumber \\
+ \frac{1}{c_s^6}\,\bm{H}^{(3)}_{i,xyz}\,a^{[3],neq}_{xyz}
\Bigr],
\label{eq:fneq_reconstruction_3d}
\end{align}
}%
\textcolor{smBlue}{with the D3Q27 Hermite tensors}
{\color{smBlue}
\begin{align}
\bm{H}^{(2)}_{i,\alpha\beta} &= c_{i\alpha}c_{i\beta} - c_s^2\delta_{\alpha\beta}, \nonumber\\
\bm{H}^{(3)}_{i,\alpha\alpha\beta} &= c_{i\beta}\,(c_{i\alpha}^2 - c_s^2), \qquad \alpha\neq\beta, \\
\bm{H}^{(3)}_{i,xyz} &= c_{ix}c_{iy}c_{iz}. \nonumber
\end{align}
}%
\textcolor{smBlue}{The prefactors $1/(2c_s^6)$ and $1/c_s^6$ account for the three and six index permutations of the $\alpha\alpha\beta$ and $xyz$ components; the $\alpha\alpha\alpha$ components vanish identically because $c_{i\alpha}^3 - 3c_s^2c_{i\alpha} = c_{i\alpha}(1-3c_s^2) = 0$. Unlike D2Q9, the $xyz$ component does not vanish on D3Q27 (the corner velocities have $c_{ix}c_{iy}c_{iz} = \pm1$).}

\textcolor{smBlue}{Collecting the coefficients in Voigt form,}
{\color{smBlue}
\begin{equation}
\label{eq:a3_3d_neq}
\bm{a}^{[2],neq} =
\begin{bmatrix}
a^{[2],neq}_{xx} \\ a^{[2],neq}_{yy} \\ a^{[2],neq}_{zz} \\ 2a^{[2],neq}_{xy} \\ 2a^{[2],neq}_{xz} \\ 2a^{[2],neq}_{yz}
\end{bmatrix}\in\mathbb{R}^{6\times1},
\qquad
\bm{a}^{[3],neq} =
\begin{bmatrix}
a^{[3],neq}_{xxx} \\ a^{[3],neq}_{yyy} \\ a^{[3],neq}_{zzz} \\ a^{[3],neq}_{xxy} \\ a^{[3],neq}_{xxz} \\ a^{[3],neq}_{xyy} \\ a^{[3],neq}_{yyz} \\ a^{[3],neq}_{xzz} \\ a^{[3],neq}_{yzz} \\ a^{[3],neq}_{xyz}
\end{bmatrix}\in\mathbb{R}^{10\times1},
\end{equation}
}%
\textcolor{smBlue}{\autoref{eq:fneq_reconstruction_3d} becomes $\bm{f}^{\mathrm{neq}} = \mathbf{A}_2\,\bm{a}^{[2],neq} + \mathbf{A}_3\,\bm{a}^{[3],neq}$ with $\mathbf{A}_2\in\mathbb{R}^{27\times6}$ and $\mathbf{A}_3\in\mathbb{R}^{27\times10}$,}
{\color{smBlue}
\begin{equation}
\begin{aligned}
(\mathbf{A}_2)_{i,:}
= \frac{w_i}{2c_s^{4}}\bigl[
&c_{ix}^2-c_s^2,\,
 c_{iy}^2-c_s^2,\,
 c_{iz}^2-c_s^2,\\
&c_{ix}c_{iy},\,
 c_{ix}c_{iz},\,
 c_{iy}c_{iz}
\bigr].
\end{aligned}
\end{equation}

\begin{equation}
\begin{aligned}
(\mathbf{A}_3)_{i,:}
= \frac{w_i}{2c_s^{6}}\bigl[
&\tfrac13 H_{xxx},\,
 \tfrac13 H_{yyy},\,
 \tfrac13 H_{zzz},\\
&H_{xxy},\,H_{xxz},\,H_{xyy},\,H_{yyz},\\
&H_{xzz},\,H_{yzz},\,2c_{ix}c_{iy}c_{iz}
\bigr],
\end{aligned}
\end{equation}
}%
where \textcolor{smBlue}{ $H_{\alpha\alpha\alpha} = c_{i\alpha}^3 - 3c_s^2c_{i\alpha}$ and $H_{\alpha\alpha\beta} = c_{i\beta}(c_{i\alpha}^2 - c_s^2)$ abbreviate the entries of $\bm{H}^{(3)}_i$ (the $\alpha\alpha\alpha$ columns are identically zero on the lattice and are kept only to match the ordering of $\bm{a}^{[3],neq}$). 
As in two dimensions, the last three columns of $ \mathbf{A}_2$ act on the doubled off-diagonal entries $2a^{[2],neq}_{\alpha\beta}$, which reproduces the double contraction $\bm{H}^{(2)}_i : a^{[2],neq}$. Writing \autoref{eq:a3_neq_3d} in matrix form, $\bm{a}^{[3],neq} = \mathbf{R}_{\bm{u}}\,\bm{a}^{[2],neq}$ with}
{\color{smBlue}
\[
\mathbf{R}_{\bm{u}} =
\begin{bmatrix}
3u_x & 0 & 0 & 0 & 0 & 0\\[2pt]
0 & 3u_y & 0 & 0 & 0 & 0\\[2pt]
0 & 0 & 3u_z & 0 & 0 & 0\\[2pt]
u_y & 0 & 0 & u_x & 0 & 0\\[2pt]
u_z & 0 & 0 & 0 & u_x & 0\\[2pt]
0 & u_x & 0 & u_y & 0 & 0\\[2pt]
0 & u_z & 0 & 0 & 0 & u_y\\[2pt]
0 & 0 & u_x & 0 & u_z & 0\\[2pt]
0 & 0 & u_y & 0 & 0 & u_z\\[2pt]
0 & 0 & 0 & \tfrac12 u_z & \tfrac12 u_y & \tfrac12 u_x
\end{bmatrix}\in\mathbb{R}^{10\times6},
\]
}%
\textcolor{smBlue}{where the entries $u_x$ (instead of $2u_x$) in the $\alpha\alpha\beta$ rows and $\tfrac12$ in the $xyz$ row compensate the Voigt factor $2$ of the off-diagonal entries of $\bm{a}^{[2],neq}$. The reconstruction operator is $\mathbf{Q}_{\bm{u}} = \mathbf{A}_2 + \mathbf{A}_3\mathbf{R}_{\bm{u}}\in\mathbb{R}^{27\times6}$, so that $\bm{f}^{\mathrm{neq}} = \mathbf{Q}_{\bm{u}}\,\bm{a}^{[2],neq}$ as in \autoref{eq:non-equilibrium_compact}.}

\subsection{\textcolor{smBlue}{Explicit Form of the Matrix $\mathbf{B}$ for D3Q27}}
\label{app:Bneq_d3q27}
\textcolor{smBlue}{The second-order moment vector in Voigt form is}
{\color{smBlue}
\begin{equation}
\mathbf{S}^{\ast} =
\begin{bmatrix}
S^{\ast}_{xx} \\ S^{\ast}_{yy} \\ S^{\ast}_{zz} \\ 2S^{\ast}_{xy} \\ 2S^{\ast}_{xz} \\ 2S^{\ast}_{yz}
\end{bmatrix},
\qquad
S^{\ast}_{\alpha\beta} = \sum_i f_i\left(c_{i\alpha}c_{i\beta} - c_s^2\delta_{\alpha\beta}\right),
\qquad
\mathbf{S}^{\ast} = \mathbf{B}\,\mathbf{f},
\end{equation}
}%
\textcolor{smBlue}{with $\mathbf{B}\in\mathbb{R}^{6\times27}$ and rows $B_{1i} = c_{ix}^2 - c_s^2$, $B_{2i} = c_{iy}^2 - c_s^2$, $B_{3i} = c_{iz}^2 - c_s^2$, $B_{4i} = 2c_{ix}c_{iy}$, $B_{5i} = 2c_{ix}c_{iz}$, $B_{6i} = 2c_{iy}c_{iz}$. Since the entries depend only on which components of $\bm{c}_i$ are non-zero, we list them by velocity class in~\autoref{tab:Bneq_d3q27}; the sign of an off-diagonal entry is the product of the signs of the two components involved.}

\begin{table}[t]
\centering
\caption{\textcolor{smBlue}{Explicit form of $\mathbf{B}$ for the D3Q27 lattice by velocity class, using the Voigt ordering $\mathbf{S}^{\ast} = [S^{\ast}_{xx}, S^{\ast}_{yy}, S^{\ast}_{zz}, 2S^{\ast}_{xy}, 2S^{\ast}_{xz}, 2S^{\ast}_{yz}]^{\mathsf{T}}$. Entries $\pm2$ carry the sign of $c_{i\alpha}c_{i\beta}$; the face and edge classes are shown for the $x$ and $xy$ orientations; the other orientations follow by permutation of the indices.}}
\label{tab:Bneq_d3q27}
\scriptsize
{\color{smBlue}
\resizebox{\linewidth}{!}{
\begin{tabular}{l|cccccc}
velocity class ($\bm{c}_i$) & $B_{1i}$ & $B_{2i}$ & $B_{3i}$ & $B_{4i}$ & $B_{5i}$ & $B_{6i}$ \\ \hline
rest $(0,0,0)$          & $-c_s^2$   & $-c_s^2$   & $-c_s^2$   & $0$     & $0$     & $0$ \\
face $(\pm1,0,0)$       & $1-c_s^2$  & $-c_s^2$   & $-c_s^2$   & $0$     & $0$     & $0$ \\
edge $(\pm1,\pm1,0)$    & $1-c_s^2$  & $1-c_s^2$  & $-c_s^2$   & $\pm2$  & $0$     & $0$ \\
corner $(\pm1,\pm1,\pm1)$ & $1-c_s^2$ & $1-c_s^2$ & $1-c_s^2$  & $\pm2$  & $\pm2$  & $\pm2$
\end{tabular}
}}
\end{table}

\begin{figure*}[ht]
    \centering
    \includegraphics[width=\textwidth]{images/vn_stability_analysis_2D_3D.pdf}
     \caption{\textcolor{smOrange}{{2D and 3D VN Stability Comparison across Different Collision Models under Varying Viscosity.}} The maximum eigenvalue $\max|\lambda(\bm{G}(k))|$ of the amplification matrix is plotted as a function of streamwise wavenumber $k_x$ on D2Q9 (a--c) and D3Q27 (d--f), for $\nu = 10^{-3}$, $\nu = 10^{-4}$, and $\nu = 10^{-5}$ from left to right. We compare across four different collision models: BGK, RM-MRT, NOCM-MRT, and HOME-LBM.}
    \Description{}
    \label{fig:vn_stability_analysis}
\end{figure*}

\subsection{\textcolor{smBlue}{HOME-LBM}}
\label{app:amplification_home}
\textcolor{smBlue}{HOME-LBM applies the NOCM-MRT collision and reconstructs the post-collision populations from the resulting moments (\autoref{sec:home-lbm-analysis}),}
{\color{smBlue}
\begin{equation}
\bm{g}_{\text{HOME}}(\bm{f}) = \bm{f}^{\text{eq}} + \mathbf{Q}_{\bm{u}}\,\mathbf{B}\,\bigl[\bm{g}_{\text{NOCM}}(\bm{f}) - \bm{f}^{\text{eq}}\bigr],
\label{eq:g_home}
\end{equation}
}%
\textcolor{smBlue}{with $\bm{g}_{\text{NOCM}}(\bm{f}) = \bm{f} + \bm{\Omega}^{\mathrm{NOCM}}(\bm{f})$, $\mathbf{Q}_{\bm{u}} = \mathbf{A}_2 + \mathbf{A}_3\mathbf{R}_{\bm{u}}$ and $\mathbf{B}$ of Appendices~\ref{app:recursive_2d} and~\ref{app:Bneq_d2q9} (D2Q9) or~\ref{app:recursive_3d} and~\ref{app:Bneq_d3q27} (D3Q27). At the uniform equilibrium $\bm{g}_{\text{NOCM}}(\bar{\bm{f}}) = \bar{\bm{f}}$, so the derivative of $\mathbf{Q}_{\bm{u}}$ with respect to $\bm{f}$ does not contribute, and the linearized post-collision operator is}
{\color{smBlue}
\begin{equation}
\begin{aligned}
\bm{\Lambda}_{\text{HOME-LBM}} &= \bm{\Lambda} + \mathbf{Q}_{\bm{u}}\,\mathbf{B}\,(\bm{\Lambda}_{\text{NOCM-MRT}} - \bm{\Lambda}),\\
\bm{\Lambda}_{\text{NOCM-MRT}} &= \mathbf{I} - \mathbf{K}^{\mathrm{NOCM}}(\mathbf{I} - \bm{\Lambda}),
\end{aligned}
\label{eq:Lambda_home}
\end{equation}
}%
\textcolor{smBlue}{which is the operator inside the HOME-LBM amplification matrix of \autoref{sec:home-lbm-analysis}. Rearranged, it takes the form of \autoref{eq:G_general},}
{\color{smBlue}
\begin{equation}
\begin{aligned}
\mathbf{G}^{\text{HOME-LBM}}(\bm{k}) &= \mathbf{D}(\bm{k})\bigl[\mathbf{I} + \mathbf{K}^{\mathrm{HOME}}(\bm{\Lambda} - \mathbf{I})\bigr],\\
\mathbf{K}^{\mathrm{HOME}} &= \mathbf{I} - \mathbf{Q}_{\bm{u}}\,\mathbf{B}\,(\mathbf{I} - \mathbf{K}^{\mathrm{NOCM}}),
\end{aligned}
\label{eq:K_home}
\end{equation}
}%
\textcolor{smBlue}{so that the four collision models differ only in $\mathbf{K}$. On D2Q9, $\mathbf{Q}_{\bm{u}}\in\mathbb{R}^{9\times3}$, $\mathbf{B}\in\mathbb{R}^{3\times9}$ and $\bm{\Lambda}$ is the Jacobian of the equilibrium of Appendix~\ref{app:eq2d}; on D3Q27, $\mathbf{Q}_{\bm{u}}\in\mathbb{R}^{27\times6}$, $\mathbf{B}\in\mathbb{R}^{6\times27}$ and $\bm{\Lambda}$ follows from Appendix~\ref{app:eq3d}. This is the operator evaluated in \autoref{app:vn_analysis_results} and in the growth-rate maps of \autoref{sec:quantization}.}

\section{Numerical Spectral Analysis}
\label{app:vn_analysis_results}

To validate our theoretical stability analysis (\autoref{sec:home-lbm-analysis}) and compare the spectral behavior of different collision models, we conduct three sets of numerical experiments \textcolor{smBlue}{in both 2D and 3D using the D2Q9 and D3Q27 lattices, respectively.} We consider four collision schemes: single-relaxation-time Bhatnagar-Gross-Krook (BGK)~\cite{bhathnagor1954model, Chen-1998}, raw-moment multiple-relaxation-time (RM-MRT)~\cite{d1992generalized, lallemand2000theory}, the non-orthogonal central-moment multiple-relaxation-time (NOCM-MRT)~\cite{Rosis2017}, and HOME-LBM~\cite{li2023high}.

\begin{figure*}[ht!]
    \centering
    \includegraphics[width=\textwidth]{images/spectral_full.pdf}
    \caption{Maximal growth rate across collision models. The maximal imaginary part of the eigenvalues, $\max_j \operatorname{Im}(\omega_j)$, is visualized over the Fourier wavenumber space $(k_x, k_y)$ for four collision models (columns): BGK, RM-MRT, NOCM-MRT, and HOME-LBM, on D2Q9 (left) and D3Q27 (right). Rows (a--e) correspond to increasing background velocities $(u_x, u_y) = (0, 0)$, $(0.1, 0.1)$, $(0.2, 0.2)$, $(0.3, 0.3)$, and $(0.4, 0.4)$, with $u_z = 0$ on D3Q27. Warmer colors indicate positive growth rates (linearly unstable modes), cooler colors indicate negative growth rates (damped modes), and grey corresponds to a maximal growth rate of zero. Black contours denote the neutral stability boundary where $\max_j \operatorname{Im}(\omega_j) = 0$. \textcolor{smBlue}{For D3Q27, the plotted growth rate is additionally maximized over $k_z \in [0, \pi]$.}}
    \Description{}
    \label{fig:velocity_analysis_full}
\end{figure*}

\paragraph{Stability margin}
As shown in \autoref{fig:vn_stability_analysis}, we evaluate the maximum modulus of eigenvalues, max $|\lambda(\bm{k})|$, of the amplification matrix $\bm{G}$ as a function of the wave number $k_x$ at $\bm{u} = 0$, under three different kinematic viscosities: $\nu$ = $10^{-3}$, $10^{-4}$, and $10^{-5}$, \textcolor{smBlue}{on D2Q9 (a--c) and D3Q27 (d--f)}. According to the von Neumann criterion, stability requires $|\lambda(\bm{k})| \leq 1$ for all wave vectors, \textcolor{smBlue}{which all four models satisfy at rest on both lattices}. \textcolor{smOrange}{At $\bm{u}=\bm{0}$ and small $k_x$, the least-damped mode of every model is hydrodynamic, and its decay is set by the shear relaxation rate, which is identical for all models; consequently, RM-MRT, NOCM-MRT and HOME-LBM coincide along the $k_x$ axis and separate only near $k_x\to\pi$. BGK keeps $\max|\lambda|$ closer to unity for $k_x > 1.2$: with a single relaxation rate, its non-hydrodynamic modes are underdamped, which is the mechanism behind its early loss of stability at finite velocity discussed below.} Importantly, HOME-LBM closely overlaps with the full NOCM-MRT model across all tested cases on D2Q9 and D3Q27. \textcolor{smOrange}{The reason is structural: $\rho$ and $\bm{u}$ are conserved by the collision, and the zeroth- to second-order central moments produced by the NOCM-MRT collision depend only on the zeroth- to second-order central moments before the collision; hence the linearized HOME-LBM operator does not depend on equilibrium components beyond second order, and the recursive closure of \autoref{eq:a3_neq} (\autoref{eq:a3_neq_3d} on D3Q27) reproduces the third-order moments of the NOCM-MRT post-collision state exactly. Moment encoding and third-order Hermite reconstruction therefore introduce negligible additional instability at rest, in two or three dimensions.} This confirms that the distribution reconstruction in HOME-LBM maintains robust numerical performance without sacrificing the baseline stability properties of the underlying collision model.

\begin{figure}[ht!]
	\centering
    \includegraphics[width=0.8\columnwidth]{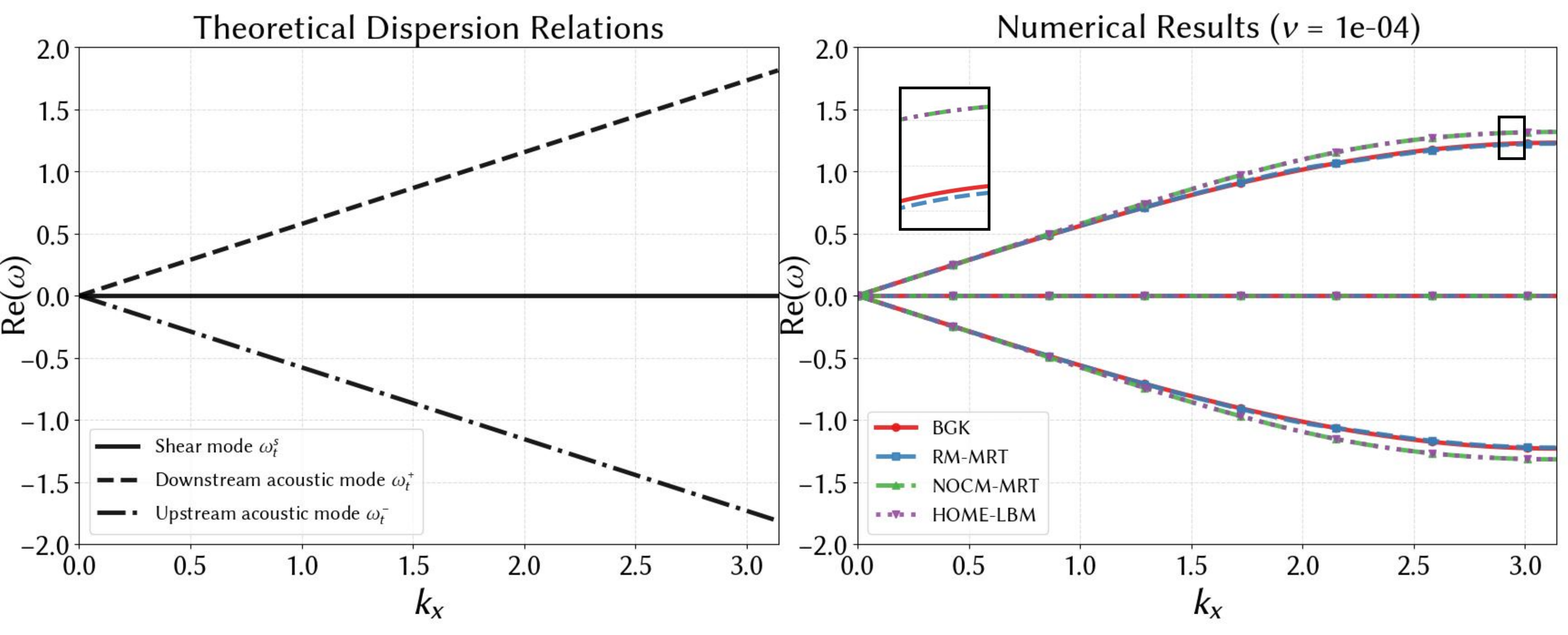} 
	\caption{{Comparison of Dispersion across Different Collision Models.} Left: theoretical dispersion relations of the shear mode and the upstream and downstream acoustic modes. Right: numerical dispersion curves on D2Q9 at $\nu = 10^{-4}$ for BGK, RM-MRT, NOCM-MRT, and HOME-LBM, obtained from eigenvalue analysis.} 
	\label{fig:dispersion_comparison}
    \Description{}
\end{figure}

\paragraph{Growth-rate maps and velocity thresholds.}
\textcolor{smBlue}{
\autoref{fig:velocity_analysis_full} shows the maximal growth rate over the wavenumber domain for increasing background velocity $\bm{u} = (a,a)$ on D2Q9, and $(a,a,0)$ on D3Q27, with $a\in\{0,0.1,0.2,0.3,0.4\}$ at $\nu = 10^{-4}$. For every eigenvalue $\lambda$ of $\mathbf{G}(\bm{k})$ we write $\omega = \mathrm{i}\ln\lambda$, so that $\operatorname{Im}(\omega) = \ln|\lambda|$ is the growth rate per time step and $\operatorname{Im}(\omega) > 0$ denotes a linearly unstable mode. On D2Q9, the wave vector is sampled on $(k_x,k_y)\in[0,\pi]^2$. On D3Q27, the two-dimensional maps report the worst case over the third component, $Z(k_x,k_y) = \max_{k_z\in[0,\pi]}\max_j\operatorname{Im}(\omega_j)$. Grey denotes a maximal growth rate of zero (neutrally stable), blue denotes damped modes, and red regions enclosed by black contours contain growing modes, with the color giving their growth rate per step.} 
\textcolor{smOrange}{As shown in \autoref{fig:velocity_analysis_full}, on both lattices, BGK and RM-MRT develop growing modes early, whereas NOCM-MRT and HOME-LBM remain linearly stable over most of the wave-number domain up to $a = 0.3$. HOME-LBM develops a few localized unstable islands at $a = 0.3$ ($|\bm{u}| = 0.42$, maximal growth $0.07$ on D2Q9 and $0.06$ on D3Q27).} The observed growth-rate maps directly provide practical reference for selecting safe quantization ranges for the stored momentum variables in \autoref{sec:quantization}. \textcolor{smOrange}{At $a = 0.3$ ($|\bm{u}| = 0.42$), the unstable islands of HOME-LBM are small, and their growth rate is modest. At $a = 0.4$ ($|\bm{u}| = 0.57$), the maps are unstable over most of the domain. The maps thus place the transition from isolated to widespread growth between $|u| = 0.42$ ($a = 0.3$) and $|u| = 0.57$ ($a = 0.4$) for a diagonal background velocity, and the empirical test of~\autoref{fig:3d_sphere_vel_tol}, with an axis-aligned inlet, is stable at $|u| = 0.4$ and diverges at 0.45, a threshold that lies inside this band. The per-component storage range $u \in [-0.4, 0.4]$ (i.e., the momentum range $[-0.4\rho_{\max}, 0.4\rho_{\max}]$ of Sec.~5.2) therefore covers all velocities of magnitude up to 0.4, which lie on the slow-growth side of the transition; velocities with all components near 0.4 simultaneously lie beyond it and are not sustained by the scheme regardless of storage precision. The weak growth that the linear analysis predicts for isolated wave vectors between $|\bm{u}|\approx0.3$ and $0.4$ is not observed in practice because, in a turbulent flow, such velocities occur only locally and transiently, whereas the linear prediction assumes a uniform background flow sustained indefinitely.} \textcolor{smBlue}{The D3Q19 Yeahright simulation (\autoref{fig:yeahright_3d}, right) also runs stably with the same storage ranges.}

\paragraph{Dispersion relations of hydrodynamic modes}
We further compare the numerical dispersion relations with the theoretical hydrodynamic modes obtained by linearizing the Navier--Stokes equations around a uniform background flow~\cite{jovanovic2001modeling}. The resulting eigenmodes consist of one shear mode $\omega_t^{s}$ \textcolor{smBlue}{(two in 3D)} and two acoustic modes $\omega_t^{\pm}$, whose analytical expressions are given by:
\begin{equation}
\omega_t^{s}
= |\bm{k}|\,|\bm{u}|\cos\phi
\;-\; i\,|\bm{k}|^2\,\nu,
\label{eq:shear_mode}
\end{equation}
\begin{equation}
\omega_t^{\pm}
= |\bm{k}|\bigl(|\bm{u}|\cos\phi \pm c_s\bigr)
\;-\; i\,|\bm{k}|^2
\left(
\frac{D-1}{D}\,\nu + \frac{\mu_v}{2}
\right),
\label{eq:acoustic_modes}
\end{equation}
where $\bm{k}$ denotes the wave vector, $\phi$ is the angle between the
perturbation direction and the background velocity $\bm{u}$, $c_s$ is the sound
speed, $\nu$ is the kinematic viscosity, $\mu_v$ is the bulk viscosity, and $D$
is the spatial dimension.

As shown in \autoref{fig:dispersion_comparison}~(right), we extract the numerical modes from the D2Q9 amplification matrix and plot $\mathrm{Re}(\omega)$ versus wave number $k_x$ for $\nu=10^{-4}$. The theoretical reference (left) consists of a zero-frequency shear branch and two acoustic branches with slopes $\pm c_s$ in the long-wavelength limit, given by~\autoref{eq:shear_mode} and~\autoref{eq:acoustic_modes}. Consistent with this prediction, all schemes agree closely at low wave numbers, indicating negligible dispersion error in the hydrodynamic regime. As $k_x$ increases, the
acoustic branches exhibit increasing deviation from the linear theory, reflecting
the expected numerical dispersion at high-frequency scales.

Across the full spectrum, HOME-LBM closely matches the NOCM-MRT baseline, with nearly indistinguishable curves, confirming that the third-order Hermite reconstruction does not introduce additional dispersion beyond that of the underlying collision model \textcolor{smBlue}{at $\bm{u}=\bm{0}$.}

\clearpage

\end{document}